\documentclass[12pt]{article}
\usepackage{mathrsfs}
\usepackage{fullpage}
\usepackage{amsfonts}
\usepackage{graphicx}
\usepackage{amsmath}
\usepackage{amssymb}
\usepackage{float}
\usepackage{subfig}
\usepackage{rotating}
\usepackage{xcolor}
\usepackage{natbib}
\usepackage{epstopdf}
\usepackage{enumerate}
\usepackage{url}
\usepackage[space]{grffile}
\usepackage{latexsym}
\usepackage{textcomp}
\usepackage{longtable}
\usepackage{tabulary}
\usepackage{booktabs,array,multirow}
\usepackage{epstopdf}
\usepackage{epsfig,fancyheadings}
\usepackage[]{hyperref}
\usepackage{sectsty, secdot}
\usepackage{amsthm}
\usepackage{bm}
\usepackage{setspace}
\usepackage{geometry,algorithm,algpseudocode}

\newtheorem{theorem}{Theorem}
\newtheorem{lemma}{Lemma}

\def\boxit#1{\vbox{\hrule\hbox{\vrule\kern6pt  \vbox{\kern6pt#1\kern6pt}\kern6pt\vrule}\hrule}}
\def\bse{\begin{eqnarray*}}
\def\ese{\end{eqnarray*}}
\def\be{\begin{eqnarray}}
\def\ee{\end{eqnarray}}
\def\bsq{\begin{equation*}}
\def\esq{\end{equation*}}
\def\bq{\begin{equation}}
\def\eq{\end{equation}}

\def\var{\hbox{var}}
\def\cov{\hbox{cov}}

\def\n{\nonumber}

\def\cov{\mbox{cov}}

\def\diag{\mbox{diag}}
\def\tr{\mbox{tr}}

\def\trans{^\top}

\def\bmv{{\boldsymbol\varepsilon}}

\def\B{{\bf B}}

\def\D{{\bf D}}

\def\f{{\bm f}}
\def\F{{\bf F}}

\def\I{{\bf I}}
\def\M{{\bf M}}

\def\K{{\bf K}}

\def\R{{\bf R}}

\def\U{{\bf U}}

\def\W{{\bf W}}

\def\H{{\bf H}}

\def\Y{{\bf Y}}

\def\Z{{\bf Z}}

\def\bms{{\bf \Sigma}}
\def\G{{\bf \Gamma}}

\def\diag{\hbox{diag}}

\def\diag{\hbox{diag}}
\def\log{\hbox{log}}

\def\squarebox#1{\hbox to #1{\hfill\vbox to #1{\vfill}}}

\def\cp{\mathop{\rightarrow}\limits^{p}}
\def\cd{\mathop{\rightarrow}\limits^{d}}
\def\0{{\bf 0}}
\def\1{{\bf 1}}

\def\E{\mathbb{E}}

\def\mR{\mathbb R}

\def\var{\hbox{var}}
\def\cov{\hbox{cov}}

\def\diag{\hbox{diag}}
\def\log{\hbox{log}}

\def\diag{\hbox{diag}}

\def\var{\mathrm {Var}}
\def\cov{\mathrm {Cov}}

\def\mA{\mathcal {A}}

\def\M{\mathcal{M}}
\def\diag{\mbox{diag}}

\def\I{{\bf I}}

\def\H{\mathbf{H}}

\def\cp{\mathop{\rightarrow}\limits^{p}}
\def\var{\mathrm {Var}}

\def\cov{\mathrm {Cov}}

\def\E{\mathbb {E}}

\def\hat{\widehat}

\allowdisplaybreaks[4]

\begin{document}

\title{Robust Mutual Fund Selection with False Discovery Rate Control}
\author{Hongfei Wang$^1$,  Long Feng$^2$, Ping Zhao$^2$ and Zhaojun Wang$^2$\\
$^1$Department of Statistics, Nanjing Audit University\\
$^2$School of Statistics and Data Science, KLMDASR, LEBPS, and LPMC,\\
 Nankai University}

\maketitle

\begin{abstract}
In this article, we address the challenge of identifying skilled mutual funds among a large pool of candidates, utilizing the linear factor pricing model. Assuming observable factors with a weak correlation structure for the idiosyncratic error, we propose a spatial-sign based multiple testing procedure (SS-BH). When latent factors are present, we first extract them using the elliptical principle component method \citep{he2022large} and then propose a factor-adjusted spatial-sign based multiple testing procedure (FSS-BH). Simulation studies demonstrate that our proposed FSS-BH procedure performs exceptionally well across various applications and exhibits robustness to variations in the covariance structure and the distribution of the error term. Additionally, real data application further highlights the superiority of the FSS-BH procedure.

\end{abstract}

\section{Introduction}
For investors and fund managers aiming to select funds with superior performance, the critical challenge lies in identifying those likely to maintain their performance in the future, given the vast pool of candidates. While many studies suggest active investment can be a ``negative-sum game" concerning investors' net returns, recent research \citep{barras2010false} demonstrates that some funds exhibit effective stock selection capabilities. Thus, measuring fund performance has garnered significant attention from both investors and academic researchers \citep{barras2010false,Fama1993Common}.

Pioneering works by \cite{jensen1968the} and \cite{sharpe1994the} introduced the Jensen alpha and Sharpe ratio criteria, which classify funds based on historical return data to identify those with exceptional performance. However, the effectiveness of these methods hinges on the assumption that historically well-performing funds will continue their superior performance. Traditional literature posits that fund performance is typically gauged by the adjusted intercept term of the factor model, alpha \citep{baks2001should,kosowski2006can}. 
To assess a single fund's performance, one can regress the fund's return series against a benchmark factor model and test whether the intercept term, alpha, is zero. If alpha exceeds zero and the associated $t$-test $p$-value is below the significance level $\gamma$, the fund is considered to have stock-picking ability. Moreover, \cite{barras2010false} highlighted that standard methods often overlook that the outstanding performance of some funds may be attributable to luck, raising questions about the validity of existing performance persistence tests \citep{fama2010luck}. Then, \cite{barras2010false} defined a ``lucky fund" as one whose estimated alpha is significant, but the real alpha is zero. Therefore, in practical applications, discerning whether a fund's superior performance stems from effective stock selection rather than random chance remains a vital area of investigation.

 Identifying skilled funds among numerous candidates necessitates hypothesis testing for each fund. However, \cite{berk2004mutual} noted that only a small fraction of funds have significantly non-zero alpha values, making false discoveries more likely in multiple hypothesis testing. For example, if there are 100 funds, and only 10 have genuine stock-picking abilities, setting a Type I error probability at 5\% could result in approximately 4 funds being mistakenly identified as skilled. Consequently, of the 14 funds considered skilled, nearly a third may be due to luck.

To identify truly predictive models from numerous candidates, researchers must employ a multiple hypothesis testing framework to address biases such as "data snooping" or "p-hacking," a significant challenge in the social sciences \citep{Sullivan1999,SULLIVAN2001,White2000,Hansen2005}. \cite{barras2010false}, \cite{bajgrowicz2012technical} and \cite{Harvey2015} proposed methods to control the false discovery rate (FDR), which is the proportion of incorrectly identified predictive models. To effectively control FDR, many classical methods assume that multiple test statistics are weakly correlated, implying that fund returns can be well explained by factor models and errors are weakly correlated among different funds \citep{Chamberlain1983}. Thus, several factor models have been proposed \citep{Sharpe1964,Lintner1965,Fama1993Common,Fama2015A}. However, in the past 20 years, there has been debate over the sufficiency of the linear asset pricing model in explaining fund returns \citep{kleibergen2015unexplained}. 

To address this issue, \cite{Giglio2018} proposed an improved BH procedure using matrix completion to detect positive alpha, which is more robust to omitted factors and missing data. Compared to the BH procedure, \cite{lan2019a} employed PCA to eliminate the influence of latent factors in FDR control, based on \cite{storey2004strong}'s multiple testing procedure, which is a more conservative approach. \cite{wang2023skilled} introduced a three-part mixed model to capture the dependence and non-normality in test statistics through a multiple testing optimal procedure based on each fund's probability of being unskilled, rather than $p$-values. However, these methods may not perform well with heavy-tailed distributions. To address this, \cite{Fengfactor} enhanced the ORSDA technique \citep{du2020false} to detect potential alpha and improve robustness against weak/strong factors and various error distributions. Nonetheless, ORSDA may lose power due to sample splitting, particularly when sample correlation is strong. Therefore, a multiple testing procedure more resilient to potential factors' strength and heavy-tailed distributions is necessary.

Recently, \cite{zhao2024} proposed a spatial sign-based test statistic for the global alpha test under the linear pricing factor model. Building on this method,  we developed a spatial sign-based multiple testing procedure (SS-BH) that assumes a weak correlation structure among the observable factors of idiosyncratic errors. Compared to the sample mean method, the spatial sign-based method is better suited for heavy-tailed distributions and exhibits greater robustness. Additionally, when unobservable factors are present in the model, we utilize the elliptical principal component method proposed by \cite{he2022large} to extract latent factors and loadings, and propose a spatial sign-based factor adjustment multiple testing procedure (FSS-BH). This approach mitigates the issue of poor PCA results due to heavy-tailed distributions.
We validate the advantages of these two proposed methods through extensive Monte Carlo experiments. The empirical analysis results indicate that the second method can identify a small number of technically sophisticated funds whose performance is more sustainable than that of their main competitors.

The remainder of the paper is structured as follows. In Section \ref{sec2}, we describe the multiple testing problem under LFPM and introduce the spatial sign-based multiple testing procedure (SS-BH), detailing its theoretical properties. In Section \ref{sec3}, we propose a factor-adjusted spatial sign-based multiple testing procedure based on the elliptical principal components method (FSS-BH) when a latent factor structure exists, and examine its theoretical properties. Extensive Monte Carlo experiments are performed in Section \ref{sec:simu}, followed by an empirical application in Section \ref{sec:real}. We conclude the paper with discussions in Section \ref{sec:conclusion}, and provide the technical proofs in the Appendix.

\textsc{Notations.}
For a random vector $\boldsymbol{X}=\left(X_1, \ldots, X_p\right)^{\top}$ following an elliptical distribution, denoted by $\boldsymbol{X} \sim E D(\boldsymbol{\mu}, \boldsymbol{\Sigma}, \zeta)$, we mean that
$
\boldsymbol{X}$ has the same distribution as $\boldsymbol{\mu}+\zeta \mathbf{A} \boldsymbol{U},$ where $\boldsymbol{\mu} \in \mathbb{R}^p, \boldsymbol{U}$ is a random vector uniformly distributed on the unit sphere $S^{q-1}$ in $\mathbb{R}^q, \zeta \geq 0$ is a scalar random variable independent of $\boldsymbol{U}, \mathbf{A} \in \mathbb{R}^{p \times q}$ is a deterministic matrix satisfying $\mathbf{A A}^{\top}=\boldsymbol{\Sigma}$ with $\boldsymbol{\Sigma}$ called scatter matrix whose rank is $q$. For any square matrix $\mathbf{A}\in\mathbb{R}^{p\times p}$, let $\lambda_{\min}(\mathbf{A})$ and $\lambda_{\max}(\mathbf{A})$ denote the smallest and largest eigenvalues of the matrix $\mathbf{A}$, respectively.
Additionally, let $\lambda_j(\mathbf{A})$ represent the $j$-th largest eigenvalue of the matrix $\mathbf{A}$.
\section{Robust Mutual Fund Selection Procedure}\label{sec2}
Let $N$ be the number of funds and $T$ be the time dimension of
the return series of each fund. Let the
$t$-th return of the $i$-th fund be $Y_{it}$. The Linear Factor Pricing Model (LFPM) is formulated as:
\begin{align}\label{eq:model}
Y_{it}=\alpha_i+\bm\beta_i^{\top} \f_t+\varepsilon_{it}
\end{align}
for $i=1,\dots, N$, $t=1, \dots, T$,
where
 $\f_t=(f_{t1}, \cdots, f_{tp})^{\top}\in\mR^p$ contains
 $p$ economic factors at time
 $t$,  $\alpha_i$ is a scalar representing the fund specific
intercept, $\bm\beta_i=(\beta_{i1},\cdots,\beta_{ip})^{\top}\in{\mR}^p$
is a vector of multiple regression betas with respect
to the $p$ factors and $\varepsilon_{it}$ is the corresponding
idiosyncratic error term with $\cov(\bmv_{t})=\bms=(\sigma_{ij})_{N\times N}$, where
$\bmv_{t}=(\varepsilon_{1t}, \cdots,
\varepsilon_{Nt})^{\top}\in\mR^N$ for each $t=1, \dots, T$ and $\bmv_{t}$'s
are independent and identically distributed.

The intercept term $\alpha_i$ in (\ref{eq:model}) represents the excessive return of the $i$-th fund. Apart from the returns associated with overall market factors, certain funds may exhibit systematic positive or negative returns due to their unique characteristics, which are referred to as excess returns. According to the definition provided in \cite{barras2010false}, a fund $i$ is considered skilled if it possesses a positive $\alpha_i$ \citep{Sharpe1964,Fama1993Common}. From an investor's perspective, identifying skilled funds holds significant investment value. To identify the skilled funds, the statistical problem of interest is to conduct $N$
simultaneous hypothesis testing problems for $N$ funds:
\begin{align}\label{eq:one}
H_{0i}: \alpha_i\le 0 ~~\text{versus}~~H_{1i}:\alpha>0,~~i=1,\cdots, N.
\end{align}
To address the aforementioned testing problem, numerous multiple testing procedures have been developed in the literature. These procedures employ statistical methods that directly control either the family-wise error rate or the False Discovery Rate (FDR) \citep{benjamini1995controlling,barras2010false,bajgrowicz2012technical}.

Specifically, we define $\theta_i = \mathbb{I}(\alpha_i > 0)$ for $i = 1, \cdots, N$, where $\mathbb{I}(\cdot)$ represents the indicator function. Here, $\theta_i = 0/1$ corresponds to a null/non-null variable. Let $\mA = \{i \in \{1, \cdots, N\}:\alpha_i > 0\}$ denote the set of non-nulls, and $\mA^c = \{1, \cdots, N\} \setminus \mA$ denote the null set. We also define $\bm\delta = (\delta_1, \cdots, \delta_N)^{\top}$ with $\delta_i \in \{0, 1\}$ as a decision rule of a multiple testing procedure, where $\delta_i = 1$ indicates that $H_{0i}$ is rejected and $\delta_i = 0$ otherwise. The set of funds with excess returns selected by the multiple testing procedure is denoted as $\widehat{\mA} = \{i \in \{1, \cdots, N\}: \delta_i = 1\}$. The False Discovery Proportion (FDP) and the True Discovery Proportion (TDP) are defined as follows:
\[
\text{FDP} = \frac{\sum_{i=1}^N (1-\theta_i)\delta_i}{(\sum_{i=1}^N\delta_j) \vee 1}, \quad\quad \text{TDP} = \frac{\sum_{i=1}^N \theta_i\delta_i}{(\sum_{i=1}^N\theta_i) \vee 1},
\]
where $a \vee b \equiv \max(a, b)$. The False Discovery Rate (FDR) is defined as the expectation of the FDP, i.e., $\text{FDR} \equiv \E(\text{FDP})$. The average power is defined as $\E(\text{TDP})$. A multiple testing procedure is considered superior to its competitor if it has a larger average power at the same FDR level.

For controlling the False Discovery Rate (FDR) in testing $N$ hypotheses as presented in \eqref{eq:one}, existing multiple testing procedures typically consist of two steps: (1) applying a simple $t$-test to the testing problem of each individual fund $i$ to obtain the corresponding $p$-value, and (2) applying a procedure to these $p$-values to control the FDR.

By the traditional least-square method, we could obtain an estimator of $\bm \alpha$:
\begin{align*}
\hat{\bm \alpha}=\omega_T^{-1} \Y \M_{\F} \1
\end{align*}
where $\Y=(\Y_1,\cdots,\Y_T)$, $\Y_t=(Y_{1t},\cdots,Y_{Nt})^\top$, $\f=(\f_1,\cdots,\f_T)^\top$, $\M_{\F}=\I_T-\f(\f^\top \f)^{-1}\f^\top$, $\1_T=(1,\cdots,1)^\top$ and $\omega_T=\1_T^\top \M_{\F}\1_T$. According to the Central Limit Theorem, we have
\begin{align}
{ \omega_T^{1/2}}(\hat{\bm\alpha}-\bm \alpha)\cd N(\bm 0,\bms)
\end{align}
Thus, the corresponding $t$-test statistic for hypothesis (\ref{eq:one}) is defined as
\begin{align*}
T^u_i=\frac{\omega_T^{1/2}\hat\alpha_i}{\hat \sigma_i}
\end{align*}
where $\sigma_i^2$ is the $i-$th diagonal element of $\hat{\bms}=T^{-1}\Y\M_{\tilde \f}\Y^\top$ with $\tilde \f=(\bm 1_T,\f)$ and $\M_{\tilde \f}=\I_T-\tilde \f(\tilde \f^\top \tilde \f)^{-1}\tilde \f^\top$. And the corresponding $p$-value for $H_{0i}$ versus $H_{1i}$ is $p^u_i=1-\Phi(T_i)$. Using the BH procedure directly for the FDR control, we
reject the null hypothesis $H_{0i}$ if $p_i^u\le
p^u_{(\widehat{k})}$. Here,  $\widehat{k}=\max\{i\in \{1,\cdots, N\}:
p^u_{(i)}\le \gamma i/N\}$, $p_{(i)}^u$ is the $i$-th order statistic
of $\{p_1^u,\cdots,p_N^u\}$ and $\gamma$ is the predetermined FDR
level.  This procedure is called the D-BH procedure hereafter,
where ``D'' stands for ``direct''.

Note that $T^{-1}\omega_T\hat{\bm \alpha}$ is the sample mean of $\{\Z_t\}_{t=1}^T$, $\Z=(\Z_1,\cdots,\Z_T)^\top=\M_{\F}\Y^\top $ where $\omega_T=\bm 1_T^\top \M_{\F}\bm
  1_T$. As known to all, the sample mean has good performance for light-tailed distributions, but is not very robust for heavy-tailed distributions. In contrast, the spatial median has very good performance for heavy-tailed distributions \citep{Oja2010Multivariate}.

So similar to \citet{Feng2016Multivariate}, we first estimate the spatial median and diagonal matrix $\D$ with the sample  $\{\Z_t\}_{t=1}^T$ by solving the following equations:
\begin{align}
&\frac{1}{T}\sum_{t=1}^T U(\D^{-1/2}(\Z_t-\bm \theta))=0\\
&\frac{1}{T}\sum_{t=1}^T\diag\{ U(\D^{-1/2}(\Z_t-\bm \theta))U(\D^{-1/2}(\Z_t-\bm \theta))^\top\}=\frac{1}{N}\I_N
\end{align}
Similary, we adopt the following algorithm to solve the above equations:
\begin{algorithm}
\caption{Spatial median estimator of $\Z$}\label{algo1}
\begin{algorithmic}[1]
\State Initialize $\bm{\theta}$ and $\mathbf{D}$ as the sample mean and variance of $\{\Z_t\}_{t=1}^T$;
\State $\boldsymbol  \xi_t \gets \D^{-1/2}(\Z_{t}-\bm \theta)$,
~~$t=1,\cdots,T$;
\State $\bm \theta \gets \bm \theta+\frac{\D^{1/2}\sum_{t=1}^T U(\bm\xi_t)}{\sum_{t=1}^T ||\bm\xi_t||^{-1}}$;
\State $\D \gets N
\D^{1/2}\diag\{T^{-1}\sum_{t=1}^{T}U(\boldsymbol  \xi_t)U(\boldsymbol  \xi_t)^\top \}\D^{1/2}$;
\State Repeat Steps 2-4 until convergence.
\end{algorithmic}
\end{algorithm}

Denote the result estimator as $\hat{\bm \theta}=(\hat{\theta}_1,\cdots,\hat{\theta}_N)^\top$ and $\hat{\D}^{1/2}=\diag\{\hat d_1,\cdots,\hat d_N\}$.

According to \cite{zhao2024}, we have
\begin{align}
T^{1/2}{ \hat{\varsigma}}^{1/2}\hat{\D}^{-1/2}(\hat{\bm \theta}-\hat{\omega} \bm \alpha)\cd N(\bm 0, N{\bf \Xi})
\end{align}
where ${\bf \Xi}=E(\U_t\U_t^\top)$, 
	$\U_t=U(\D^{-1/2}\bmv_t)$, $\hat{\omega}=T^{-1}\sum_{t=1}^{T}||\D^{-1/2}\bmv_t||^{-1}\vartheta_t/E(||\D^{-1/2}\bmv_t||^{-1})$, $\vartheta_t=1-\1_T^{\top} \f(\f^\top \f)^{-1}\f_t$,
	{ $\hat{\varsigma}=N(\hat{\zeta}_{-1})^2/\{1-2(1-\omega_T/T)\hat{\zeta}_{-1}\hat{\zeta}_{1}+(1-\omega_T/T)\hat{\zeta}_{2}(\hat{\zeta}_{-1})^2\}$, 
			$\hat{\zeta}_{-1}=T^{-1}\sum_{t=1}^T ||\tilde{\bmv}_{t}||^{-1}$,
			$\hat{\zeta}_1=T^{-1}\sum_{t=1}^T ||\tilde{\bmv}_{t}||$,
			$\hat{\zeta}_2=T^{-1}\sum_{t=1}^T ||\tilde{\bmv}_{t}||^{2}
$
	and $\tilde{\bm \varepsilon}_{t}=\hat\D^{-1/2}(\Z_t-\hat{\bm \theta})$}. Thus, we construct the following test statistic for each hypothesis $H_{0i}$ versus $H_{1i}$:
\begin{align}
T_i^s=T^{1/2}\hat \varsigma^{1/2}\hat{\theta}_i/\hat{d}_i
\end{align}
Then the corresponding $p$-value for $H_{0i}$ versus $H_{1i}$ is $p^s_i=1-\Phi(T_i^s)$. We also Use the Benjamini-Hochberg procedure for controlling FDR, we reject the null hypothesis $H_{0i}$ if $p_i^s \leq p^s_{(\widehat{k})}$. Here, $\widehat{k} = \max\{i \in \{1, \ldots, N\} : p^s_{(i)} \leq \gamma i/N\}$, where $p_{(i)}^s$ is the $i$-th order statistic of $\{p_1^s, \ldots, p_N^s\}$ and $\gamma$ is the predetermined FDR level. This procedure is called the SS-BH procedure hereafter,
where ``SS'' stands for ``Spatial-sign''.

Denote the true number of securities with positive alpha is $N_0$, and define $T^{-1}\sum_{t=1}^{T}\vartheta_t\cp \omega$. We need the following conditions:
\begin{itemize}
	\item[(C1)] The $p$-dimensional vector of common factors $\f_t$ is strictly stationary and
	distributed independently of the errors $\varepsilon_{it'}$,
	for all  $i=1,\cdots,N$ and all $t,t'=1,\cdots,T$. The
	number of factors $p$ is fixed
	and $\f_t\trans \f_t\le K<+\infty$, for a constant $K$ and all
	$t=1,\cdots,T$.
	The matrix $T^{-1}\tilde \f\trans\tilde \f$
	is positive definite, and as $T\to \infty$,
	$T^{-1}\1_T\trans\M_\F\1_T>\tau_{\min}$ for some positive constant
	$\tau_{\min}$.
	\item[(C2)] We consider the following model for error term:
	\begin{align}\label{modelx}
		\bmv_{ t}=v_t\mathbf{L} \boldsymbol{V}_t,
	\end{align}
	where $\boldsymbol V_t$ is a
	p-dimensional random vector with independent components, $\mathbb{E}(\boldsymbol V_t)=0$, $\mathbf \Sigma= \mathbf{L}\mathbf{L}^\top$, $v_i$ is a nonnegative univariate random variable and is  independent with the spatial sign of $\boldsymbol V_t$.
	\item[(C3)] (i) $V_{t, 1}, \ldots, V_{t, N}$ are i.i.d. symmetric random variables with $\mathbb{E}\left(V_{t, j}\right)=0, \mathbb{E}\left(V_{t, j}^2\right)=$ 1 , and $\left\|V_{t, j}\right\|_{\psi_\alpha} \leqslant c_0$ with some constant $c_0>0$ and $1 \leqslant \alpha \leqslant 2$. (ii) Let $\D$ is the diagonal matrix of $\bms$ and $r_t=||\D^{-1/2}\bmv_{ t}||$. The moments $\zeta_{-k}=\mathbb{E}\left(r_t^{-k}\right)$ for $k=1,2,3,4$ exist for large enough $N$. In addition, there exist two positive constants $\underline{b}$ and $\bar{B}$ such that $\underline{b} \leqslant \lim \sup_N \mathbb{E}\left(r_t / \sqrt{N}\right)^{-k} \leqslant \bar{B}$ for $k=1,2,3,4$.
	\item[(C4)] (i) The shape matrix $\R=\mathbf D^{-1/2}{ \mathbf{L}\mathbf{L}}^\top \mathbf D^{-1/2}=\left(\sigma_{j \ell}\right)_{N \times N}$ satisfies $\|\R\|_1 \leqslant a_0(N).$ We assume that $a_0(N)\asymp N^{1-\delta}$, ${\color{black}0<\delta\leq1/2}$,  $\log N=o(T^{1/5})$ and $\log T=o(N^{1/3 \wedge \delta})$. In addition, $\lim\inf_{N\rightarrow \infty}\min_{j=1,2,\cdots,N}{\color{black}d}_j>\underline{d}$ for some constant $\underline d>0$, where $\mathbf D=\operatorname{diag}\{d_1^2,d_2^2,\cdots,d_N^2\}$. (ii)   For some $\varrho \in(0,1)$, assume $|\sigma_{ij}|\leq \varrho$ for all $1\leq i<j\leq N$ and $N
	\geq 2$. Suppose $\left\{\delta_N ; N \geq 1\right\}$ and $\left\{\kappa_N ; N \geq 1\right\}$ are positive constants with $\delta_N=o(1 / \log N)$ and $\kappa=\kappa_N \rightarrow 0$ as $N \rightarrow \infty$. For $1 \leq i \leq N$, define $B_{N, i}=\left\{1 \leq j \leq N ;\left|\sigma_{i j}\right| \geq \delta_N\right\}$ and $C_N=\left\{1 \leq i \leq N ;\left|B_{N, i}\right| \geq N^\kappa\right\}$. We assume that $\left|C_N\right| / N \rightarrow 0$ as $N \rightarrow \infty$. (iii) In addition, $\sum_{j=1}^N \mathbb{I}\left(\sigma_{j \ell}\neq 0\right)=O\left(N^\eta\right)$ for some constant $0<\eta<(1-\varrho) /(1+\varrho)$.
\end{itemize}

\begin{theorem}\label{th1}
Suppose Conditions (C1)-(C4) and $\|\bm\alpha\|^2=O_p(NT^{-1}\mathrm{log} N)$ hold, and there exists $\mathcal{H} \subset\{1, \ldots, N\}$ such that $\mathcal{H}=\left\{i:T^{1/2} \varsigma^{1/2}\omega  \alpha_i/{d}_i \geqslant2\sqrt {\mathrm{log} N}\right\}$ and $|\mathcal{H}| \geqslant \mathrm{log} \mathrm{log} N \rightarrow \infty$ as $N \rightarrow \infty$, where $\varsigma=N\{E(r_t^{-1})\}^2/\eta_{\omega}$, $\eta_{\omega}=1-2(1-\omega) E(r_t^{-1})E(r_t)+(1-\omega) [E(r_s^{-1})]^2E(r_t^2)$. Assume that the number of false null hypotheses $N_1 \leqslant N^{\varpi}$ for some $0<\varpi<1$. Then, $FDR_{SS-BH}{ \leq \gamma N_0/N\leq \gamma}$ as $T \rightarrow \infty$.
\end{theorem}

\section{Factor-adjusted multiple testing Procedure}\label{sec3}
The D-BH procedure is only suitable for independent or weakly
correlated cases. Once the covariance matrix $\bms$ has large
off-diagonal elements, which reflects strong correlation between
different funds other than those explained by the common $\f_t$'s,
the resulting FDR control may be too conservative. To overcome this issue,
\cite{Giglio2018} and \cite{lan2019a} proposed factor-adjusted multiple
tests under the latent
factor models \citep{fan2012estimating}, where the error $\bmv_t$ is
assumed to have a latent factor structure, i.e.
\begin{align}\label{efactor}
\bmv_t=\G \W_t+\bm
\eta_{t}.
\end{align}
Here, $\W_t=(W_{1t},\cdots,W_{rt})^{\top} \in \mR^r$ is an
unknown low dimensional factor vector with the identification
restriction $\cov(\W_t)=\I_r$, $\G=(\bm \gamma_1,\cdots,\bm
\gamma_N)^{\top} \in \mR^{N\times r}$ is the corresponding factor
loading matrix composed of the unknown factor loadings $\bm
\gamma_i$'s and $\bm \eta_t=(\eta_{1t},\cdots,\eta_{Nt})^{\top} \in
\mR^N$ is a random error that is independent of $\W_t$ and
$\f_t$.

In \cite{lan2019a}, $\W_t$ and $\G$ are estimated by
$\tilde\W_t$ and $\tilde\G$ respectively, using the principal
component analysis method. Specifically, they first obtain the residual matrix
$\Z=\M_{\f}\Y^{\top}$. Let
$\tilde{\lambda}_k$ denote the $k$-th largest eigenvalue of scaled sample covariance matrix
$(TN)^{-1}\Z\Z^{\top}$,
and let $\tilde{\xi}_k$ denote the corresponding eigenvector. Then,
let $\tilde\W=(\tilde\W_1,\cdots,\tilde\W_T)^{\top}\equiv
T^{1/2}(\tilde{\xi}_1,\cdots,\tilde{\xi}_{\tilde{r}})$ and
$\tilde{\G}\equiv \tilde{\bm \varepsilon}^{\top} \tilde{\W}
(\tilde{\W}^{\top}\tilde{\W})^{-1}$, where $\tilde{r}$ is selected
by maximizing the eigenvalue ratios as $\tilde{r}=\arg\max_{k\le
  K_{\max}}\tilde{\lambda}_k/\tilde{\lambda}_{k+1}$ and $K_{\max}$ is
some pre-specified maximum possible order. In addition, the variance
of $\eta_{it}$, i.e. $\sigma_{\bm \eta, i}$, can be estimated by
$\tilde{\sigma}_{\bm \eta,i}=T^{-1}\tilde{\bm
  \eta_{i}}^{\top}\tilde{\bm \eta}_{i}$, where $\tilde{\bm
  \eta_{i}}$ is the $i$-th column of
$\M_{\tilde{\W}}\tilde{\bm \varepsilon}$, $\M_{\tilde \W}=\I_T-\tilde \W(\tilde \W^\top \tilde \W)^{-1}\tilde \W^\top$.

According to this latent factor structure, \cite{Giglio2018} and \cite{lan2019a}
used the factor-adjusted test statistic
$$T_i^a\equiv\left\{\omega_T^{1/2}\widehat{\alpha}_i-\omega_T^{-1/2}\bm
  1_T^{\top} \M_{\f}\widetilde{\W}\widetilde{\bm
    \gamma}_i\right\}/\widehat{\sigma}_{\eta,i},$$
where $\widetilde{\sigma}_{\eta,i}^2$, $\widetilde{\W}_i$ and $\widetilde{\bm
  \gamma}_i$ are the estimators of ${\sigma}_{\eta,i}^2$, ${\W}_i$ and
${\bm \gamma}_i$, respectively, and
$\widetilde{\W}=(\widetilde{\W}_1,\cdots,\widetilde{\W}_T)^{\top}$.
Then, \cite{Giglio2018} applied the BH procedure to the $p$-values
$p_i^a=1-\Phi(T_i^a)$ for the FDR control. This procedure is called
the F-BH procedure hereafter, where ``F'' stands for ``factor-adjusted''.

PCA exhibits notable drawbacks when dealing with heavy-tailed distributions. Firstly, it is highly sensitive to outliers, which are more prevalent in such distributions. These outliers can significantly distort the principal components, leading to biased results that fail to accurately reflect the underlying data structure. Secondly, variance is often higher in heavy-tailed distributions, and PCA may overestimate the importance of certain dimensions, potentially selecting irrelevant or noisy features. This can impact the accuracy and interpretability of the analysis. Furthermore, PCA is based on the assumption of multivariate normality, which heavy-tailed distributions clearly violate. This deviation can undermine the theoretical foundations of PCA, resulting in principal components that may not optimally capture the latent structure of the data. Lastly, the estimation of eigenvalues can become unstable when handling heavy-tailed distributions, affecting the accuracy of the derived principal components and compromising the robustness of PCA as a dimensionality reduction technique.

So we adopt the robust two-step estimation procedure proposed by \cite{he2022large}. First, we estimate the spatial Kendall's tau matrix of $\Z_t$ by
\begin{align*}
\K_Z=\frac{2}{T(T-1)}\sum_{i<j}U(\Z_i-\Z_j)U(\Z_i-\Z_j)^\top.
\end{align*}
The estimator of the factor loading matrix $\G$ as $\hat \G=\sqrt{N}(\hat{\xi}_1,\cdots,\hat{\xi}_{\hat{r}})$ where $\{\hat{\xi}_1,\cdots,\hat{\xi}_{\hat{r}}\}$ is the leading $\hat{r}$ eigenvectors of $\K_Z$. Here $\hat{r}$ is selected
by maximizing the eigenvalue ratios as $\hat{r}=\arg\max_{k\le
  K_{\max}}\hat{\lambda}_k/\hat{\lambda}_{k+1}$ and $\hat{\lambda}_k$ is the $k$-th largest eigenvalues of $\K_Z$ \citep{yu2019robust}. The factors in (\ref{efactor}) are estimated by the following least-square optimization
\begin{equation}
\widehat{\boldsymbol{W}_t}=\underset{\boldsymbol{\beta}_t \in \mathbb{R}^{\hat{r}}}{\arg \min } \sum_{i=1}^N\left(Z_{i t}-\widehat{\boldsymbol{\gamma}}_i^{\top} \boldsymbol{\beta}_t\right)^2, \quad t=1, \ldots, T,
\end{equation}
where $\hat{\bm \gamma}_i$ is the $i-$the row of $\hat{\G}$, i.e. $\hat{\bm \gamma}_i=\sqrt{N}\hat{\xi}_i$.

Secondly, we subtract the latent factor $\G \W_t$ from $\Y_t$, resulting in $\breve{\Y}_t = \Y_t - \hat{\G} \hat{\W}_t$. Define $\breve{\Z} = \M_{\F}\breve{\Y}^\top$, where $\breve{\Y} = (\breve{\Y}_1, \cdots, \breve{\Y}_T)$. Note that the test statistic $T_i^a$ corresponds to the $t$-test statistic of the $i$-th column of $\breve{Y}_t$. Akin to Section 2, the classical $t$-test is not robust against heavy-tailed distributions. Therefore, we also employ a spatial-sign-based test procedure to construct a more robust multiple testing procedure.

Then we estimate the spatial median and diagonal matrix $\B$ of the scatter matrix of $\bm \eta_t$ by Algorithm \ref{algo1}, denoted as $\breve{\bm \theta}=(\breve \theta_1,\cdots,\breve \theta_N)^\top$ and $\breve{\bm \B}=(\breve d_1,\cdots,\breve d_N)$. Thus, we construct the following test statistic for each hypothesis $H_{0i}$ versus $H_{1i}$:
\begin{align}
T_i^f=T^{1/2}\breve \varsigma^{1/2}\breve{\theta}_i/\breve{d}_i^{1/2}
\end{align}
where
{ $\breve{\varsigma}=N(\breve{\zeta}_{-1})^2/\{1-2(1-\omega_T/T)\breve{\zeta}_{-1}\breve{\zeta}_1+(1-\omega_T/T)\breve{\zeta}_2(\breve{\zeta}_{-1})^2\}$ and
	$\breve{\zeta}_2=T^{-1}\sum_{t=1}^T ||\breve{\bmv}_{t}||^2,$
		$\breve{\zeta}_{-1}=T^{-1}\sum_{t=1}^T ||\breve{\bmv}_{t}||^{-1},$
		$\breve{\zeta}_{1}=T^{-1}\sum_{t=1}^T ||\breve{\bmv}_{t}||$}
and $\breve{\bm \varepsilon}_{t}=\breve\D^{-1/2}({ \breve{\Z}_t}-\breve{\bm \theta})$. Then the corresponding $p$-value for $H_{0i}$ versus $H_{1i}$ is $p^s_i=1-\Phi(T_i^s)$. We also Use the Benjamini-Hochberg procedure for controlling FDR, we reject the null hypothesis $H_{0i}$ if $p_i^f \leq p^f_{(\widehat{k})}$. Here, $\breve{k} = \max\{i \in \{1, \ldots, N\} : p^f_{(i)} \leq \gamma i/N\}$, where $p_{(i)}^f$ is the $i$-th order statistic of $\{p_1^f, \ldots, p_N^f\}$ and $\gamma$ is the predetermined FDR level. This procedure is called the FSS-BH procedure hereafter,
where ``FSS'' stands for ``Factor adjusted Spatial-sign''.

We need the	following technical conditions.
\begin{itemize}
	\item[(C5)] We assume that
	 \begin{align*}
		&\binom{\boldsymbol{W}_t}{\boldsymbol{\eta}_t}=\nu_t\left(\begin{array}{cc}
			\mathbf{I}_r & \mathbf{0} \\
			\mathbf{0} & \mathbf{A}
		\end{array}\right) \frac{\boldsymbol{g}_t}{\left\|\boldsymbol{g}_t\right\|}
	\end{align*}
	where $\nu_t$ 's are independent samples of a scalar random variable $\nu$, and $\boldsymbol{g}_t$ 's are independent Gaussian samples from $\boldsymbol{g} \sim$ $\mathcal{N}\left(\mathbf{0}, \mathbf{I}_{r+N}\right) . m$ is fixed. Further, $\nu$ and $\boldsymbol{g}$ are independent and $\nu / \sqrt{N}=O_p(1)$ as $N\rightarrow \infty$. Therefore, $\left(\boldsymbol{W}_t^{\top}, \boldsymbol{\eta}_t^{\top}\right)^{\top}$ are independent samples from $\operatorname{ED}\left(\mathbf{0}, \boldsymbol{\Sigma}_0, \nu\right)$ for $t=1, \ldots, T$ where $$\boldsymbol{\Sigma}_0=\left(\begin{array}{cc}\mathbf{I}_r & \mathbf{0} \\ \mathbf{0} & \boldsymbol{\Sigma}_\eta\end{array}\right),$$ and $\boldsymbol{\Sigma}_\eta=\mathbf{A A}^{\top}$. To make the model identifiable, we further assume that $\left\|\operatorname{diag}\left(\boldsymbol{\Sigma}_0\right)\right\|_{\infty}=1$.
	
	\item[(C6)] Assume $\mathbf{\Gamma}^{\top} \mathbf{\Gamma} / N\rightarrow \mathbf{V}$ as $p \rightarrow \infty$, where $\mathbf{V}$ is a positive definite matrix. There exist positive constants $c_1, c_2$ such that $c_2 \leq \lambda_m(\mathbf{V})<\cdots<\lambda_1(\mathbf{V}) \leq c_1$ and $c_2 \leq \lambda_{\min }\left(\boldsymbol{\Sigma}_\eta\right) \leq \lambda_{\max }\left(\boldsymbol{\Sigma}_\eta\right) \leq c_1$. { Define $m_{t,s}=||\bm\varepsilon_{ t}-\bm\varepsilon_{ s}||$. The moments $\mathbb{E}\left(m_{t,s}^{-k}\right)$ for $k=2,4$ exist for large enough $N$. In addition, $ \mathbb{E}\left(m_{t,s} / \sqrt{N}\right)^{-k} \asymp 1$ for $k=2,4$. }
	
	\item[(C7)] We consider the following model for error term:
	\begin{align}\label{modelx}
		\bm \eta_{ t}=v_t\mathbf{L} \boldsymbol{V}_t,
	\end{align}
	where $\boldsymbol V_t$ is a
	p-dimensional random vector with independent components, $\mathbb{E}(\boldsymbol V_t)=0$, $\mathbf \Sigma= \mathbf{L}\mathbf{L}^\top$, $v_i$ is a nonnegative univariate random variable and is  independent with the spatial sign of $\boldsymbol V_t$.
	\item[(C8)] (i) $V_{t, 1}, \ldots, V_{t, N}$ are i.i.d. symmetric random variables with $\mathbb{E}\left(V_{t, j}\right)=0, \mathbb{E}\left(V_{t, j}^2\right)=$ 1 , and $\left\|V_{t, j}\right\|_{\psi_\alpha} \leqslant c_0$ with some constant $c_0>0$ and $1 \leqslant \alpha \leqslant 2$. (ii) Let $\D$ is the diagonal matrix of $\bms$ and $r_t=||\D^{-1/2}\bm \eta_{ t}||$. The moments $\zeta_{-k}=\mathbb{E}\left(r_t^{-k}\right)$ for $k=1,2,3,4$ exist for large enough $N$. In addition, there exist two positive constants $\underline{b}$ and $\bar{B}$ such that $\underline{b} \leqslant \lim \sup_N \mathbb{E}\left(r_t / \sqrt{N}\right)^{-k} \leqslant \bar{B}$ for $k=1,2,3,4$.
	\item[(C9)] Assume that $T\log N=o(N)$. For any constant $C_\alpha>0$, $\|\boldsymbol{\alpha}\| \leq C_\alpha  N^{1 / 2} / T^{1 / 2}/\sqrt{\log N}$. In addition, there exists some positive constant $\gamma_{\max}$ such that $\max_i \|\bm \gamma_i\|^2\leq \gamma_{\max}$. 
	\end{itemize}

\begin{theorem}\label{th2}
Suppose Conditions (C1), (C4) and (C5)-(C8) hold, and there exists $\mathcal{H} \subset\{1, \ldots, N\}$ such that $\mathcal{H}=\left\{i:T^{1/2} \varsigma^{1/2}\omega \alpha_i/{d}_i \geqslant2\sqrt {\mathrm{log} N}\right\}$ and $|\mathcal{H}| \geqslant \mathrm{log} \mathrm{log} N \rightarrow \infty$ as $N \rightarrow \infty$, where $\varsigma=N\{E(r_t^{-1})\}^2/\eta_{\omega}$, $\eta_{\omega}=1-2(1-\omega) E(r_t^{-1})E(r_t)+(1-\omega) [E(r_s^{-1})]^2E(r_t^2)$. Assume that the number of false null hypotheses $N_1 \leqslant N^{\varpi}$ for some $0<\varpi<1$. Then, $FDR_{FSS-BH}\leq { \gamma N_0/N\leq \gamma}$ as $T \rightarrow \infty$. 
\end{theorem}

\section{Simulation}\label{sec:simu}
The Monte Carlo experiments are designed to mimic the commonly used Fama-French three-factor model \citep{Fama1993Common}. The response variables $Y_{it}$ are generated according to the Linear Factor Pricing Model (LFPM) in (\ref{eq:model}) with $p=3$, as follows:
\begin{align*}
Y_{it}=\alpha_i+\sum_{j=1}^p \beta_{ij} f_{tj} +\varepsilon_{it},
\end{align*}
where the three factors, $f_{t1}$, $f_{t2}$, and $f_{t3}$, correspond to the Fama-French three factors. The factor loadings $\f_t=(f_{1t},f_{2t},f_{3t})^\top$ are generated from a multivariate normal distribution $N(\bm \mu_{\f},\bms_{\f})$, where $\bm \mu_{\f}=(\mu_{f1},\mu_{f2},\mu_{f3})^\top$ with $\mu_{fi}\sim U(0,1)$ for $i=1,2,3$, and $\bms_{\f}=(0.5^{|i-j|})_{1\le i,j\le 3}$. The coefficients $\beta_{i1}$, $\beta_{i2}$, and $\beta_{i3}$ corresponding to the three factors are generated independently from $U(0.2,2)$, $U(-1,1.5)$, and $U(-1.5,1.5)$, respectively.

Three scenarios of covariance structures for $\bm \varepsilon_{t}$ are considered:
\begin{itemize}
\item[(I)] (no latent factor) $\bms=(\rho^{|i-j|})_{1\le i,j\le N}$;
\item[(II)] (one latent factor) $\bms=\frac{1}{\sqrt{2}}(\bms_1+\bms_2)$, where $\bms_1=(\rho^{|i-j|})_{1\le i,j\le N}$ and $\bms_2=(1-\rho)\I_N+\rho \bm 1_N\bm 1_N^\top$.
\item[(III)] (two latent factor with one weak factor)  $\varepsilon_{it}=0.5z_{1t}+\zeta_iz_{2t}+\varepsilon_{it}$, where
  $z_{1t}\sim N(0,1)$, $z_{2t}\sim t(3)/\sqrt{3}$, $\zeta_i\sim
  U(0,1)$ for $i\in A$, $\zeta_i=0$ for $i\not\in A$, $A$ is a random subset of $\{1,\cdots,N\}$ with $|A|=[N^{0.5}]$. The covariance matrix of $\bmv_t$ is $\bms=(\rho^{|i-j|})_{1\le i,j\le N}$.
\end{itemize}
For Scenario I and II, we set $\varepsilon_{it}=\varepsilon_{it}$ for all $i, t$.
The error terms $\bm \varepsilon_{t}$ are generated from four different distributions:
\begin{itemize}
\item[(i)] Multivariate normal distribution: $\bm \varepsilon_t\sim N(\bm 0,\bms)$.
\item[(ii)] Multivariate $t$-distribution: $\bm \varepsilon_{t}$ are generated from standardized $t_{{ N},3}/\sqrt{3}$ with mean zero and scatter matrix $\bms$.
\item[(iii)] Multivariate mixture normal distribution: $\bm \varepsilon_{t}$ are generated from standardized $[\kappa N(\bm 0,\bms)+(1-\kappa)N(\bm 0,9\bms)]/\sqrt{\kappa+9(1-\kappa)}$, denoted by $\mbox{MN}_{p_n,\gamma,9}$, with $\kappa=0.8$.
\item[(iv)] Independent Component Model: $\bm \varepsilon_t=\bms^{1/2}\bmv_t$, where $\bmv_t=(\varepsilon_{t1},\cdots,\varepsilon_{tN})^\top$ and $\varepsilon_{ti}, i=1,\cdots,N$ are all independent and identically distributed as $(3-\chi_3^2)/\sqrt{6}$.
\end{itemize}

We consider the setting $(N,\pi_0,\gamma)=(200,0.1,0.1)$, where $\pi_0=\sum_{i=1}^N \theta_i/N$. Specifically, we randomly choose $\mA_{+}, \mA_{-}\subseteq \{1,\cdots,N\}$ with $|\mA_{+}|=|\mA_{-}|=\pi_0N$. We consider two sample sizes, $T=60$ and $T=120$. The intercepts $\alpha_i$ are generated independently for each $i\in \mA_{+}$ from $U(0.15,0.3)+\delta$ and for each $i\in \mA_{-}$ from $U(-0.3,-0.2)$.

\begin{figure}[ht]
\centering
\caption{The FDP and TDP with varying $\delta$ under Scenario I (no latent factor). \label{fig1}}
\subfloat[$T=60,N=200$]{\includegraphics[width=1\textwidth]{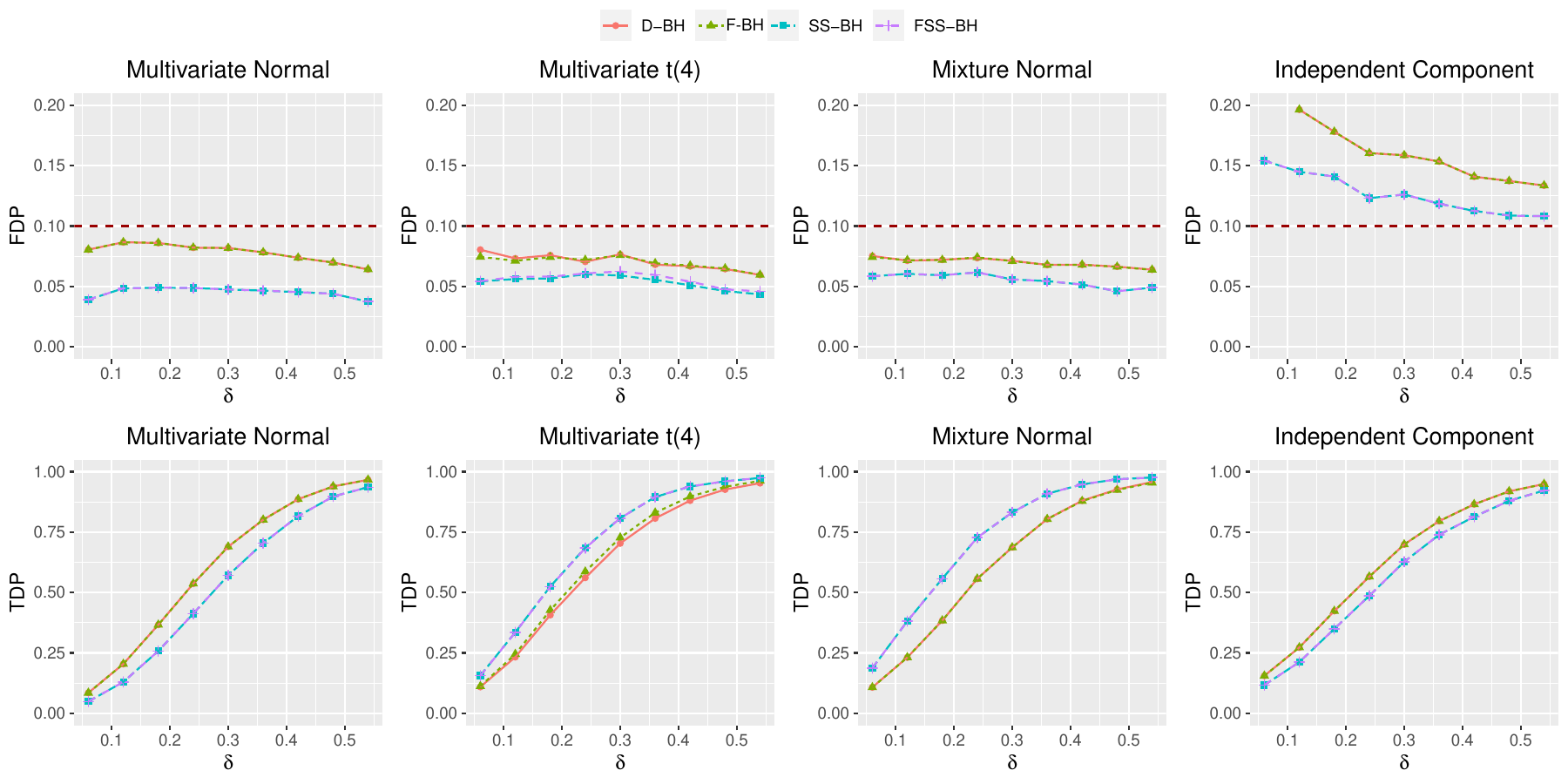}}\\
\subfloat[$T=120,N=200$]{\includegraphics[width=1\textwidth]{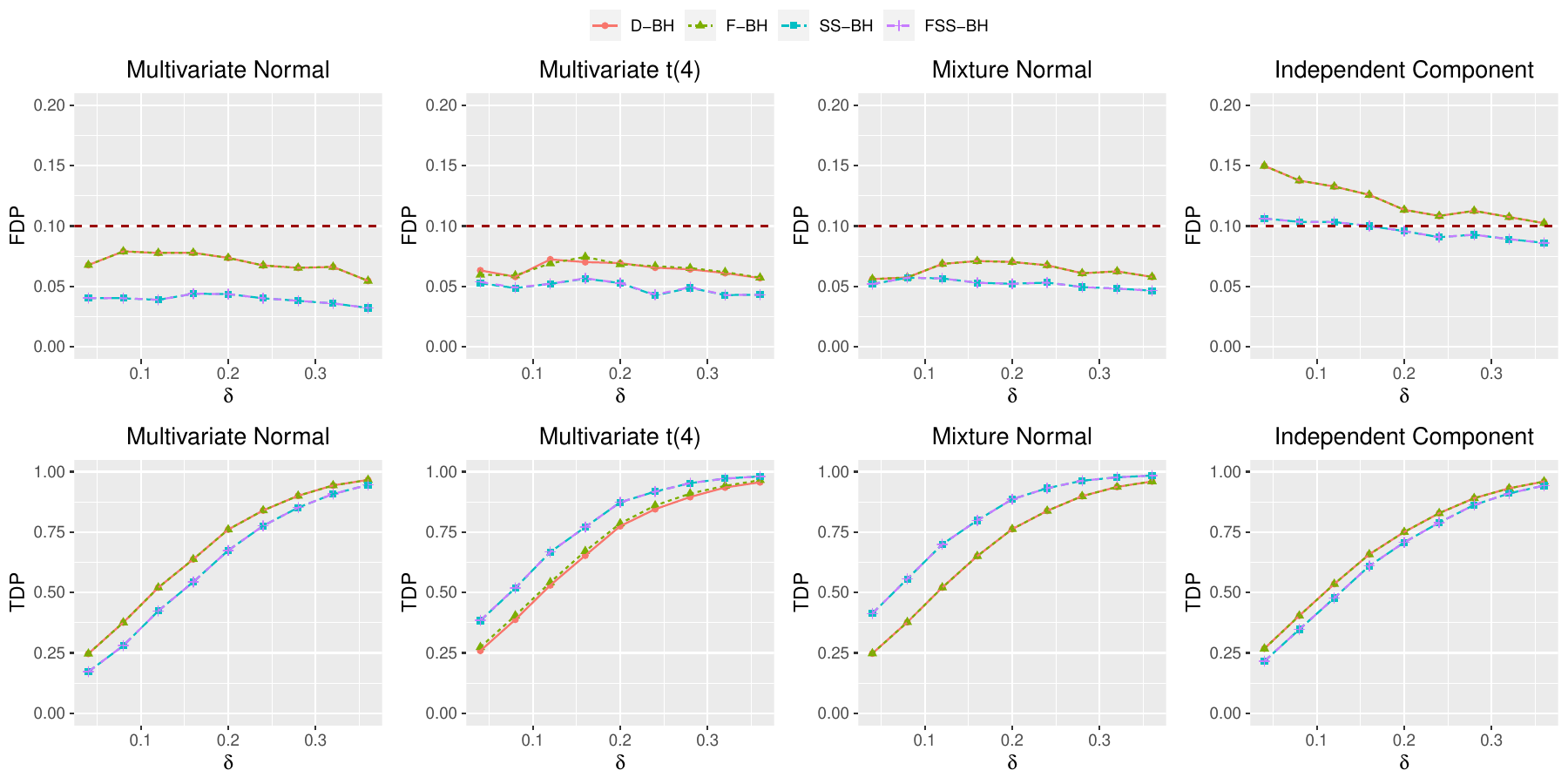}}
\end{figure}

Figures \ref{fig1} presents the FDP and TDP of the four methods with $\rho=0.5$ under Scenario I. In the absence of latent factors, the factor-adjusted methods, F-BH and FSS-BH, exhibit performance comparable to their direct counterparts, D-BH and SS-BH, respectively. This outcome is expected. For multivariate normal distributions, both D-BH and SS-BH effectively control the FDP in all instances, with D-BH demonstrating a slightly higher TDP than SS-BH. This is unsurprising, given that the D-BH method is optimized for normal distributions. However, in the context of heavy-tailed distributions, such as the multivariate t(4) distribution and the mixture normal distribution, our proposed SS-BH method exhibits a higher TDP than D-BH, highlighting the robustness of our approach. Lastly, when dealing with an independent component model characterized by a nonsymmetric distribution, D-BH maintains control over the FDP in all scenarios. Nevertheless, our proposed SS-BH method is capable of controlling the FDP as the signal intensity increases (large $\delta$ or $T$). Consequently, our method proves applicable even to nonsymmetric distributions, despite our theoretical assumptions being grounded in the symmetry of the error term.

\begin{figure}[ht]
\centering
\caption{The FDP and TDP with varying $\delta$ under Scenario II (one latent factor). \label{fig2}}
\subfloat[$T=60,N=200$]{\includegraphics[width=1\textwidth]{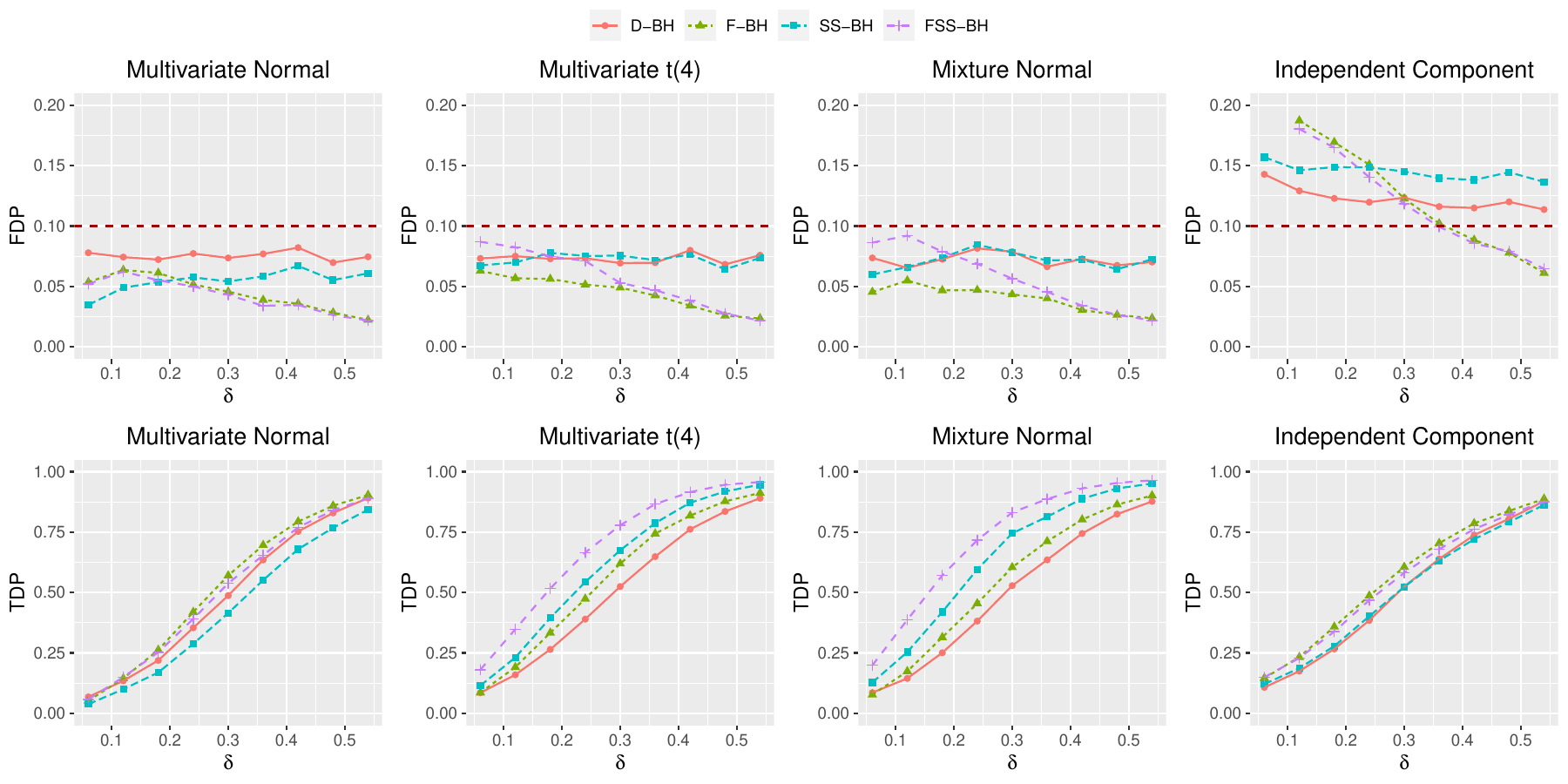}}\\
\subfloat[$T=120,N=200$]{\includegraphics[width=1\textwidth]{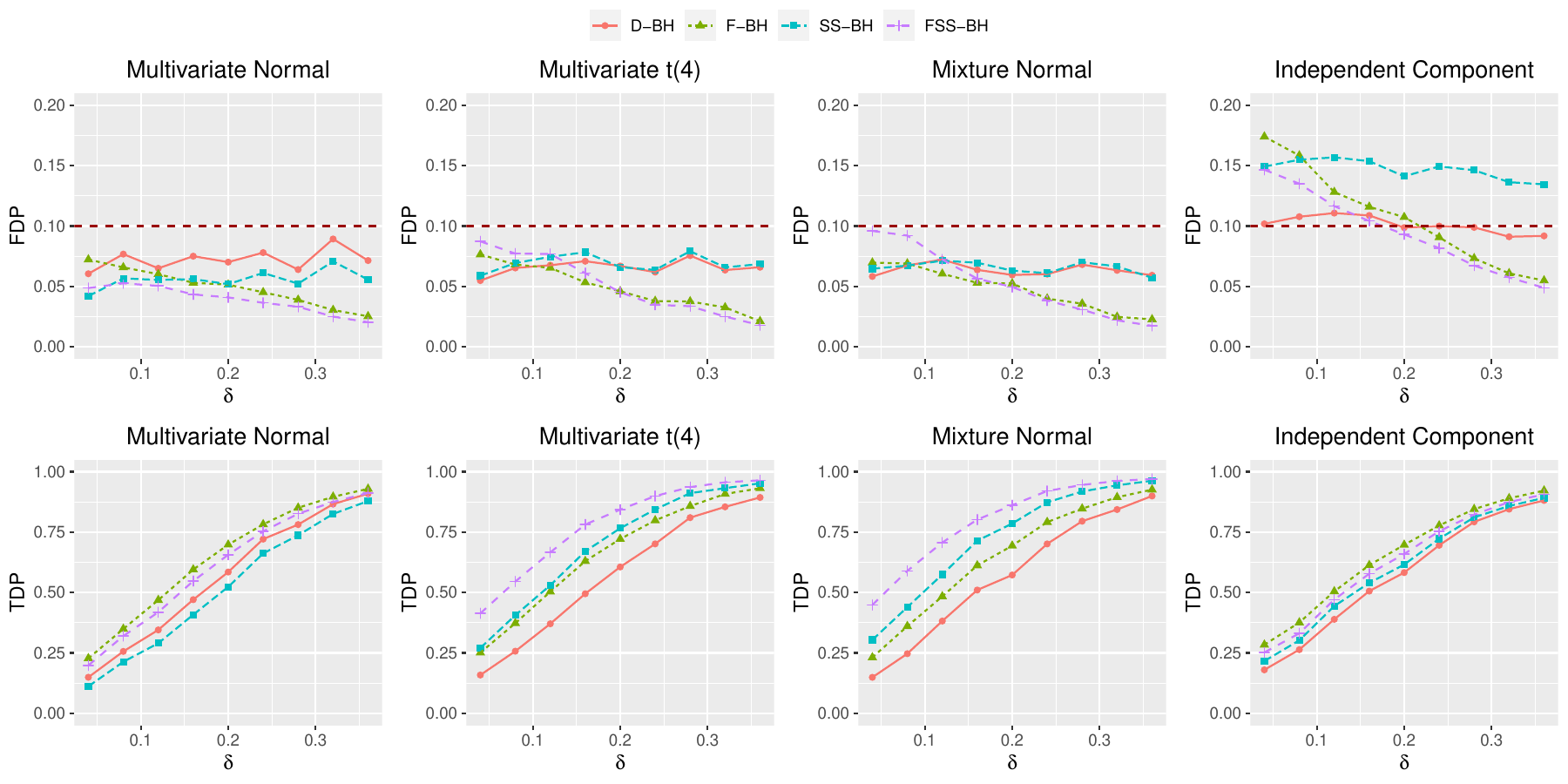}}
\end{figure}

Figure \ref{fig2} presents the FDP and TDP of the four methods under Scenario II with $\rho=0.5$. In the presence of latent factors, factor-adjusted methods outperform their direct counterparts. Similar to Scenario I, the spatial sign-based methods SS-BH and FSS-BH exhibit higher TDP than D-BH and F-BH for heavy-tailed distributions in cases (ii) and (iii), with FSS-BH achieving the highest TDP in these two cases. In case (iv), the direct methods D-BH and SS-BH may not control the FDP in some cases, while the factor-adjusted methods F-BH and FSS-BH can control the FDP when the signal gets larger. The TDP of FSS-BH has similar performance as F-BH in all cases. This shows that, even when the model assumptions are not fully met, our proposed FSS-BH still exhibits very remarkable performance, demonstrating the robustness of our methods.

\begin{figure}[ht]
\centering
\caption{The FDP and TDP with varying $\delta$ under Scenario III (two latent factors with one weak factor). \label{fig22}}
\subfloat[$T=60,N=200$]{\includegraphics[width=1\textwidth]{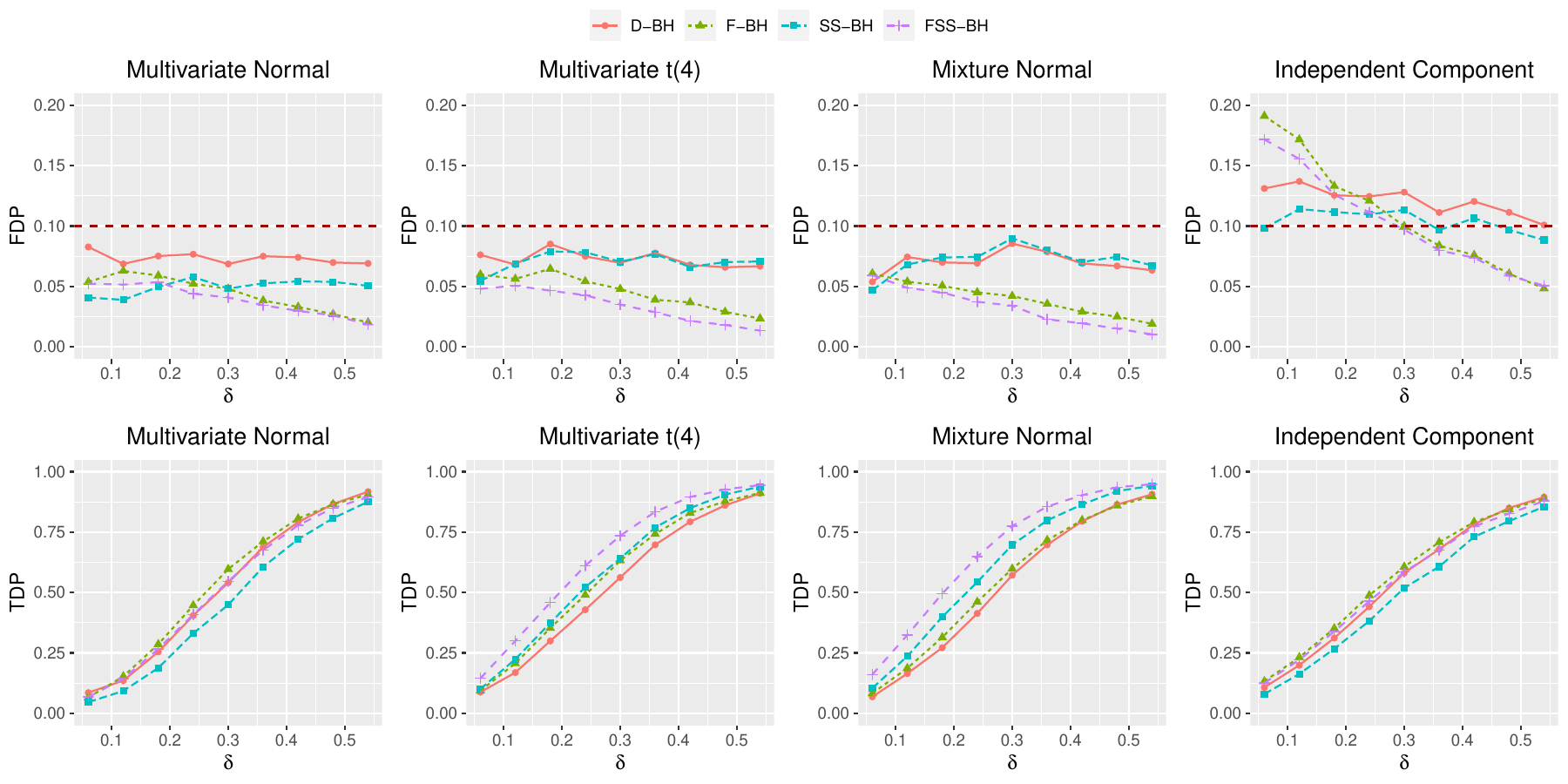}}\\
\subfloat[$T=120,N=200$]{\includegraphics[width=1\textwidth]{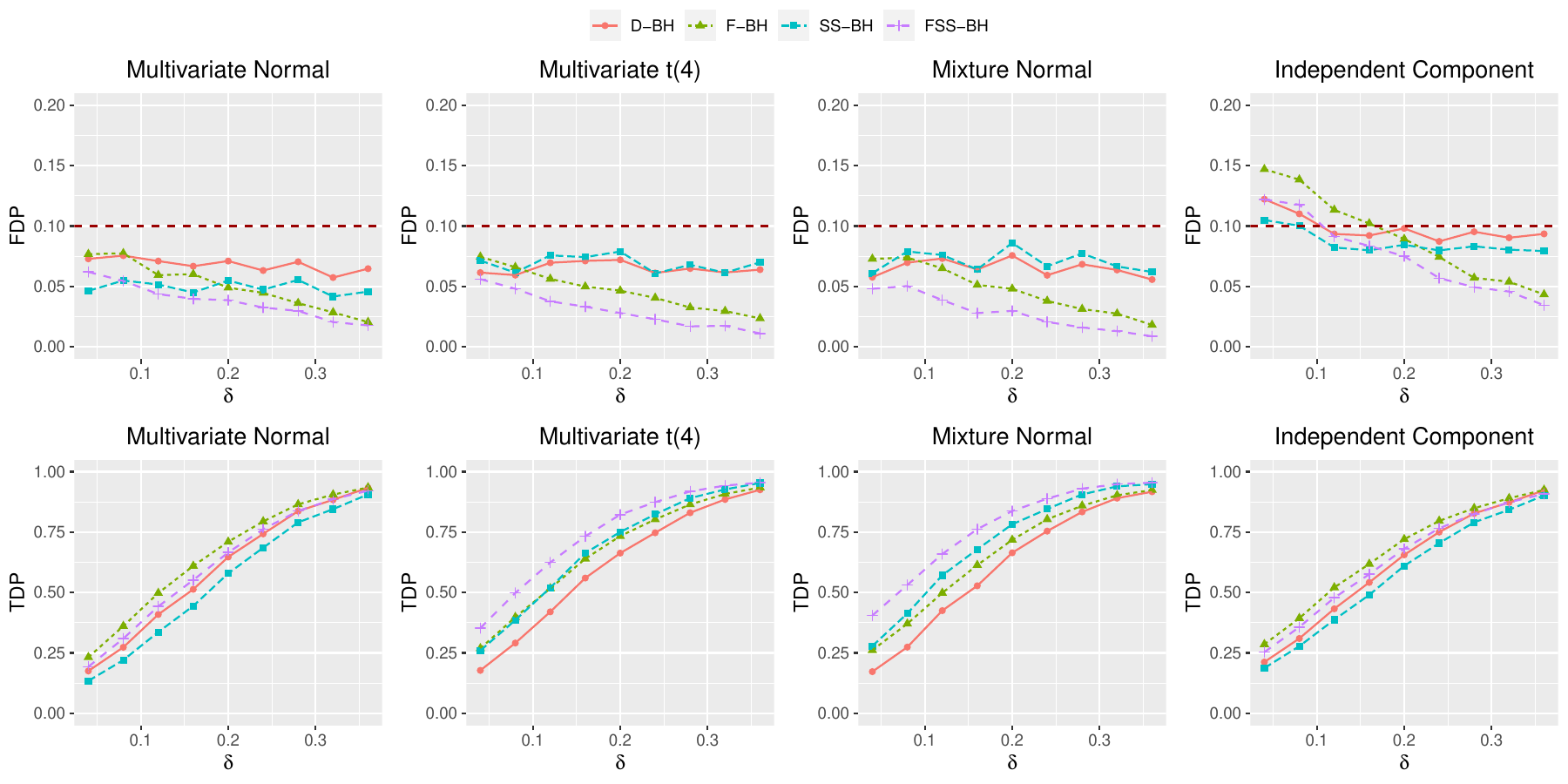}}
\end{figure}

Figure \ref{fig22} presents the FDP and TDP of the four methods under Scenario III with $\rho=0.5$. When there is one weak factor, the performance of each method is similar to Scenario II. Notably, our SS-BH method can control the FDP in case (iv) in most cases. It is important to note that the errors in Scenario III still do not satisfy the model assumptions. This demonstrates that the performance of our methods is not impacted by the weak factor, indicating that our methods can work effectively in a wide range of applications.

\begin{figure}[!ht]
\centering
\caption{The FDP and TDP with varying $\rho$ (different correlation) and $T=60,N=200$ under Scenario I. \label{fig3}}
{\includegraphics[width=1\textwidth]{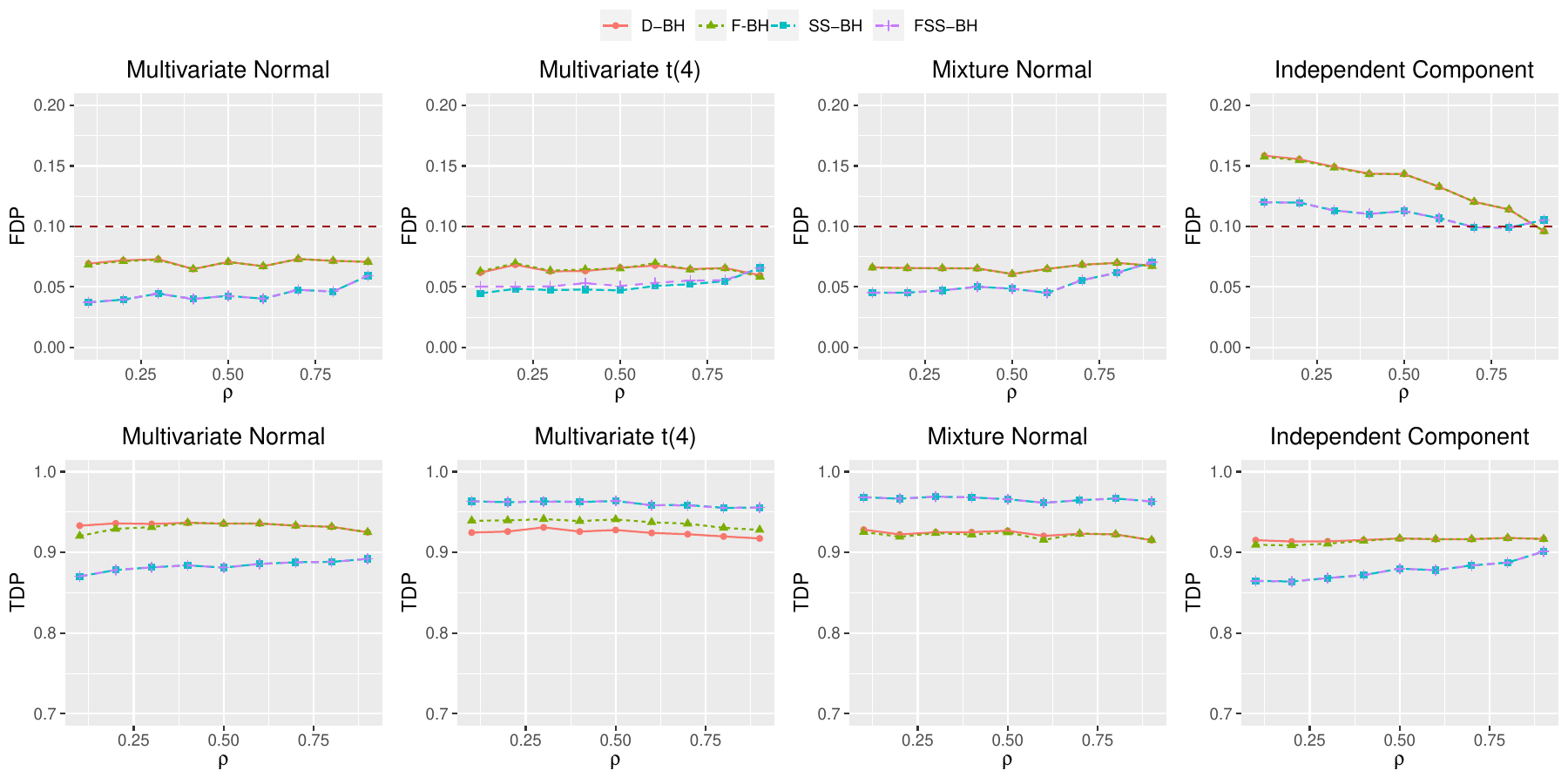}}
\end{figure}

\begin{figure}[!ht]
\centering
\caption{The FDP and TDP with varying $\rho$ (different correlation) and $T=60,N=200$ under Scenario II. \label{fig4}}
{\includegraphics[width=1\textwidth]{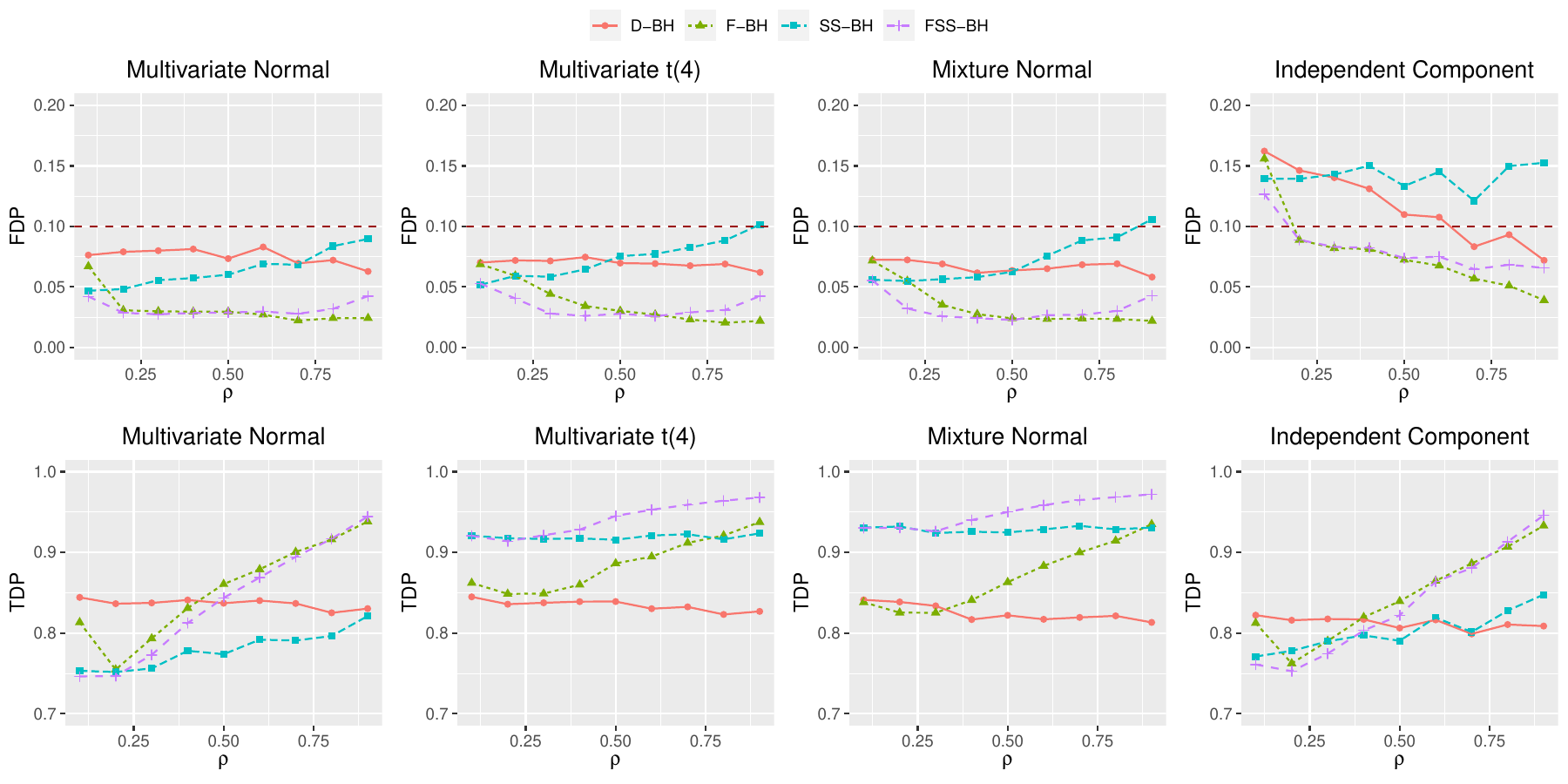}}
\end{figure}

\begin{figure}[!ht]
\centering
\caption{The FDP and TDP with varying $\rho$ (different correlation) and $T=60,N=200$ under Scenario III. \label{fig5}}
{\includegraphics[width=1\textwidth]{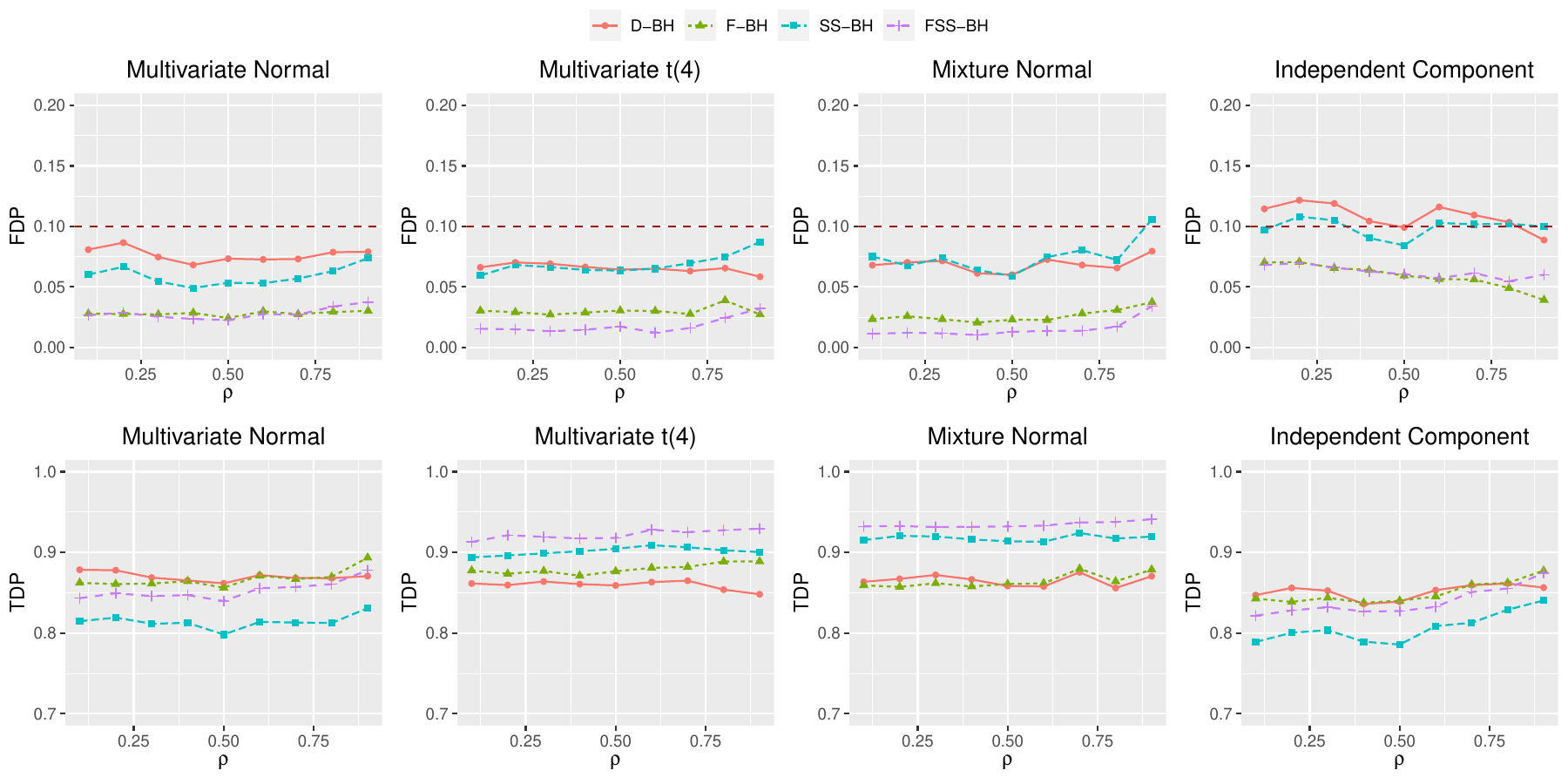}}
\end{figure}

To assess the impact of correlation on performance, we set $T=60$ and $\delta=0.48$. Figures \ref{fig3}-\ref{fig5} present the FDP and TDP of the four procedures as $\rho$ varies within the interval $(0,1)$ under Scenarios I, II, and III, respectively. Our findings indicate that the results of all procedures are comparable to the case where $\rho=0.5$ under Scenarios I and III. In Scenario II, both the factor-adjusted procedures, F-BH and FSS-BH, exhibit increased TDP as correlation strengthens, particularly for heavy-tailed distributions in cases (ii) and (iii). Notably, our proposed FSS-BH method demonstrates robust performance across various correlation levels, highlighting its robustness in handling different degrees of correlation.

All the results consistently demonstrate that our proposed FSS-BH multiple testing procedure exhibits excellent performance across a wide array of applications and demonstrates robustness to variations in the covariance structure and the distribution of the error term.

\section{Real Data Application}\label{sec:real}
In this section, we conduct a comparison of the aforementioned methods within a real-data application. We obtained data on active U.S. equity mutual funds from the Center for Research in Security Prices (CRSP) Survivor-Bias Free Mutual Fund database, spanning the period from February 1987 to September 2017, while excluding passive index funds as per \citep{harvey2018detecting}. Given the monthly collection of data, the sample size is $T = 368$. Our analysis includes only those funds with initial total net assets exceeding 10 million and having more than 80\% of their holdings in equity markets. It is worth noting that this dataset is the same as the one used in \citet{gao2023higher}.

We employ the widely recognized Fama-French three-factor model to analyze this dataset. The explicit formulation of these factor models is provided as follows:

\[
Y_{it} = \alpha_i + \beta_{i1} (r_{mt} - r_{ft}) + \beta_{i2} SMB_t + \beta_{i3} HML_t + \varepsilon_{it},
\]

where $i = 1, \cdots, N$ represents the funds, $t = 1, \cdots, T$ represents the time periods, $Y_{it}$ denotes the monthly return of the $i$-th fund at time $t$, $r_{ft}$ is the risk-free rate at time $t$ approximated by the monthly 30-day T-bill yield at the beginning of the month, and $r_{mt}$ is the market return at time $t$ approximated by the value-weighted return on all NYSE, AMEX, and NASDAQ stocks obtained from CRSP. Additionally, $SMB_t$ represents the average return on the three small portfolios minus the average return on the three big portfolios, and $HML_t$ represents the average return on the two value portfolios minus the average return on the two growth portfolios, both calculated based on the stocks listed on the NYSE, AMEX, and NASDAQ. These factor data can be downloaded from Kenneth R. French's data library.

\begin{figure}[!ht]
\centering
\caption{The histogram of p-values of the Shaprio-Wilk test of the original data and residuals of all funds under 3-factor model, repectively. \label{figdn}}
{\includegraphics[width=1\textwidth]{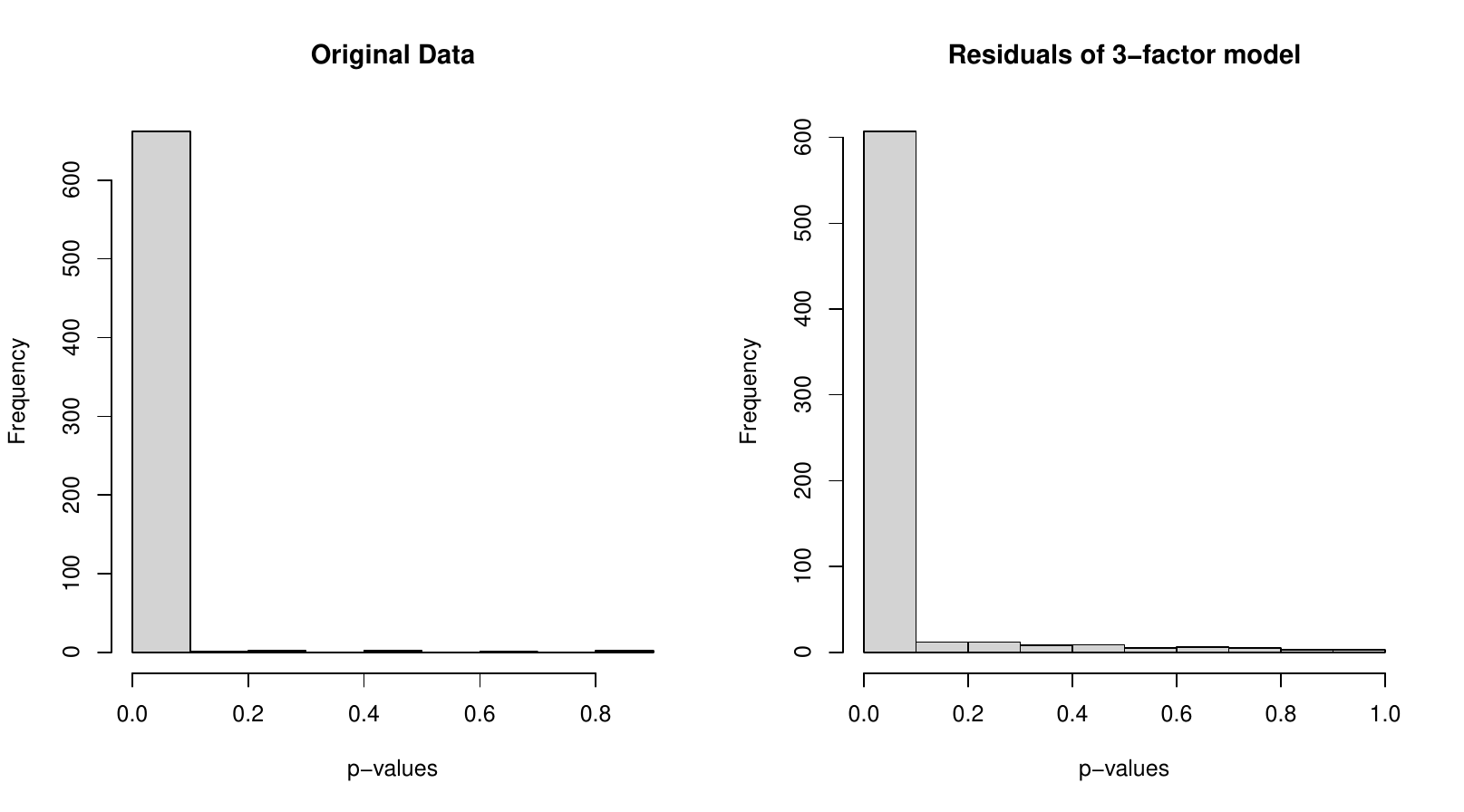}}
\end{figure}

\begin{figure}[!ht]
\centering
\caption{The scree plots of the eigenvalue and cumulative variance contribution of the spatial multivariate Kendall' tau matrix based on the original data and the residuals. \label{figdataana}}
\subfloat[Original Data]{\includegraphics[width=0.48\textwidth]{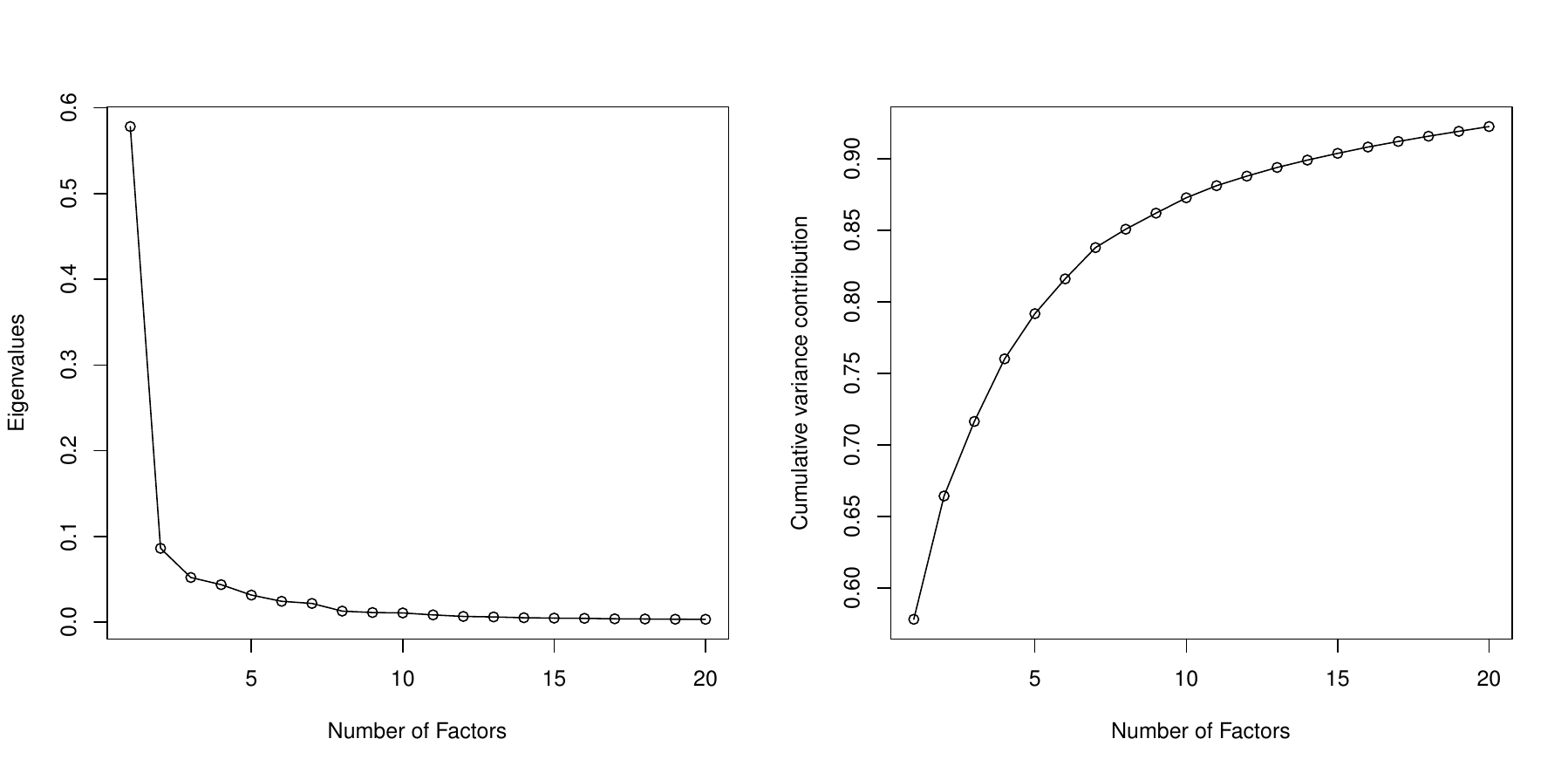}}\hfill
\subfloat[Residuals of 3-factor model]{\includegraphics[width=0.48\textwidth]{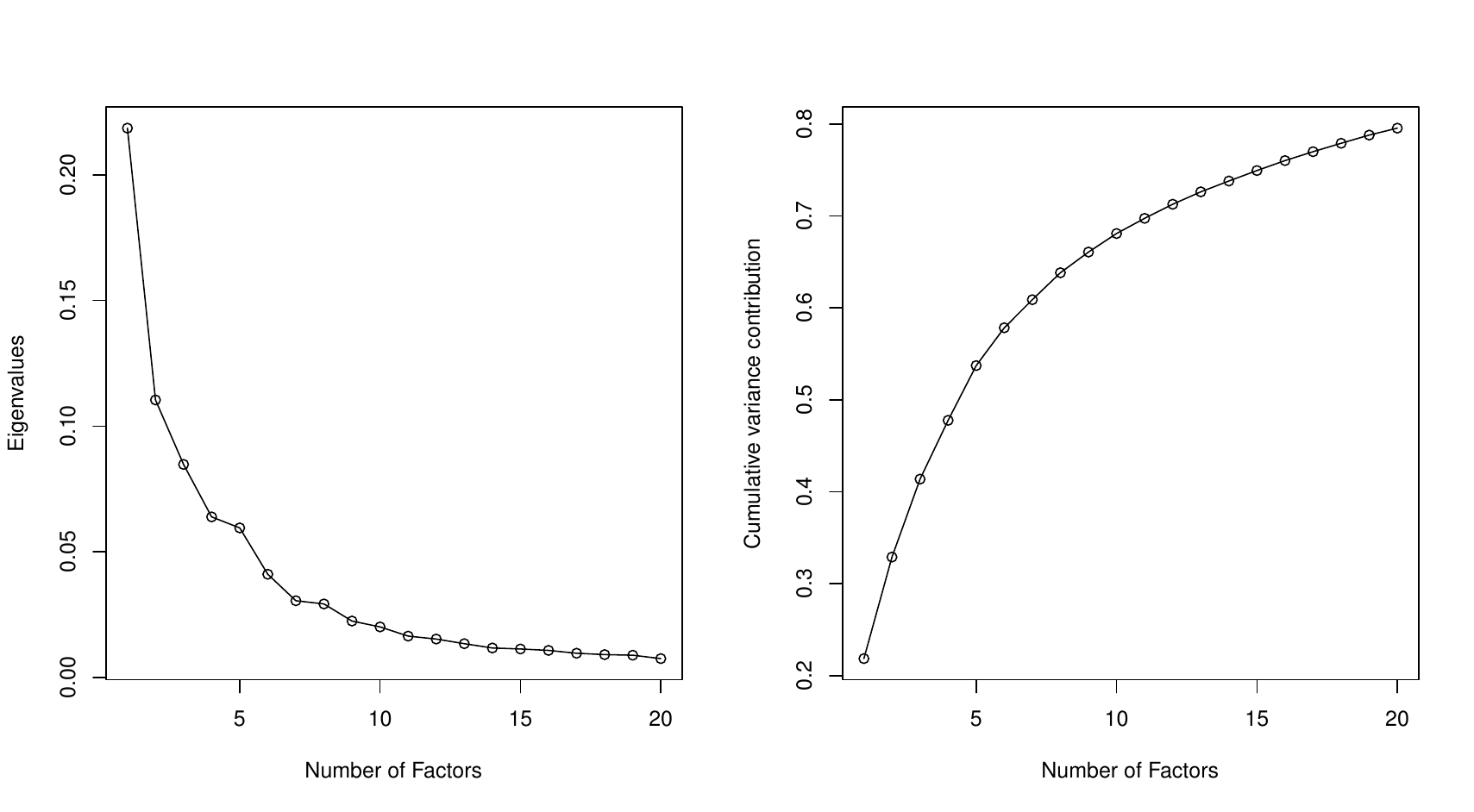}}
\end{figure}

Before demonstrating the utility of our methods, we initially assess the normality of the original data and the residuals of the three-factor model. Figure \ref{figdn} presents the p-values of the Shapiro-Wilk test for each security, based on both the original data and the residuals of the three-factor model. Our observations indicate that most of the data do not follow a normal distribution. Therefore, we opt to use robust procedures for analyzing this dataset. Next, we examine the latent factor structure of the original data and the residuals of the three-factor model. Figure \ref{figdataana} displays the scree plots of the eigenvalue and cumulative variance contribution of the spatial multivariate Kendall's tau matrix for both the original data and the residuals of the three-factor model. Our findings reveal that the original data exhibit both strong and non-strong factors. The first three eigenvalues contribute more than 70\% of the total variance. However, the residuals of the three-factor model do not exhibit strong factors, with the first eigenvalue contributing only 20\% of the total variance. Consequently, we conclude that a three-factor model is adequate for fitting the dataset.

\begin{figure}[!ht]
\centering
\caption{The curves of net value for each procedure with FDR level $\gamma=0.1$. \label{figdata2}}
\subfloat[$T=120$]{\includegraphics[width=0.48\textwidth]{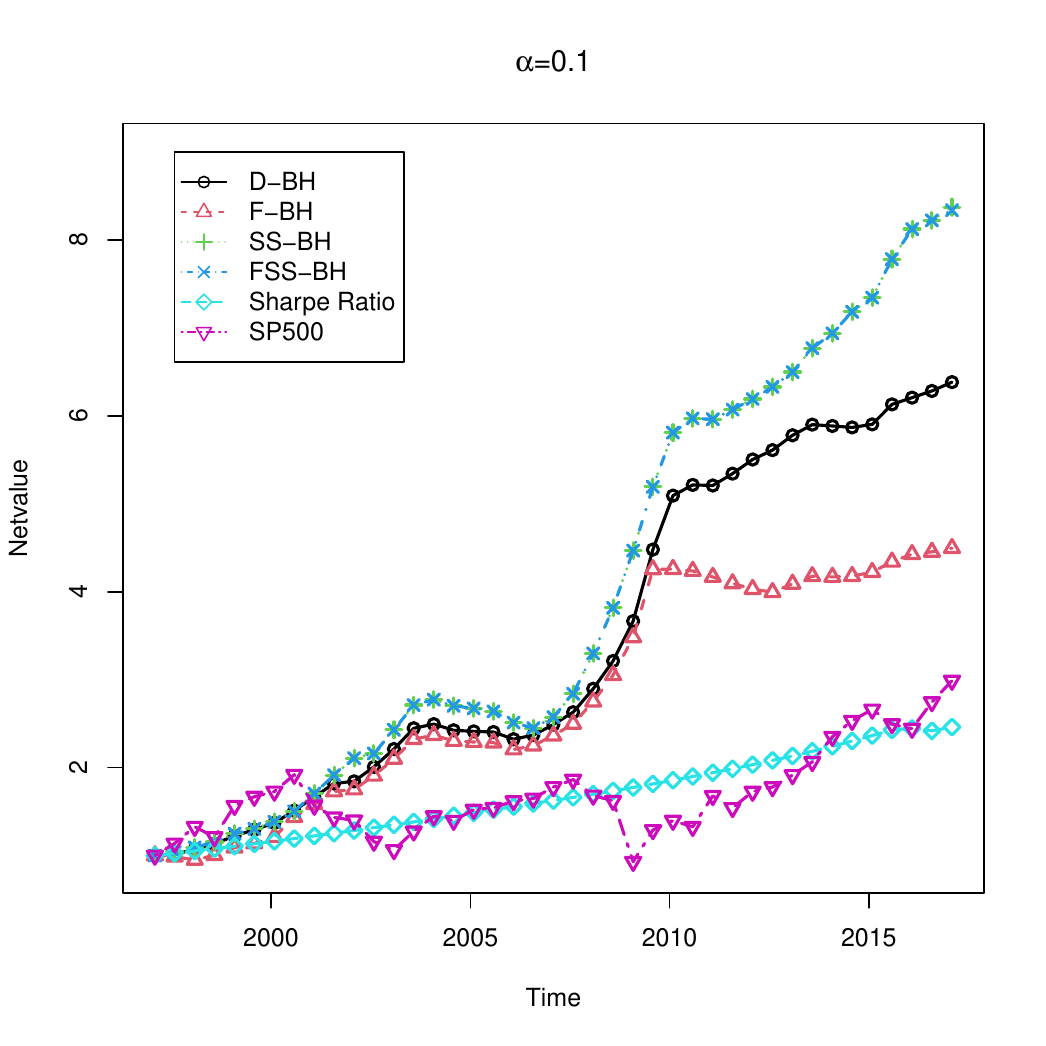}}\hfill
\subfloat[$T=180$]{\includegraphics[width=0.48\textwidth]{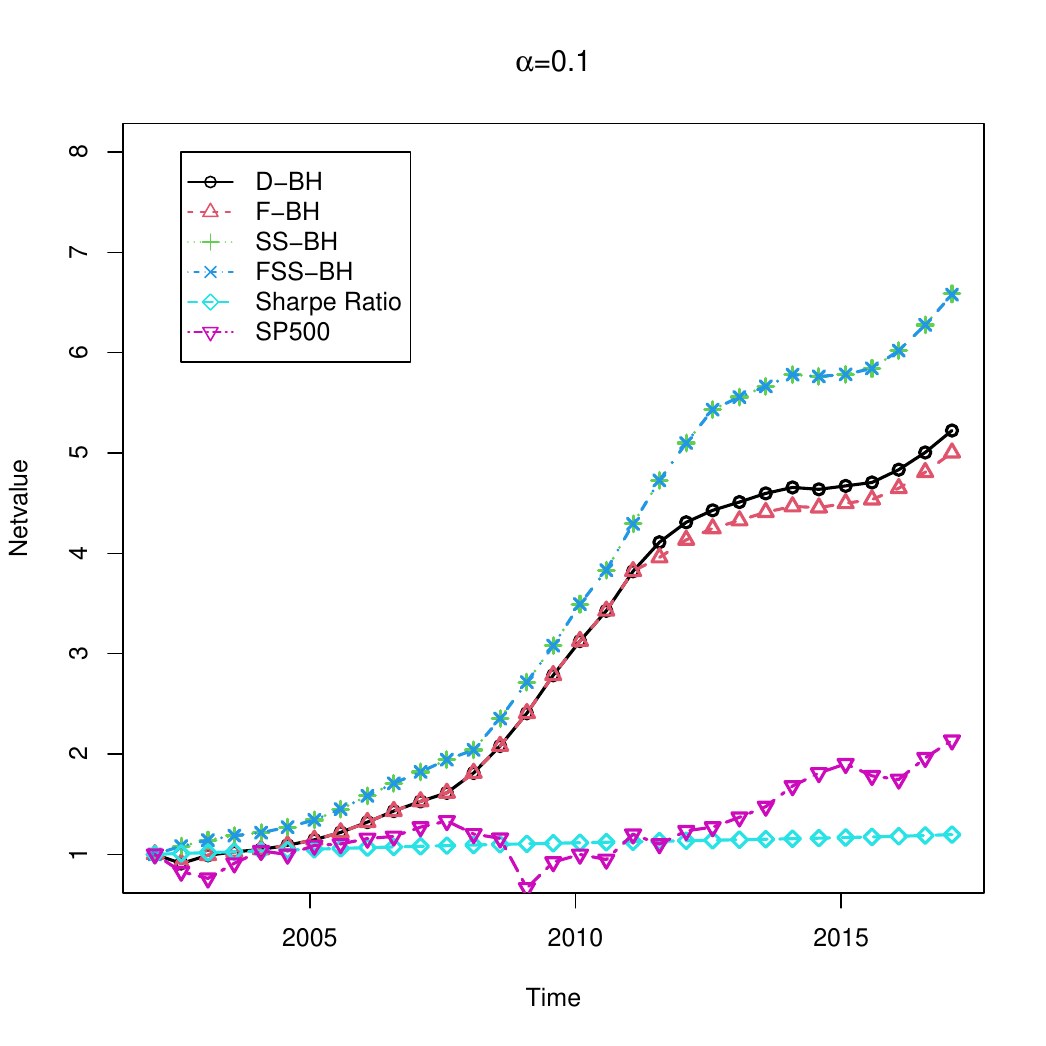}}
\end{figure}

To assess the effectiveness of the proposed spatial-sign based procedures, we must verify that the funds selected by our methods have the potential to generate higher returns in the future. To achieve this, we employ the following rolling window procedure. Specifically, we adopt a trading strategy as follows: using the first $L$ months of data, we identify skilled funds with $\alpha > 0$ using each method, applying a pre-determined FDR level $\gamma$. Subsequently, we create a fund of funds by investing in the selected funds with equal weighting. We then hold this portfolio for six months. After this period, we reselect the funds based on the last $L$ months of data (referred to as the selection period). We repeat this process until September 2017. Due to missing values during some periods, for each selection process, we only consider funds without missing values over the $L$ months. Additionally, we consider the Sharpe Ratio method, selecting skilled funds with a Sharpe ratio greater than one during the selection period. We use the return of the S\&P 500 index as a benchmark for comparison. Figure \ref{figdata2} present the curves of the net values for each procedure with FDR level $\gamma=0.1$ under $L=120$ (10 years) and $L=180$ (15 years).

First, all four methods rooted in the linear factor pricing model have demonstrated a pronounced ability to generate significantly higher returns compared to the widely benchmarked S\&P 500 index and the Sharpe Ratio method. This outstanding performance highlights the inherent advantages of the factor-based pricing paradigm, which systematically identifies and exploits systematic risk premiums across various asset classes. The linear factor pricing model, at its core, provides a structured framework for understanding and quantifying the sources of asset returns. By isolating and quantifying the impact of various factors such as market risk, size, and value, it enables investors to construct portfolios that are specifically designed to harness these premiums. This targeted approach, as evidenced by our findings, can lead to substantial improvements in risk-adjusted returns over more traditional, broad-based investment strategies.

In our comprehensive analysis, we have observed that the spatial-sign based methods, specifically SS-BH and FSS-BH, exhibit remarkably similar performance profiles, surpassing the conventional D-BH and F-BH approaches. This finding underscores the efficacy and robustness of the spatial-sign framework in capturing intricate market dynamics and translating them into superior investment outcomes. The superior performance of the spatial-sign based methods within this factor-pricing context further underscores the potential benefits of integrating those robust techniques into traditional financial models. By leveraging the unique properties of spatial-sign representations, these methods are able to capture heavy-tailed distributions that may be overlooked by more conventional approaches.

In conclusion, our study underscores the importance of continuous innovation and the integration of advanced analytics into financial decision-making. The spatial-sign based methods, when combined with the linear factor pricing model, represent a powerful combination that has the potential to deliver superior investment outcomes for investors. As the financial landscape continues to evolve, we believe that the exploration and implementation of such cutting-edge techniques will be critical in unlocking new sources of alpha and enhancing overall portfolio performance.

\section{Conclusion}\label{sec:conclusion}
Driven by the task of selecting skilled mutual funds, we propose a spatial sign-based multiple testing procedure (SS-BH), inspired by \cite{zhao2024}, to simultaneously test the positive alpha of all funds within a linear factor pricing model. In scenarios where unobservable factors may exist, we introduce factor-adjusted spatial sign-based multiple testing procedures (FSS-BH) and examine their theoretical properties. Our analysis verifies the significant advantages of both SS-BH and FSS-BH in controlling the false discovery rate (FDR). Additionally, empirical results demonstrate that FSS-BH excels in identifying skilled funds.
\subsection*{Acknowledgment}

\section{Appendix}
\subsection{Proof of Theorem \ref{th1}}
According to the proof of Theorem 2.1 in \cite{zhao2024}, we have 
\begin{align}\label{h1theata=}
	{ T^{1 / 2} \hat{\mathbf{D}}^{-1 / 2}\hat{\boldsymbol{\theta}}-T^{1/2}\hat{\omega}\hat{\D}^{-1/2}\bm \alpha=T^{-1 / 2}  \sum_{t=1}^T\zeta_1^{-1} (1-\sum_{s=1}^Tr_s^{-1}r_tV_{st})\U_t+\bm C_T},
\end{align}
where $\|\bm C_T\|_{\infty}=o_p(1/\sqrt{\log N})$.
Hence, we have
$$T^{1/2}\hat{\varsigma}^{1/2}\hat{\D}^{-1/2}(\hat{\bm \theta}-\hat{\omega}\bm \alpha)\cd N(\bm 0, N{\bf \Xi}).$$
Under the conditions in Theorem \ref{th1}, we have $(N{\bf \Xi})_{ij}=R_{ij}+O(N^{-\delta/2})$ for any $1\leq i,j\leq N$ by the proof of Theorem 2.1 in \cite{zhao2024}, where $(N{\bf \Xi})_{ij}$ and $R_{ij}$ are the element in row $i$ and column $j$ of matrices $N{\bf \Xi}$ and $\R$, respectively. Similar to the proof of Proposition 2.1 in \cite{zhao2024}, we have 
$$\hat{\varsigma}^{1/2}/{\varsigma}^{1/2}=1+O_p(T^{-1/2}),$$
where $\varsigma=N\{E(r_t^{-1})\}^2/\eta_{\omega}$, $\eta_{\omega}=1-2\eta E(r_t^{-1})E(r_t)+\eta [E(r_s^{-1})]^2E(r_t^2)$ and $T^{-1}\omega_{T}\cp \omega=1-\eta$.
Hence, we can obtain that \begin{align*}
	&\max_{1\leq i\leq N}|T^{1/2}\hat{\varsigma}^{1/2}\hat{d}_i^{-1}(\hat{\bm \theta}_i-\hat{\omega}  \alpha_i)-T^{1/2}{\varsigma}^{1/2}\hat{d}_i^{-1}(\hat{\bm \theta}_i-\hat{\omega}  \alpha_i)|\\
	\leq &\max_{1\leq i\leq N}|T^{1/2}{\varsigma}^{1/2}\hat{d}_i^{-1}(\hat{\bm \theta}_i-\hat{\omega } \alpha_i)|\times\Big|\frac{\hat{\varsigma}^{1/2}}{{\varsigma}^{1/2}}-1\Big|\\
	=&O_p(T^{-1/2}\log N) .
\end{align*}
Hence, we have ${ T_i^s=T^{1/2}\hat{\varsigma}\hat{d}_i^{-1}\hat{\omega}  \alpha_i+T^{-1/2}\sum_{t=1}^TN^{1/2}(1-\sum_{s=1}^Tr_s^{-1}r_tV_{st})\eta_{\omega}^{-1/2}U_{t,i}+C_{Ti}}$,
where $$\max_{1\leq i\leq N}|C_{Ti}|=O_p\left(T^{-1 / 4} \log ^{1 / 2}(NT)+N^{-(1 / 6 \wedge \delta / 2)} \log ^{1 / 2}(NT)+T^{-1 / 2}(\log N)^{1 / 2} \log ^{1 / 2}(NT)\right) ,$$ and $ \bm U_t=U(\D^{-1/2}\bmv_t)=(U_{t,1},\dots,U_{t,N})$ for $1\leq t\leq T.$
By assumption in Theorem \ref{th1}, there is $\mathcal{H} \subset\{1, \ldots, N\}$ so that $|\mathcal{H}| \rightarrow \infty$ and
$$
T^{1/2} \varsigma^{1/2} \omega\alpha_i/{d}_i^{1/2} \geq 2 \sqrt{\log N}, \forall i \in \mathcal{H}.
$$
Next, let $\mathcal{H}_0$ denote the index set of all the true null hypotheses. Also, let $\Psi(x):=1-\Phi(x)$, where $\Phi(x)$ is the cumulative distribution function of standard normal random variable.  Our major goal is to bound the number of false rejections
$$
\mathcal{F}=\sum_{i \in \mathcal{H}_0} 1\left\{p_i^s \leq p_{(\widehat{k})}^s\right\}.
$$
According to Lemma 1 of \cite{storey2004strong}, we have an equivalent statement for rejections: $p_i^s \leq p_{(\widehat{k})}^s$ if and only if $T_i^s \geq \widehat{t}$, where
\begin{align}\label{widehatt}
\widehat{t}:=\inf \left\{x \in \mathbb{R}: \Psi(x) \leq \gamma \frac{1}{N} \max \left\{\sum_{i=1}^N 1\left\{T_i^s \geq x\right\}, 1\right\}\right\}.
\end{align}
Consequently, our goal becomes to bound $\mathcal{F}=\sum_{i \in \mathcal{H}_0} 1\left\{i \leq N: T_i^s \geq \widehat{t}\right\}$. 
The main inequality to use is: uniformly for $x \in\left[0, t^*\right]$, where $t^*=\Psi^{-1}(\gamma|\mathcal{H}| / N)$,
\begin{align}\label{9}
\frac{1}{\left|\mathcal{H}_0\right|} \sum_{i \in \mathcal{H}_0} 1\left\{T_i^s \geq x\right\} \leq \Psi(x)\left\{1+o_p(1)\right\}.
\end{align}

The remaining proof is divided into the following steps.

Step 1. Show the inequality (\ref{9}). This inequality is essentially the Gaussian approximation to the "empirical measure" of the statistics for those true null hypotheses, whose proof requires weak dependence among the statistics. 

Write ${ z_i=T^{-1/2}\sum_{t=1}^TN(1-\sum_{s=1}^Tr_s^{-1}r_tV_{st})\eta_{\omega}^{-1/2}U_{t,i}}$ for $1\leq i\leq N$. When $\alpha_i \leq 0$ and $\hat{\omega}>0$, we have $T_i^s \leq T^{1/2}\hat{\varsigma}^{1/2}\hat{d}_i^{-1}(\hat{\bm \theta}_i-\hat{\omega } \alpha_i)=z_i+C_{Ti}$ where $\max _i\left|C_{Ti}\right|=o_p(1 / \sqrt{\log N})$. Hence
{ \begin{align*}
\frac{1}{\left|\mathcal{H}_0\right|} \sum_{i \in \mathcal{H}_0} 1\left\{T_i^s \geq x\right\} \leq &\frac{1}{\left|\mathcal{H}_0\right|} \sum_{i \in \mathcal{H}_0} 1\left\{z_i \geq x-\|\bm C_{T}\|_{\infty}\right\}1\{\hat{\omega}>0\}\\
&+\frac{1}{\left|\mathcal{H}_0\right|} \sum_{i \in \mathcal{H}_0} 1\left\{z_i \geq x-\|\bm C_{T}\|_{\infty}-\|T^{1/2}\hat{\varsigma}^{1/2}\hat{\D}^{-1/2}(\hat{\omega}-\omega)  \bm\alpha\|_{\infty}\right\}1\{\hat{\omega}<0\},
\end{align*}
where $T^{-1}\1_T^{\top}\M_{\F}\1_T\cp \omega>0$,} $\bm C_{T}=(C_{T1},\dots, C_{TN})$ and the right hand side does not depend on $\bm\alpha$ because $z_i$ is centered and independent of $\bm\alpha$.
Note that for all $x>0,1-\Phi(x) \leq e^{-x^2 / 2}$ by Proposition 2.5 in \cite{dudley2014uniform}. 
Therefore, as long as $|\mathcal{H}| \geq 1 / \gamma$, which holds for all $T$ large enough,
$$
 \gamma|\mathcal{H}| / N \geq 1 / N=\exp \left(-\frac{(\sqrt{2 \log N})^2}{2}\right) \geq(1-\Phi(\sqrt{2 \log N}))=\Psi(\sqrt{2 \log N}).
$$
Hence, we have $t^*\leq \sqrt{2 \log N}.$
On the other hand, there is $\eta_x \in\left[0,\|\bm C_{T}\|_{\infty}\right]$ so that for some universe constant $C>0$, uniformly for $0<x \leq t^*$
\begin{align*}
	\left|\Psi(x)-\Psi\left(x-\|\bm C_{T}\|_{\infty}\right)\right| & \leq \phi\left(x+\eta_x\right)\|\bm C_{T}\|_{\infty} \leq \phi(x)\|\bm C_{T}\|_{\infty} \frac{\phi\left(x+\eta_x\right)}{\phi(x)} \\
	& \leq C \phi(x)\|\bm C_{T}\|_{\infty} \exp \left(C \eta_x\left(\eta_x+t^*\right)\right) \\
	& \leq C x \Psi(x)\|\bm C_{T}\|_{\infty}(1+o(1)) \leq C t^* \Psi(x)\|\bm C_{T}\|_{\infty}(1+o(1)) \\
	& \leq o(1) \Psi(x),
\end{align*}
where $o(1)$ is a uniform term because $\eta_x t^* \leq\|\bm C_{T}\|_{\infty} t^* \leq o_p(1 / \sqrt{\log N}) \sqrt{2 \log N}=o(1)$. This proves $\Psi(x)=\Psi\left(x-\|\bm C_{T}\|_{\infty}\right)(1+o(1))$. Also,
$$
\Psi\left(x-\|\bm C_{T}\|_{\infty}\right)=\Psi(x)(1+o(1)) \geq \Psi\left(t^*\right)(1+o(1)) \geq(1+o(1)) \gamma|\mathcal{H}| / N \geq \gamma|\mathcal{H}| /(2 N). 
$$
So, $x-\|\bm C_{T}\|_{\infty} \leq \Psi^{-1}(\gamma|\mathcal{H}| /(2 N))$ when $0<x \leq t^*$.
{ Obviously, we have $\mathbb{E}(1\{\hat{\omega}<0\})=P(\hat{\omega}<0)=O(T^{-1})$ and $\Psi\left(x-\|\bm C_{T}\|_{\infty}-\|T^{1/2}\hat{\varsigma}^{1/2}\hat{\D}^{-1/2}(\hat{\omega}-\omega)  \bm\alpha\|_{\infty}\right)1\{\hat{\omega}<0\}\leq o_p(1)\Psi(x)$ due to $\|T^{1/2}\hat{\varsigma}^{1/2}\hat{\D}^{-1/2}(\hat{\omega}-\omega)  \bm\alpha\|_{\infty}=O_p(1)$ for $0<x<t^{*}$.}
 Additionally, similar to Equation (13) of \cite{liuandshao}, we have uniformly for $0<x\leq  t^{*}$,
$$\frac{1}{\left|\mathcal{H}_0\right|} \sum_{i \in \mathcal{H}_0} 1\left\{z_i \geq x\right\} \leq \Psi\left(x\right)\{1+o(1)\}.$$
Combing the above results, we have
{ \begin{align*}
&\frac{1}{\left|\mathcal{H}_0\right|} \sum_{i \in \mathcal{H}_0} 1\left\{T_i^s \geq x\right\}\\
 \leq &\frac{1}{\left|\mathcal{H}_0\right|} \sum_{i \in \mathcal{H}_0} 1\left\{z_i \geq x-\|\bm C_{T}\|_{\infty}\right\}\\
 &+\frac{1}{\left|\mathcal{H}_0\right|} \sum_{i \in \mathcal{H}_0} 1\left\{z_i \geq x-\|\bm C_{T}\|_{\infty}-\|T^{1/2}\hat{\varsigma}^{1/2}\hat{\D}^{-1/2}(\hat{\omega}-\omega)  \bm\alpha\|_{\infty}\right\}1\{\hat{\omega}<0\}\\
  \leq &\Psi\left(x-\|\bm C_{T}\|_{\infty}\right)\left(1+o_p(1)\right)+o_p(1)\Psi(x)=\Psi(x)\left(1+o_p(1)\right).
\end{align*}}

Step 2. To use inequality (\ref{9}), we then aim to prove that $\widehat{t} \leq t^*$ with probability converging to one. %
Hence, by the definition of $\widehat{t}$, proving $\widehat{t} \leq t^*$ is equivalent to
$$
\gamma|\mathcal{H}| / N=\Psi(t^*) \leq \gamma \frac{1}{N} \max \left\{\sum_{i=1}^N 1\left\{T_i^s \geq t^*\right\}, 1\right\},
$$
which is also equivalent to $$\sum_{i=1}^N 1\left\{T_i^s \geq t^*\right\}\geq |\mathcal{H}|.$$
In words, the number of rejections is at least $|\mathcal{H}|$. Next, to prove the above inequality, we will prove $$\mathbb{P}\left(\forall j \in \mathcal{H}, T_j^s \geq \sqrt{2 \log N}\right) \rightarrow 1.$$ It then implies
$$
\sum_{i=1}^N 1\left\{T_i^s \geq t^{*}\right\} \geq\sum_{i=1}^N 1\left\{T_i^s \geq \sqrt{2 \log N}\right\}\geq|\mathcal{H}|.
$$
Recall that $z_i=T^{-1/2}\sum_{t=1}^TN(1-\sum_{s=1}^Tr_s^{-1}r_tV_{st})\eta_{\omega}^{-1/2}U_{t,i}$ and $T^{1/2}\hat{\varsigma}^{1/2}\hat{d}_i^{-1}(\hat{\bm \theta}_i-\hat{\omega}  \alpha_i)=z_i+C_{Ti}$. So,
{ $$
T_i^s \geq T^{1/2}\hat{\varsigma}^{1/2}\hat{d}_i^{-1}\omega  \alpha_i-\left|z_i\right|-|C_{Ti}|-|T^{1/2}\hat{\varsigma}^{1/2}\hat{d}_i^{-1}(\hat{\omega}-\omega)  \alpha_i|,
$$}
where $|T^{1/2}\hat{\varsigma}^{1/2}\hat{d}_i^{-1}(\hat{\omega}-\omega)  \alpha_i|=O_p(1)$.
Now note that $T^{1/2}{\varsigma}^{1/2}{d}_i^{-1}\omega  \alpha_i\geq 2\sqrt{\log N}$ for all $i \in \mathcal{H}$. So for all $i \in \mathcal{H}$, we have 
\begin{align*}
&T^{1/2}\hat{\varsigma}^{1/2}{d}_i^{-1}\omega  \alpha_i\\
=& T^{1/2}{\varsigma}^{1/2}{d}_i^{-1}\omega  \alpha_i+T^{1/2}\hat{\varsigma}^{1/2}{d}_i^{-1}\omega\alpha_i-T^{1/2}{\varsigma}^{1/2}{d}_i^{-1}\omega \alpha_i\\
= &T^{1/2}{\varsigma}^{1/2}{d}_i^{-1}\omega  \alpha_i\Big(1+\frac{\hat{\varsigma}^{1/2}}{{\varsigma}^{1/2}}-1\Big)\\
= &T^{1/2}{\varsigma}^{1/2}{d}_i^{-1}\omega  \alpha_i\big\{1+o_p(\log^{-1}(N))\Big\}.
\end{align*}
So due to $P(\max_{1\leq i\leq N}|z_{i}|\leq 3\log N)\rightarrow 1$, uniformly for these $i$,
{ $$
T_i^s \geq 2\sqrt{\log N}+o_p(1)-\sqrt{3 \log N}-o_p(1)-O_p(1) \geq \sqrt{2 \log N}.
$$}
So, we have $\widehat{t} \leq t^*$ and thus $\Psi(\widehat{t}) \geq \Psi(t^*)= \gamma|\mathcal{H}| / N$. In addition, by the definition of $\widehat{t}$, we have
$$
\Psi(\widehat{t})=\gamma \frac{1}{N} \sum_{i=1}^N 1\left\{T_i^s \geq \widehat{t}\right\} \geq \gamma \frac{|\mathcal{H}|}{N}.
$$
We have proved that $\widehat{t} \leq t^*$ with probability converging to one, then by (\ref{9}),
$\mathcal{F}\leq \Psi(\widehat{t})\left|\mathcal{H}_0\right|+o_p(1)\Psi(\widehat{t})\left|\mathcal{H}_0\right|$. Also by $\Psi(\widehat{t})=\gamma \frac{1}{N} \max\{\sum_{i=1}^N 1\left\{T_i^s \geq \widehat{t}\right\},1\}$, we have
$$
\mathcal{R}=\max \left\{\sum_{i=1}^N 1\left\{t_i \geq \widehat{t}\right\}, 1\right\}=\Psi(\widehat{t}) N / \gamma.
$$
It then gives, for some $X=o_p(1)$, and $|X| \leq 1$ almost surely, $\frac{\mathcal{F}}{\mathcal{R}} \leq \gamma \frac{\left|\mathcal{H}_0\right|}{N}+X$, on the event $\widehat{t} \leq t^*$. Hence
$$
\mathrm{FDP}_{FSS-BH} \leq \gamma+o_p(1).
$$
Together, for any $\epsilon>0$,
\begin{align*}
	\mathrm{FDR}_{FSS-BH} & \leq \mathbb{E}\left(\left.\gamma \frac{\left|\mathcal{H}_0\right|}{N}+X \right\rvert\, \widehat{t} \leq t^*\right)+\mathbb{P}\left(\widehat{t}>t^*\right) \\
	& \leq \gamma \frac{\left|\mathcal{H}_0\right|}{N}+\epsilon+\mathbb{P}(|X| \geq \epsilon \mid \mathcal{R} \geq 1)+o(1).
\end{align*}
Since $\epsilon$ is chosen arbitrarily, $\mathrm{FDR}_{FSS-BH} \leq \gamma \frac{\left|\mathcal{H}_0\right|}{N}+o(1)$.
 \hfill$\Box$
\subsection{Proof of Theorem \ref{th2}}

\begin{lemma}\label{KZeigen}
Under Conditions (C5)-(C9), as $\min \{N, T\} \rightarrow \infty$, we have
	$$\left\{\begin{array}{cl}
			\lambda_j\left({\mathbf{K}}_Z\right) \asymp r^{-1}, \,\,\,\,\,\,\,\,  j \leq r, \\
			 \lambda_j\left({\mathbf{K}}_Z\right)=o_p(1), \,\,\, j>r .
		\end{array}\right.$$
\end{lemma}
\begin{lemma}\label{deltagammaH}
Under Conditions (C5)-(C9), there exist a series of matrices $\widehat{\mathbf{H}}$ (dependent on $T, N$ and $\widehat{\mathbf{\Gamma}})$ so that $\widehat{\mathbf{H}}^{\top} \mathbf{V} \widehat{\mathbf{H}} \xrightarrow{p} \mathbf{I}_r$ and
	$$
	\frac{1}{N}\|\widehat{\mathbf{\Gamma}}-\mathbf{\Gamma} \widehat{\mathbf{H}}\|_F^2=O_p\left(\frac{1}{T}\right) .
	$$
\end{lemma}
\begin{lemma}\label{deltaWH}
 Assume that Conditions (C5)-(C9) hold, then for any $t \leq T$,
$$
\left\|\widehat{\mathbf{H}} \widehat{\boldsymbol{W}_t}-\boldsymbol{W}_t\right\|^2=O_p\left(\frac{1}{N}+\frac{1}{T}\right) .
$$
\end{lemma}
\begin{lemma}\label{maxdeltagammaH}
Under Conditions (C5)-(C9),  we have 
$$\max_{1\leq i\leq N }\|\bm{\gamma}_i^{\top}-\hat{\gamma}_i^\top\hat{\H}^{-1}\|^2=o_p(1/{\rm log} N).$$
\end{lemma}
According to the definition, we have $\breve{\Y}_t = \Y_t - \hat{\G} \hat{\W}_t$, $\breve{\Z} = \M_{\F}\breve{\Y}^\top$ and $\breve{\Y} = (\breve{\Y}_1, \cdots, \breve{\Y}_T)$.
Let $\breve{\Z}=(\breve{\Z}_1, \cdots, \breve{\Z}_T)$ and ${\bf W}=({\bf W}_1\dots, {\bf W}_T)^{\top}$. Then, we have
$\breve{\Z}_t=\bm{\eta}_{ t}+\vartheta_t\bm \alpha-\bm{\eta}\bm {V}_t+\mathbf{\Gamma}\bm{W}_t-\hat{\mathbf{\Gamma}}\hat{\bm{W}}_t-\mathbf{\Gamma}{\bf{W}}^{\top}\bm V_t+\hat{\mathbf{\Gamma}}\hat{{\bf{W}}}^{\top}\bm V_t$, where $\bm{\eta}=(\bm{\eta}_{1},\dots,\bm{\eta}_{T})$ and $\bm V_t=\f(\f^{\top}\f)^{-1}\f_t$. Define
$\breve{\U}_t=U(\breve{\D}^{-1/2}\bm{\eta}_{t})$, $\breve{r}_t=\|\breve{\D}^{-1/2}\bm{\eta}_{ t}\|$ and $\bm{\Delta}_t=\mathbf{\Gamma}\bm{W}_t-\hat{\mathbf{\Gamma}}\hat{\bm{W}}_t-\mathbf{\Gamma}{\bf{W}}^{\top}\bm V_t+\hat{\mathbf{\Gamma}}\hat{{\bf{W}}}^{\top}\bm V_t$.
The estimator $\tilde{\bm \theta}=\breve{\bm \theta}-\omega\bm{\alpha}$ satisfies $\sum_{t=1}^TU(\hat{\D}^{-1/2}(\Z_t-\omega\bm{\alpha}-\title{\bm \theta}))=\mathbf{0}$, which is equivalent to 
\begin{align*}
	&\frac{1}{T}\sum_{t=1}^T( \breve{\U}_t- \breve{r}^{-1}_t \breve{\D}^{-1/2}\tilde{\bm\theta}- \breve{r}^{-1}_t \breve{\D}^{-1/2}\bm{\eta}\bm V_t+\breve{r}_t^{-1} \breve{\D}^{-1/2}(\vartheta_t-\omega)\bm\alpha+\breve{r}^{-1}_t\breve{\D}^{-1/2}\bm{\Delta}_t)\\
	&\times(1+ \breve{r}^{-2}_t\| \breve{\D}^{-1/2}\tilde{\bm\theta}\|^2+ \breve{r}^{-2}_t\| \breve{\D}^{-1/2}\bm{\eta}\bm V_t\|^2-2 \breve{r}^{-1}_t \breve{\U}^\top_t \breve{\D}^{-1/2}\tilde{\bm\theta}\\
	&\quad\quad-2 \breve{r}^{-1}_t \breve{\U}^\top_t \breve{\D}^{-1/2}\bm{\eta}\bm V_t+ \breve{r}_t^{-2}\| \breve{\D}^{-1/2}(\vartheta_t-\omega)\bm\alpha\|^2+2 \breve{r}_t^{-1} \breve{\U}_t^{\top} \breve{\D}^{-1/2}(\vartheta_t-\omega)\bm\alpha\\
	&\quad\quad+2 \breve{r}^{-2}_t\tilde{\bm\theta}^\top \breve{\D}^{-1}\bm{\eta}\bm V_t-2 \breve{r}_t^{-2}\bm V^\top_t\bm{\eta}^\top \breve{\D}^{-1}(\vartheta_t-\omega)\bm\alpha-2 \breve{r}_t^{-2}\tilde{\bm\theta}^{\top} \breve{\D}^{-1}(\vartheta_t-\omega)\bm\alpha\\
	&\quad\quad+2\breve{r}^{-1}_t\breve{\U}^\top_t \breve{\D}^{-1/2}\bm{\Delta}_t+2\breve{r}^{-2}_t(\vartheta_t-\omega){\bm\alpha}^\top \breve{\D}^{-1}\bm{\Delta}_t-2\breve{r}^{-2}_t\tilde{\bm\theta}^\top \breve{\D}^{-1}\bm{\Delta}_t\\
	&\quad\quad-2\breve{r}_t^{-2}\bm{V}_t^{\top}\bm{\eta}^{\top}\breve{\D}^{-1}\bm{\Delta}_t
	+\breve{r}_t^{-2}\bm{\Delta}_t^{\top}\breve{\D}^{-1}\bm{\Delta}_t
	)^{-1/2}=\mathbf{0}.
\end{align*}
First, we focus on $\|\bm{\Delta}_t\|^2$. 
By Lemmas \ref{deltagammaH} and \ref{deltaWH}, we have 
$N^{-1}\|\widehat{\mathbf{\Gamma}}-\mathbf{\Gamma} \widehat{\mathbf{H}}\|_F^2=O_p\left(1/T+N^{-2}\right)$, and $\left\|\widehat{\mathbf{H}} \widehat{\boldsymbol{W}}_t-\boldsymbol{W}_t\right\|^2=O_p\left(N^{-1}+1/{T}\right)$ for any $t \leq T$.
According to the proof of Theorem 3.2 in \cite{he2022large}, we have $E\|\bm W_t\|^2=O(1)$.
Hence, by triangular inequality and Cauchy-Schwartz inequality and $\|\bm W_t\|^2=O_p(1)$, we have for any $t \leq T$,
\begin{align*}
	\frac{1}{N}\left\|\widehat{\mathbf{\Gamma}} \widehat{\boldsymbol{W}}_t-\mathbf{\Gamma} \boldsymbol{W}_t\right\|^2 & =\frac{1}{N}\left\|\widehat{\mathbf{\Gamma}} \widehat{\mathbf{H}}^{-1} \widehat{\mathbf{H}} \widehat{\boldsymbol{W}}_t-\widehat{\mathbf{\Gamma}} \widehat{\mathbf{H}}^{-1} \boldsymbol{W}_t+\widehat{\mathbf{\Gamma}} \widehat{\mathbf{H}}^{-1} \boldsymbol{W}_t-\mathbf{\Gamma} \boldsymbol{W}_t\right\|^2 \\
	& \leq\left(\frac{2}{N}\left\|\widehat{\mathbf{\Gamma}} \widehat{\mathbf{H}}^{-1}\right\|_F^2\right)\left(\left\|\widehat{\mathbf{H}} \widehat{\boldsymbol{W}}_t-\boldsymbol{W}_t\right\|^2\right)+\left(\frac{2}{N}\left\|\widehat{\mathbf{\Gamma}} \widehat{\mathbf{H}}^{-1}-\mathbf{\Gamma}\right\|_F^2\right)\left\|\boldsymbol{W}_t\right\|^2 \\
	& =O_p\left(\frac{1}{T}+\frac{1}{N}\right),
\end{align*}
which concludes $$\left\|\widehat{\mathbf{\Gamma}} \widehat{\boldsymbol{W}}_t-\mathbf{\Gamma} \boldsymbol{W}_t\right\|^2 =O_p(NT^{-1} ).$$
Moreover, we have $\|\mathbf{\Gamma}{\bf{W}}^{\top}\bm V_t-\hat{\mathbf{\Gamma}}\hat{{\bf{W}}}^{\top}\bm V_t\|^2=O_p(NT^{-1} )$ due to $\bm V_t^{\top}\bm V_t=O(T^{-1})$. 
Hence, we can obtain that $\|\bm{\Delta}_t\|^2=O_p(NT^{-1})$.

Then, according to the proof of Theorem 2 in \cite{zhao2024}, we just need to prove that 
$$\|T^{-1}\sum_{t=1}^T\breve{r}_t^{-1}\breve{\D}^{-1/2}\bm{\Delta}_t\|_{\infty}=o_p(N^{-1/2}T^{-1/2}/\sqrt{\log N}).$$
We mainly consider $\|T^{-1}\sum_{t=1}^T\bm{\Delta}_t\|_{\infty}$. 
We have 
\begin{align*}
&\|T^{-1}\sum_{t=1}^T\bm{\Delta}_t\|_{\infty}\\
=&\|T^{-1}(\mathbf{\Gamma}{\bf{W}}^{\top}-\hat{\mathbf{\Gamma}}\hat{\bf{W}}^{\top})\M_{\F} {\1}_{T}\|_{\infty}\\
=&\max_{1\leq i\leq N }T^{-1}|(\bm{\gamma}_i^{\top}{\bf{W}}^{\top}-\hat{\bm \gamma_i}^{\top}\hat{\bf{W}}^{\top})\M_{\F}{\1}_{T}|\\
\leq& T^{-1}\max_{1\leq i\leq N }\|(\bm{\gamma}_i^{\top}{\bf{W}}^{\top}-\hat{\gamma}_i^\top\hat{\H}^{-1}{\bf{W}}^{\top})\M_{\F}{\1}_{T}\|+
T^{-1}\max_{1\leq i\leq N }\|\hat{\bm{\gamma}}_i^{\top}\hat{\H}^{-1}\|\|(\hat{\H}\hat{\bf{W}}^{\top}-{\bf{W}}^{\top})\M_{\F}{\1}_{T}\|.
\end{align*}
Next, we first consider the first term of the above inequality. Due to $E({\bf W}_t^{\top}{\bf W}_t)=O(1)$ and Lemma \ref{maxdeltagammaH}, we have 
\begin{align*}
	&\max_{1\leq i\leq N }\|(\bm{\gamma}_i^{\top}{\bf{W}}^{\top}-\hat{\gamma}_i^\top\hat{\H}^{-1}{\bf{W}}^{\top})\M_{\F}{\1}_{T}\|^2\\
	\leq &\max_{1\leq i\leq N }(\bm{\gamma}_i^{\top}{\bf{W}}^{\top}-\hat{\gamma}_i^\top\hat{\H}^{-1}{\bf{W}}^{\top})\M_{\F}{\1}_{T}{\1}_{T}^{\top}\M_{\F}(\bm{\gamma}_i^{\top}{\bf{W}}^{\top}-\hat{\gamma}_i^\top\hat{\H}^{-1}{\bf{W}}^{\top})^{\top}\\
	\leq &\max_{1\leq i\leq N }\|\bm{\gamma}_i^{\top}{\bf{W}}^{\top}-\hat{\gamma}_i^\top\hat{\H}^{-1}{\bf{W}}^{\top}\|\M_{\F}{\1}_{T}{\1}_{T}^{\top}\M_{\F}(\bm{\gamma}_i^{\top}{\bf{W}}^{\top}-\hat{\gamma}_i^\top\hat{\H}^{-1}{\bf{W}}^{\top})^{\top}\\
	\leq &\max_{1\leq i\leq N }\|\bm{\gamma}_i^{\top}-\hat{\gamma}_i^\top\hat{\H}^{-1}\|^2\lambda_{\max}({\bf{W}}^{\top}\M_{\F}{\1}_{T}{\1}_{T}^{\top}\M_{\F}{\bf W})\\
		\leq & O_p(Tr)\max_{1\leq i\leq N }\|\bm{\gamma}_i^{\top}-\hat{\gamma}_i^\top\hat{\H}^{-1}\|^2=o_p(T/\log N).
\end{align*}
Next, we will prove that 
$ T^{-1/2}\|(\hat{\H}\hat{\bf{W}}^{\top}-{\bf{W}}^{\top})\M_{\F}{\1}_{T}\|=o_p(1/\sqrt{\log N})$.
Note that 
$ T^{-1}\|(\hat{\H}\hat{\bf{W}}^{\top}-{\bf{W}}^{\top})\M_{\F}{\1}_{T}\|^2\leq T^{-1} \1_T^{\top}\M_{\F}{\1}_{T}\|\hat{\H}\hat{\bf{W}}^{\top}-{\bf{W}}^{\top}\|^2$ and
\begin{align*}
	\widehat{\bf{W}}^{\top}-\widehat{\mathbf{H}}^{-1} {\bf W}^{\top}=&-\frac{1}{N} \widehat{\mathbf{\Gamma}}^{\top}(\widehat{\mathbf{\Gamma}}-\mathbf{\Gamma} \widehat{\mathbf{H}}) \widehat{\mathbf{H}}^{-1} {\bf W}^{\top}+\frac{1}{N}(\widehat{\mathbf{\Gamma}}-\mathbf{\Gamma} \widehat{\mathbf{H}})^{\top} {\bm \eta}+\frac{1}{N} \widehat{\mathbf{H}}^{\top} \mathbf{\Gamma}^{\top} {\bm \eta}\\
	&+\frac{1}{N} \widehat{\mathbf{\Gamma}}^{\top}\bm\alpha{\1}_T^{\top}\M_{\F} -\frac{1}{N} \widehat{\mathbf{\Gamma}}^{\top} \bmv \f( \f\trans \f)^{-1}\f^{\top}.
	\end{align*}
and $ \widehat{\mathbf{\Gamma}}^{\top}\bmv \f( \f\trans \f)^{-1}\f^{\top}\M_{\F}{\1}_{T}=\bm 0$.
Moreover, $T^{-1}N^{-2}\|\widehat{\mathbf{\Gamma}}^{\top}\bm\alpha{\1}_T^{\top}\M_{\F}{\1}_T\|^2=o_p(1/\log N).$ 
Hence, we just need to prove that 
$$\left\|-\frac{1}{N} \widehat{\mathbf{\Gamma}}^{\top}(\widehat{\mathbf{\Gamma}}-\mathbf{\Gamma} \widehat{\mathbf{H}}) \widehat{\mathbf{H}}^{-1} {\bf W}^{\top}+\frac{1}{N}(\widehat{\mathbf{\Gamma}}-\mathbf{\Gamma} \widehat{\mathbf{H}})^{\top} {\bm \eta}+\frac{1}{N} \widehat{\mathbf{H}}^{\top} \mathbf{\Gamma}^{\top} {\bm \eta} \right\|^2=o_p(1/\log N).$$
By the proof of Theorem 3.2 in \cite{he2022large}, we have $$\left\|-\frac{1}{N} \widehat{\mathbf{\Gamma}}^{\top}(\widehat{\mathbf{\Gamma}}-\mathbf{\Gamma} \widehat{\mathbf{H}}) \widehat{\mathbf{H}}^{-1} {\bm W}_t+\frac{1}{N}(\widehat{\mathbf{\Gamma}}-\mathbf{\Gamma} \widehat{\mathbf{H}})^{\top} {\bm \eta}_t+\frac{1}{N} \widehat{\mathbf{H}}^{\top} \mathbf{\Gamma}^{\top} {\bm \eta}_t \right\|^2=O_p(T^{-2}+N^{-1}).$$
Hence, we have \begin{align*}
&\left\|-\frac{1}{N} \widehat{\mathbf{\Gamma}}^{\top}(\widehat{\mathbf{\Gamma}}-\mathbf{\Gamma} \widehat{\mathbf{H}}) \widehat{\mathbf{H}}^{-1} {\bf W}^{\top}+\frac{1}{N}(\widehat{\mathbf{\Gamma}}-\mathbf{\Gamma} \widehat{\mathbf{H}})^{\top} {\bm \eta}+\frac{1}{N} \widehat{\mathbf{H}}^{\top} \mathbf{\Gamma}^{\top} {\bm \eta} \right\|^2\\
\leq &\left\|-\frac{1}{N} \widehat{\mathbf{\Gamma}}^{\top}(\widehat{\mathbf{\Gamma}}-\mathbf{\Gamma} \widehat{\mathbf{H}}) \widehat{\mathbf{H}}^{-1} {\bf W}^{\top}+\frac{1}{N}(\widehat{\mathbf{\Gamma}}-\mathbf{\Gamma} \widehat{\mathbf{H}})^{\top} {\bm \eta}+\frac{1}{N} \widehat{\mathbf{H}}^{\top} \mathbf{\Gamma}^{\top} {\bm \eta} \right\|_F^2\\
\leq &T\left\|-\frac{1}{N} \widehat{\mathbf{\Gamma}}^{\top}(\widehat{\mathbf{\Gamma}}-\mathbf{\Gamma} \widehat{\mathbf{H}}) \widehat{\mathbf{H}}^{-1} {\bm W}_t+\frac{1}{N}(\widehat{\mathbf{\Gamma}}-\mathbf{\Gamma} \widehat{\mathbf{H}})^{\top} {\bm \eta}_t+\frac{1}{N} \widehat{\mathbf{H}}^{\top} \mathbf{\Gamma}^{\top} {\bm \eta}_t \right\|^2\\
=&O_p(T^{-1}+T/N)=o_p(1/\log N).
\end{align*}
Hence, we have $\|T^{-1}\sum_{t=1}^T\bm{\Delta}_t\|_{\infty}=o_p(1/\log N)$.
Finally, it is easy to prove that $$\|T^{-1}\breve{ r}_t^{-1}\sum_{t=1}^T\breve{\D}^{-1/2}\bm{\Delta}_t\|_{\infty}=o_p(1/\sqrt{\log N}).$$
Hence, according to the proof of Theorem 2 in \cite{zhao2024}, we have $T^{1/2}\breve \varsigma^{1/2} \breve{\theta}_i/\breve{d}_i^{1/2}\cd N(\bm 0, {\R})$. Similar to the proof of Theorem \ref{th1}, we can obtain the desired results.
 \hfill$\Box$
 
\subsection{Proof of Lemma \ref{KZeigen}}
Define 
\begin{align*}
	\K_\varepsilon=\frac{2}{T(T-1)}\sum_{i<j}U(\bmv_i-\bmv_j)U(\bmv_i-\bmv_j)^\top.
\end{align*}
Note that according to Lemma A.3 in \cite{he2022large}, we have $\lambda_{j}(\K_\varepsilon)\asymp r^{-1}$, if $j\leq r$ and $\lambda_{j}(\K_\varepsilon)=o_p(1)$, if $j\geq r$.

First, we will prove that $\|\K_\varepsilon-\K_Z\|=o_p(1)$. Let $\bmv=(\bmv_1,\dots,\bmv_T)\in\mathbb{R}^{N\times T}$ and $\vartheta_t=1-\bm 1_{T}^{\top}  \f( \f\trans \f)^{-1}\f_t$ for any $1\leq t\leq T$. Without loss of generality, we assume $T$ is even and $\bar{T}=T / 2$, otherwise we can delete the last observation. For any permutation $\sigma$ of $\{1, \ldots, T\}$, define $\vartheta_{t}^{\sigma}$, $\f_t^{\sigma},$ $\boldsymbol{Z}_t^\sigma$ and $\boldsymbol{\varepsilon}_t^\sigma$ as the corresponding $t$-th observation after permutation. Define ${\mathbf{K}}_Z^\sigma=\bar{T}^{-1} \sum_{s=1}^{\bar{T}} \left(\boldsymbol{Z}_{2 s-1}^\sigma-\boldsymbol{Z}_{2 s}^\sigma\right) /\left\|\boldsymbol{Z}_{2 s-1}^\sigma-\boldsymbol{Z}_{2 s}^\sigma\right\|$ and $${\mathbf{K}}_\varepsilon^\sigma=\bar{T}^{-1} \sum_{s=1}^{\bar{T}} \left(\boldsymbol{\varepsilon}_{2 s-1}^\sigma-\boldsymbol{\varepsilon}_{2 s}^\sigma\right)\left(\boldsymbol{\varepsilon}_{2 s-1}^\sigma-\boldsymbol{\varepsilon}_{2 s}^\sigma\right)^{\top} /\left\|\boldsymbol{\varepsilon}_{2 s-1}^\sigma-\boldsymbol{\varepsilon}_{2 s}^\sigma\right\|^2,$$ then we have
$$
\sum_{\sigma \in \mathcal{S}_T} \bar{T} {\mathbf{K}}_Z^\sigma=T \times(T-2)!\times \frac{T(T-1)}{2}{\mathbf{K}}_Z \Rightarrow {\mathbf{K}}_Z=\frac{1}{T!} \sum_{\sigma \in \mathcal{S}_T} {\mathbf{K}}_Z^\sigma,$$
and $$
\sum_{\sigma \in \mathcal{S}_T} \bar{T} {\mathbf{K}}_\varepsilon^\sigma=T \times(T-2)!\times \frac{T(T-1)}{2}{\mathbf{K}}_\varepsilon \Rightarrow {\mathbf{K}}_\varepsilon=\frac{1}{T!} \sum_{\sigma \in \mathcal{S}_T} {\mathbf{K}}_\varepsilon^\sigma,\\
$$
where $\mathcal{S}_T$ is the permutation group of $\{1, \ldots, T\}$. Let $m_{s,\sigma}=\|\bmv_{2s-1}^{\sigma}-\bmv_{2s}^{\sigma}\|$.
Note that \begin{align*}
	&\|\bm Z_{2s-1}^{\sigma}-\bm Z_{2s}^{\sigma}\|^2\\
	=&\|\bmv_{2s-1}^{\sigma}-\bmv_{2s}^{\sigma}\|^2+\|(\vartheta_{2s-1}^{\sigma}-\vartheta_{2s}^{\sigma})\bm\alpha\|^2+\|\bmv \f( \f\trans \f)^{-1}(\f_{2s-1}^{\sigma}-\f_{2s}^{\sigma})\|^2+2(\vartheta_{2s-1}^{\sigma}-\vartheta_{2s}^{\sigma})(\bmv_{2s-1}^{\sigma}-\bmv_{2s}^{\sigma})^{\top}\bm\alpha\\
	&+2(\bmv_{2s-1}^{\sigma}-\bmv_{2s}^{\sigma})^{\top}\bmv \f( \f\trans \f)^{-1}(\f_{2s-1}^{\sigma}-\f_{2s}^{\sigma})+2(\vartheta_{2s-1}^{\sigma}-\vartheta_{2s}^{\sigma})\bm\alpha^{\top}\bmv \f( \f\trans \f)^{-1}(\f_{2s-1}^{\sigma}-\f_{2s}^{\sigma})\\
	=&\|\bmv_{2s-1}^{\sigma}-\bmv_{2s}^{\sigma}\|^2(1+m_{s,\sigma}^{-2}\|(\vartheta_{2s-1}^{\sigma}-\vartheta_{2s}^{\sigma})\bm\alpha\|^2+m_{s,\sigma}^{-2}\|\bmv \f( \f\trans \f)^{-1}(\f_{2s-1}^{\sigma}-\f_{2s}^{\sigma})\|^2\\
	&+2m_{s,\sigma}^{-2}(\vartheta_{2s-1}^{\sigma}-\vartheta_{2s}^{\sigma})(\bmv_{2s-1}^{\sigma}-\bmv_{2s}^{\sigma})^{\top}\bm\alpha+2m_{s,\sigma}^{-2}(\bmv_{2s-1}^{\sigma}-\bmv_{2s}^{\sigma})^{\top}\bmv \f( \f\trans \f)^{-1}(\f_{2s-1}^{\sigma}-\f_{2s}^{\sigma})\\
	&+2m_{s,\sigma}^{-2}(\vartheta_{2s-1}^{\sigma}-\vartheta_{2s}^{\sigma})\bm\alpha^{\top}\bmv \f( \f\trans \f)^{-1}(\f_{2s-1}^{\sigma}-\f_{2s}^{\sigma})).
\end{align*}
Hence, by the first-order Taylor expansion, we have
\begin{align*}
	&\|\bm Z_{2s-1}^{\sigma}-\bm Z_{2s}^{\sigma}\|^{-2}\\
	=&\|\bmv_{2s-1}^{\sigma}-\bmv_{2s}^{\sigma}\|^{-2}(1+m_{s,\sigma}^{-2}\|(\vartheta_{2s-1}^{\sigma}-\vartheta_{2s}^{\sigma})\bm\alpha\|^2+m_{s,\sigma}^{-2}\|\bmv \f( \f\trans \f)^{-1}(\f_{2s-1}^{\sigma}-\f_{2s}^{\sigma})\|^2\\
	&+2m_{s,\sigma}^{-2}(\vartheta_{2s-1}^{\sigma}-\vartheta_{2s}^{\sigma})(\bmv_{2s-1}^{\sigma}-\bmv_{2s}^{\sigma})^{\top}\bm\alpha+2m_{s,\sigma}^{-2}(\bmv_{2s-1}^{\sigma}-\bmv_{2s}^{\sigma})^{\top}\bmv \f( \f\trans \f)^{-1}(\f_{2s-1}^{\sigma}-\f_{2s}^{\sigma})\\
	&+2m_{s,\sigma}^{-2}(\vartheta_{2s-1}^{\sigma}-\vartheta_{2s}^{\sigma})\bm\alpha^{\top}\bmv \f( \f\trans \f)^{-1}(\f_{2s-1}^{\sigma}-\f_{2s}^{\sigma}))^{-1}\\
	=&\|\bmv_{2s-1}^{\sigma}-\bmv_{2s}^{\sigma}\|^{-2}(1-m_{s,\sigma}^{-2}\|(\vartheta_{2s-1}^{\sigma}-\vartheta_{2s}^{\sigma})\bm\alpha\|^2-m_{s,\sigma}^{-2}\|\bmv \f( \f\trans \f)^{-1}(\f_{2s-1}^{\sigma}-\f_{2s}^{\sigma})\|^2\\
	&-2m_{s,\sigma}^{-2}(\vartheta_{2s-1}^{\sigma}-\vartheta_{2s}^{\sigma})(\bmv_{2s-1}^{\sigma}-\bmv_{2s}^{\sigma})^{\top}\bm\alpha-2m_{s,\sigma}^{-2}(\bmv_{2s-1}^{\sigma}-\bmv_{2s}^{\sigma})^{\top}\bmv \f( \f\trans \f)^{-1}(\f_{2s-1}^{\sigma}-\f_{2s}^{\sigma})\\
	&-2m_{s,\sigma}^{-2}(\vartheta_{2s-1}^{\sigma}-\vartheta_{2s}^{\sigma})\bm\alpha^{\top}\bmv \f( \f\trans \f)^{-1}(\f_{2s-1}^{\sigma}-\f_{2s}^{\sigma})+\delta_{1s}),
\end{align*}
where $\delta_{1s}=O_p(m_{s,\sigma}^{-2}\|(\vartheta_{2s-1}^{\sigma}-\vartheta_{2s}^{\sigma})\bm\alpha\|^2+m_{s,\sigma}^{-2}\|\bmv \f( \f\trans \f)^{-1}(\f_{2s-1}^{\sigma}-\f_{2s}^{\sigma})\|^2+2m_{s,\sigma}^{-2}(\vartheta_{2s-1}^{\sigma}-\vartheta_{2s}^{\sigma})(\bmv_{2s-1}^{\sigma}-\bmv_{2s}^{\sigma})^{\top}\bm\alpha+2m_{s,\sigma}^{-2}(\bmv_{2s-1}^{\sigma}-\bmv_{2s}^{\sigma})^{\top}\bmv \f( \f\trans \f)^{-1}(\f_{2s-1}^{\sigma}-\f_{2s}^{\sigma})+2m_{s,\sigma}^{-2}(\vartheta_{2s-1}^{\sigma}-\vartheta_{2s}^{\sigma})\bm\alpha^{\top}\bmv \f( \f\trans \f)^{-1}(\f_{2s-1}^{\sigma}-\f_{2s}^{\sigma}))^2=O_p(T^{-1})$.
Next, we have
\begin{align*}
	&\max_{s}|\frac{\|\bmv_{2s-1}^{\sigma}-\bmv_{2s}^{\sigma}\|^2}{\|\bm Z_{2s-1}^{\sigma}-\bm Z_{2s}^{\sigma}\|^2}-1|\\
	=&\max_{s}|-m_{s,\sigma}^{-2}\|(\vartheta_{2s-1}^{\sigma}-\vartheta_{2s}^{\sigma})\bm\alpha\|^2-m_{s,\sigma}^{-2}\|\bmv \f( \f\trans \f)^{-1}(\f_{2s-1}^{\sigma}-\f_{2s}^{\sigma})\|^2\\
	&-2m_{s,\sigma}^{-2}(\vartheta_{2s-1}^{\sigma}-\vartheta_{2s}^{\sigma})(\bmv_{2s-1}^{\sigma}-\bmv_{2s}^{\sigma})^{\top}\bm\alpha-2m_{s,\sigma}^{-2}(\bmv_{2s-1}^{\sigma}-\bmv_{2s}^{\sigma})^{\top}\bmv \f( \f\trans \f)^{-1}(\f_{2s-1}^{\sigma}-\f_{2s}^{\sigma})\\
	&-2m_{s,\sigma}^{-2}(\vartheta_{2s-1}^{\sigma}-\vartheta_{2s}^{\sigma})\bm\alpha^{\top}\bmv \f( \f\trans \f)^{-1}(\f_{2s-1}^{\sigma}-\f_{2s}^{\sigma})+\delta_{1s}|\\
	\leq&O(1)\{\max_{s}m_{s,\sigma}^{-2}\|\bm\alpha\|^2+\max_{s}m_{s,\sigma}^{-2}\max_{s}\|\bmv \f( \f\trans \f)^{-1}(\f_{2s-1}^{\sigma}-\f_{2s}^{\sigma})\|^2\\
	&+2\max_{s}m_{s,\sigma}^{-1}|U(\bmv_{2s-1}^{\sigma}-\bmv_{2s}^{\sigma})^{\top}\bm\alpha|+2\max_{s}m_{s,\sigma}^{-1}|U(\bmv_{2s-1}^{\sigma}-\bmv_{2s}^{\sigma})^{\top}\bmv \f( \f\trans \f)^{-1}(\f_{2s-1}^{\sigma}-\f_{2s}^{\sigma})|\\
	&+2\max_{s}m_{s,\sigma}^{-2}\max_{s}|\bm\alpha^{\top}\bmv \f( \f\trans \f)^{-1}(\f_{2s-1}^{\sigma}-\f_{2s}^{\sigma})|+\max_{s}\delta_{1s}\}\\
	=&O_p(T^{-1/4}),
\end{align*}
where the last equality holds because $\max_{s}\|\bmv \f( \f\trans \f)^{-1}\f_{2s-1}^{\sigma}\|^2=O_p(NT^{-1})$ by the proof of Theorem 1 in \cite{zhao2024}, $\|\bm\alpha\|^2=O(NT^{-1})$ and $\max_{s}m_{s,\sigma}^{-2}=O_p(N^{-1}T^{1/2})$. Similarly, we can also get $\bar{T}^{-1}\sum_{s=1}^{\bar{T}}(\|\bmv_{2s-1}^{\sigma}-\bmv_{2s}^{\sigma}\|^2/\|\bm Z_{2s-1}^{\sigma}-\bm Z_{2s}^{\sigma}\|^2-1)^2=O_p(T^{-1})$ and $\max_{s}|\|\bm Z_{2s-1}^{\sigma}-\bm Z_{2s}^{\sigma}\|^2/\|\bmv_{2s-1}^{\sigma}-\bmv_{2s}^{\sigma}\|^2-1|=O_p(T^{-1/4})$.
Then, we can obtain that 
\begin{align*}
&\|\K_Z-\K_\varepsilon\|\\
=&\Big\|\frac{1}{T!} \sum_{\sigma \in \mathcal{S}_T} {\mathbf{K}}_Z^\sigma-\frac{1}{T!} \sum_{\sigma \in \mathcal{S}_T} {\mathbf{K}}_\varepsilon^\sigma\Big\|	\\
\leq  &\Big\|{\mathbf{K}}_Z^\sigma-{\mathbf{K}}_\varepsilon^\sigma\Big\|\\
=&\Big\|\bar{T}^{-1} \sum_{s=1}^{\bar{T}} \frac{(\boldsymbol{Z}_{2 s-1}^\sigma-\boldsymbol{Z}_{2 s}^\sigma) (\boldsymbol{Z}_{2 s-1}^\sigma-\boldsymbol{Z}_{2 s}^\sigma) ^{\top}}{\|\boldsymbol{Z}_{2 s-1}^\sigma-\boldsymbol{Z}_{2 s}^\sigma\|^2}-\bar{T}^{-1} \sum_{s=1}^{\bar{T}} \frac{(\boldsymbol{\varepsilon}_{2 s-1}^\sigma-\boldsymbol{\varepsilon}_{2 s}^\sigma) (\boldsymbol{\varepsilon}_{2 s-1}^\sigma-\boldsymbol{\varepsilon}_{2 s}^\sigma) ^{\top}}{\|\boldsymbol{\varepsilon}_{2 s-1}^\sigma-\boldsymbol{\varepsilon}_{2 s}^\sigma\|^2}\Big\|\\
=&\Big\|\bar{T}^{-1} \sum_{s=1}^{\bar{T}} \frac{(\boldsymbol{\varepsilon}_{2 s-1}^\sigma-\boldsymbol{\varepsilon}_{2 s}^\sigma)(\boldsymbol{\varepsilon}_{2 s-1}^\sigma-\boldsymbol{\varepsilon}_{2 s}^\sigma)^\top}{\|\boldsymbol{\varepsilon}_{2 s-1}^\sigma-\boldsymbol{\varepsilon}_{2 s}^\sigma\|^2}\Big(\frac{\|\boldsymbol{\varepsilon}_{2 s-1}^\sigma-\boldsymbol{\varepsilon}_{2 s}^\sigma\|^2}{\|\boldsymbol{Z}_{2 s-1}^\sigma-\boldsymbol{Z}_{2 s}^\sigma\|^2}-1\Big)\Big\|\\
&+\Big\|\bar{T}^{-1} \sum_{s=1}^{\bar{T}} \frac{(\boldsymbol{Z}_{2 s-1}^\sigma-\boldsymbol{Z}_{2 s}^\sigma)(\boldsymbol{Z}_{2 s-1}^\sigma-\boldsymbol{Z}_{2 s}^\sigma)^\top-(\boldsymbol{\varepsilon}_{2 s-1}^\sigma-\boldsymbol{\varepsilon}_{2 s}^\sigma)(\boldsymbol{\varepsilon}_{2 s-1}^\sigma-\boldsymbol{\varepsilon}_{2 s}^\sigma)^\top}{\|\boldsymbol{\varepsilon}_{2 s-1}^\sigma-\boldsymbol{\varepsilon}_{2 s}^\sigma\|^2}\times\frac{\|\boldsymbol{\varepsilon}_{2 s-1}^\sigma-\boldsymbol{\varepsilon}_{2 s}^\sigma\|^2}{\|\boldsymbol{Z}_{2 s-1}^\sigma-\boldsymbol{Z}_{2 s}^\sigma\|^2}\Big\|\\
\leq &O_p(T^{-1/4}),
\end{align*}
where the proof of the last inequality uses basic properties of norms, $\|\bm\alpha\|^2=O(NT^{-1})$, $\max_{s}m_{s,\sigma}^{-2}=O_p(N^{-1}T^{1/2})$ and $\max_{s}\|\bmv \f( \f\trans \f)^{-1}\f_{2s-1}^{\sigma}\|^2=O_p(NT^{-1})$. Hence, we can conclude that $\|\K_\varepsilon-\K_Z\|=o_p(1)$.Then, from Weyl’s theorem, we can obtain that $\lambda_{j}(\K_Z)\asymp r^{-1}$, if $j\leq r$ and $\lambda_{j}(\K_Z)=o_p(1)$, if $j\geq r$.
 \hfill$\Box$
\subsection{Proof of Lemma \ref{deltagammaH}}
Define $\widehat{\boldsymbol{\Lambda}}$ as the diagonal matrix composed of the leading $r$ eigenvalues of ${\mathbf{K}}_Z$. The above results implies that $\widehat{\boldsymbol{\Lambda}}$ is asymptotically invertible, $\|\widehat{\boldsymbol{\Lambda}}\|_F=O_p(1)$ and $\big\|\widehat{\boldsymbol{\Lambda}}^{-1}\big\|_F=O_p(1)$. Because $\widehat{\mathbf{\Gamma}}=\sqrt{N} \tilde{\boldsymbol{\Gamma}}$ and $\tilde{\boldsymbol{\Gamma}}$ is composed of the leading eigenvectors of ${\mathbf{K}}_Z$, we have
$$
{\mathbf{K}}_Z \widehat{\mathbf{\Gamma}}=\widehat{\mathbf{\Gamma}} \widehat{\boldsymbol{\Lambda}}.
$$
According to the definition of ${\mathbf{K}}_Z$, then
\begin{align*}
	\widehat{\mathbf{\Gamma}} \widehat{\boldsymbol{\Lambda}}=	&\frac{2}{T(T-1)} \sum_{1 \leq t<s \leq T} \frac{\left(\boldsymbol{Z}_t-\boldsymbol{Z}_s\right)\left(\boldsymbol{Z}_t-\boldsymbol{Z}_s\right)^{\top}}{\left\|\boldsymbol{Z}_t-\boldsymbol{Z}_s\right\|^2} \widehat{\mathbf{\Gamma}}\\
	=&\frac{2}{T(T-1)} \sum_{1 \leq t<s \leq T} \frac{\left(\boldsymbol{Z}_t-\boldsymbol{Z}_s\right)\left(\boldsymbol{Z}_t-\boldsymbol{Z}_s\right)^{\top}}{\left\|\boldsymbol{\bmv}_t-\boldsymbol{\bmv}_s\right\|^2} \frac{\left\|\boldsymbol{\bmv}_t-\boldsymbol{\bmv}_s\right\|^2}{\left\|\boldsymbol{Z}_t-\boldsymbol{Z}_s\right\|^2}\widehat{\mathbf{\Gamma}}\\
	=&\frac{2}{T(T-1)} \sum_{1 \leq t<s \leq T} \frac{\left(\boldsymbol{\varepsilon}_t-\boldsymbol{\varepsilon}_s\right)\left(\boldsymbol{\varepsilon}_t-\boldsymbol{\varepsilon}_s\right)^{\top}}{\left\|\boldsymbol{\bmv}_t-\boldsymbol{\bmv}_s\right\|^2} \frac{\left\|\boldsymbol{\bmv}_t-\boldsymbol{\bmv}_s\right\|^2}{\left\|\boldsymbol{Z}_t-\boldsymbol{Z}_s\right\|^2}\widehat{\mathbf{\Gamma}}\\
	&+\frac{2}{T(T-1)} \sum_{1 \leq t<s \leq T} \frac{\left(\boldsymbol{Z}_t-\boldsymbol{Z}_s\right)\left(\boldsymbol{Z}_t-\boldsymbol{Z}_s\right)^{\top}-\left(\boldsymbol{\varepsilon}_t-\boldsymbol{\varepsilon}_s\right)\left(\boldsymbol{\varepsilon}_t-\boldsymbol{\varepsilon}_s\right)^{\top}}{\left\|\boldsymbol{\bmv}_t-\boldsymbol{\bmv}_s\right\|^2} \frac{\left\|\boldsymbol{\bmv}_t-\boldsymbol{\bmv}_s\right\|^2}{\left\|\boldsymbol{Z}_t-\boldsymbol{Z}_s\right\|^2}\widehat{\mathbf{\Gamma}}.
\end{align*}
To simplify notations, we denote 
\begin{align*}
	& \mathbf{M}_1=\frac{2}{T(T-1)} \sum_{1 \leq t<s \leq T} \frac{\left(\bm W_t-\bm W_s\right)\left(\bm W_t-\boldsymbol{W}_s\right)^{\top}}{\left\|\bmv_t-\bmv_s\right\|^2}, \\
	& \mathbf{M}_2=\frac{2}{T(T-1)} \sum_{1 \leq t<s \leq T} \frac{\left(\boldsymbol{\eta}_t-\boldsymbol{\eta}_s\right)\left(\boldsymbol{W}_t-\boldsymbol{W}_s\right)^{\top}}{\left\|\bmv_t-\bmv_s\right\|^2}, \\
	& \mathbf{M}_3=\frac{2}{T(T-1)} \sum_{1 \leq t<s \leq T} \frac{\left(\boldsymbol{W}_t-\boldsymbol{W}_s\right)\left(\boldsymbol{\eta}_t-\boldsymbol{\eta}_s\right)^{\top}}{\left\|\bmv_t-\bmv_s\right\|^2}, \\
	& \mathbf{M}_4=\frac{2}{T(T-1)} \sum_{1 \leq t<s \leq T} \frac{\left(\boldsymbol{\eta}_t-\boldsymbol{\eta}_s\right)\left(\boldsymbol{\eta}_t-\boldsymbol{\eta}_s\right)^{\top}}{\left\|\bmv_t-\bmv_s\right\|^2},
\end{align*}
and let $\hat{\mathbf{H}}=\mathbf{M}_1\mathbf{\Gamma}^{\top}\hat{\mathbf{\Gamma}}\hat{\mathbf \Lambda}^{-1}.$
Next, we have 
\begin{align}\label{A.1}
	&\hat{\mathbf{\Gamma}}\hat{\mathbf \Lambda}-\mathbf{\Gamma}\hat{\mathbf{H}}\hat{\mathbf \Lambda}\n\\
	=&(\mathbf{M}_2\mathbf{\Gamma}^{\top}\hat{\mathbf{\Gamma}}+\mathbf{\Gamma}\mathbf{M}_3\hat{\mathbf{\Gamma}}+\mathbf{M}_4\hat{\mathbf{\Gamma}})\hat{\mathbf{\Lambda}}+\frac{2}{T(T-1)} \sum_{1 \leq t<s \leq T}\frac{\left(\boldsymbol{\varepsilon}_t-\boldsymbol{\varepsilon}_s\right)\left(\boldsymbol{\varepsilon}_t-\boldsymbol{\varepsilon}_s\right)^{\top}}{\left\|\boldsymbol{\bmv}_t-\boldsymbol{\bmv}_s\right\|^2}\Big( \frac{\left\|\boldsymbol{\bmv}_t-\boldsymbol{\bmv}_s\right\|^2}{\left\|\boldsymbol{Z}_t-\boldsymbol{Z}_s\right\|^2}-1\Big)\widehat{\mathbf{\Gamma}}\hat{\mathbf{\Lambda}}\n\\
	&+\frac{2}{T(T-1)} \sum_{1 \leq t<s \leq T} \frac{\left(\boldsymbol{Z}_t-\boldsymbol{Z}_s\right)\left(\boldsymbol{Z}_t-\boldsymbol{Z}_s\right)^{\top}-\left(\boldsymbol{\varepsilon}_t-\boldsymbol{\varepsilon}_s\right)\left(\boldsymbol{\varepsilon}_t-\boldsymbol{\varepsilon}_s\right)^{\top}}{\left\|\boldsymbol{\bmv}_t-\boldsymbol{\bmv}_s\right\|^2} \frac{\left\|\boldsymbol{\bmv}_t-\boldsymbol{\bmv}_s\right\|^2}{\left\|\boldsymbol{Z}_t-\boldsymbol{Z}_s\right\|^2}\widehat{\mathbf{\Gamma}}\hat{\mathbf{\Lambda}}.
\end{align}
By Lemmas S2-S4 in the supplement materials of \cite{he2022large}, we have
$$
\left\|\mathbf{M}_2\right\|_F^2=O_p\left(\frac{1}{NT}+\frac{1}{N^3}\right), \quad\left\|\mathbf{M}_3\right\|_F^2=O_p\left(\frac{1}{NT}+\frac{1}{N^3}\right).
$$
Hence, we have $N^{-1}\|(\mathbf{M}_2\mathbf{\Gamma}^{\top}\hat{\mathbf{\Gamma}}+\mathbf{\Gamma}\mathbf{M}_3\hat{\mathbf{\Gamma}})\hat{\mathbf{\Lambda}}\|_F^2=O_p(N^{-2}+T^{-1})$.
Moreover, according to decomposition $\hat{\mathbf{\Gamma}}=\hat{\mathbf{\Gamma}}-{\mathbf{\Gamma}}\hat{\mathbf{H}}+{\mathbf{\Gamma}}\hat{\mathbf{H}}$, we have
$$\frac{1}{N}\|\mathbf{M}_4\hat{\mathbf{\Gamma}}\|_{F}^2\leq \frac{1}{N}\|\mathbf{M}_4{\mathbf{\Gamma}}\|_{F}^2\|\hat{\mathbf{H}}\|_{F}^2+\|\mathbf{M}_4\|_{F}^2\times\frac{1}{N}\|\hat{\mathbf{\Gamma}}-{\mathbf{\Gamma}}\hat{\mathbf{H}}\|_F^2.$$
Thus, according to Lemma S4 in the supplement materials of \cite{he2022large}, we can obtain that 
$$\frac{1}{N}\|\mathbf{M}_4\hat{\mathbf{\Gamma}}\|_{F}^2=O_p(\frac{1}{T}+\frac{1}{N^2})+o_p(1)\times\frac{1}{N}\|\hat{\mathbf{\Gamma}}-{\mathbf{\Gamma}}\hat{\mathbf{H}}\|_F^2.$$
Hence, we have
\begin{align*}
	&\frac{1}{N}\|\hat{\mathbf{\Gamma}}-\mathbf{\Gamma}\hat{\mathbf{H}}\|_F^2\\
	=&O_p(\frac{1}{T}+\frac{1}{N^2})+o_p(1)\times	\frac{1}{N}\|\hat{\mathbf{\Gamma}}-\mathbf{\Gamma}\hat{\mathbf{H}}\|_F^2\\
	&+\frac{1}{N}\Big\|\frac{2}{T(T-1)} \sum_{1 \leq t<s \leq T}\frac{\left(\boldsymbol{\varepsilon}_t-\boldsymbol{\varepsilon}_s\right)\left(\boldsymbol{\varepsilon}_t-\boldsymbol{\varepsilon}_s\right)^{\top}}{\left\|\boldsymbol{\bmv}_t-\boldsymbol{\bmv}_s\right\|^2}\Big( \frac{\left\|\boldsymbol{\bmv}_t-\boldsymbol{\bmv}_s\right\|^2}{\left\|\boldsymbol{Z}_t-\boldsymbol{Z}_s\right\|^2}-1\Big)\widehat{\mathbf{\Gamma}}\Big\|_F^2\\
	&+\frac{1}{N}\Big\|\frac{2}{T(T-1)} \sum_{1 \leq t<s \leq T} \frac{\left(\boldsymbol{Z}_t-\boldsymbol{Z}_s\right)\left(\boldsymbol{Z}_t-\boldsymbol{Z}_s\right)^{\top}-\left(\boldsymbol{\varepsilon}_t-\boldsymbol{\varepsilon}_s\right)\left(\boldsymbol{\varepsilon}_t-\boldsymbol{\varepsilon}_s\right)^{\top}}{\left\|\boldsymbol{\bmv}_t-\boldsymbol{\bmv}_s\right\|^2} \frac{\left\|\boldsymbol{\bmv}_t-\boldsymbol{\bmv}_s\right\|^2}{\left\|\boldsymbol{Z}_t-\boldsymbol{Z}_s\right\|^2}\widehat{\mathbf{\Gamma}}\Big\|_F^2\\
	\leq&O_p(\frac{1}{T}+\frac{1}{N^2})+o_p(1)\times	\frac{1}{N}\|\hat{\mathbf{\Gamma}}-\mathbf{\Gamma}\hat{\mathbf{H}}\|_F^2\\
	&+\Big\|\frac{2}{T(T-1)} \sum_{1 \leq t<s \leq T}\frac{\left(\boldsymbol{\varepsilon}_t-\boldsymbol{\varepsilon}_s\right)\left(\boldsymbol{\varepsilon}_t-\boldsymbol{\varepsilon}_s\right)^{\top}}{\left\|\boldsymbol{\bmv}_t-\boldsymbol{\bmv}_s\right\|^2}\Big( \frac{\left\|\boldsymbol{\bmv}_t-\boldsymbol{\bmv}_s\right\|^2}{\left\|\boldsymbol{Z}_t-\boldsymbol{Z}_s\right\|^2}-1\Big)\Big\|_F^2\\
	&+\Big\|\frac{2}{T(T-1)} \sum_{1 \leq t<s \leq T} \frac{\left(\boldsymbol{Z}_t-\boldsymbol{Z}_s\right)\left(\boldsymbol{Z}_t-\boldsymbol{Z}_s\right)^{\top}-\left(\boldsymbol{\varepsilon}_t-\boldsymbol{\varepsilon}_s\right)\left(\boldsymbol{\varepsilon}_t-\boldsymbol{\varepsilon}_s\right)^{\top}}{\left\|\boldsymbol{\bmv}_t-\boldsymbol{\bmv}_s\right\|^2} \frac{\left\|\boldsymbol{\bmv}_t-\boldsymbol{\bmv}_s\right\|^2}{\left\|\boldsymbol{Z}_t-\boldsymbol{Z}_s\right\|^2}\Big\|_F^2.
\end{align*}
Note that
\begin{align*}
	&\Big\|\frac{2}{T(T-1)} \sum_{1 \leq t<s \leq T}\frac{\left(\boldsymbol{\varepsilon}_t-\boldsymbol{\varepsilon}_s\right)\left(\boldsymbol{\varepsilon}_t-\boldsymbol{\varepsilon}_s\right)^{\top}}{\left\|\boldsymbol{\bmv}_t-\boldsymbol{\bmv}_s\right\|^2}\Big( \frac{\left\|\boldsymbol{\bmv}_t-\boldsymbol{\bmv}_s\right\|^2}{\left\|\boldsymbol{Z}_t-\boldsymbol{Z}_s\right\|^2}-1\Big)\Big\|_F\\
	=&\Big\|\frac{1}{T!}\sum_{\sigma \in \mathcal{S}_T}\bar{T}^{-1} \sum_{s=1}^{\bar{T}}\frac{\left(\boldsymbol{\varepsilon}_{2s-1}^{\sigma}-\boldsymbol{\varepsilon}_{2s}^{\sigma}\right)\left(\boldsymbol{\varepsilon}_{2s-1}^{\sigma}-\boldsymbol{\varepsilon}_{2s}^{\sigma}\right)^{\top}}{\left\|\boldsymbol{\varepsilon}_{2s-1}^{\sigma}-\boldsymbol{\bmv}_{2s}^{\sigma}\right\|^2}\Big( \frac{\left\|\boldsymbol{\varepsilon}_{2s-1}^{\sigma}-\boldsymbol{\bmv}_{2s}^{\sigma}\right\|^2}{\left\|\boldsymbol{Z}_{2s-1}^{\sigma}-\boldsymbol{Z}_{2s}^{\sigma}\right\|^2}-1\Big) \Big\|_{F}\\
	\leq&\Big\|\bar{T}^{-1} \sum_{s=1}^{\bar{T}}\frac{\left(\boldsymbol{\varepsilon}_{2s-1}^{\sigma}-\boldsymbol{\varepsilon}_{2s}^{\sigma}\right)\left(\boldsymbol{\varepsilon}_{2s-1}^{\sigma}-\boldsymbol{\varepsilon}_{2s}^{\sigma}\right)^{\top}}{\left\|\boldsymbol{\varepsilon}_{2s-1}^{\sigma}-\boldsymbol{\bmv}_{2s}^{\sigma}\right\|^2}\Big( \frac{\left\|\boldsymbol{\varepsilon}_{2s-1}^{\sigma}-\boldsymbol{\bmv}_{2s}^{\sigma}\right\|^2}{\left\|\boldsymbol{Z}_{2s-1}^{\sigma}-\boldsymbol{Z}_{2s}^{\sigma}\right\|^2}-1\Big) \Big\|_{F}\\
	\leq&\bar{T}^{-1} \sum_{s=1}^{\bar{T}}\Big| \frac{\left\|\boldsymbol{\varepsilon}_{2s-1}^{\sigma}-\boldsymbol{\bmv}_{2s}^{\sigma}\right\|^2}{\left\|\boldsymbol{Z}_{2s-1}^{\sigma}-\boldsymbol{Z}_{2s}^{\sigma}\right\|^2}-1\Big|\Big\|\frac{\left(\boldsymbol{\varepsilon}_{2s-1}^{\sigma}-\boldsymbol{\varepsilon}_{2s}^{\sigma}\right)\left(\boldsymbol{\varepsilon}_{2s-1}^{\sigma}-\boldsymbol{\varepsilon}_{2s}^{\sigma}\right)^{\top}}{\left\|\boldsymbol{\varepsilon}_{2s-1}^{\sigma}-\boldsymbol{\bmv}_{2s}^{\sigma}\right\|^2}\Big\|_{F}\\
	\leq&\bar{T}^{-1} \sum_{s=1}^{\bar{T}}\Big| \frac{\left\|\boldsymbol{\varepsilon}_{2s-1}^{\sigma}-\boldsymbol{\bmv}_{2s}^{\sigma}\right\|^2}{\left\|\boldsymbol{Z}_{2s-1}^{\sigma}-\boldsymbol{Z}_{2s}^{\sigma}\right\|^2}-1\Big|=O_p(T^{-1/2}),
\end{align*}
we have 
\begin{align}\label{J1}
	\Big\|\frac{2}{T(T-1)} \sum_{1 \leq t<s \leq T}\frac{\left(\boldsymbol{\varepsilon}_t-\boldsymbol{\varepsilon}_s\right)\left(\boldsymbol{\varepsilon}_t-\boldsymbol{\varepsilon}_s\right)^{\top}}{\left\|\boldsymbol{\bmv}_t-\boldsymbol{\bmv}_s\right\|^2}\Big( \frac{\left\|\boldsymbol{\bmv}_t-\boldsymbol{\bmv}_s\right\|^2}{\left\|\boldsymbol{Z}_t-\boldsymbol{Z}_s\right\|^2}-1\Big)\Big\|_F^2=O_{p}(T^{-1}).
\end{align}
Similarly, we have 
\begin{align}\label{J2}
	&\Big\|\frac{2}{T(T-1)} \sum_{1 \leq t<s \leq T} \frac{\left(\boldsymbol{Z}_t-\boldsymbol{Z}_s\right)\left(\boldsymbol{Z}_t-\boldsymbol{Z}_s\right)^{\top}-\left(\boldsymbol{\varepsilon}_t-\boldsymbol{\varepsilon}_s\right)\left(\boldsymbol{\varepsilon}_t-\boldsymbol{\varepsilon}_s\right)^{\top}}{\left\|\boldsymbol{\bmv}_t-\boldsymbol{\bmv}_s\right\|^2} \frac{\left\|\boldsymbol{\bmv}_t-\boldsymbol{\bmv}_s\right\|^2}{\left\|\boldsymbol{Z}_t-\boldsymbol{Z}_s\right\|^2}\Big\|_F\n\\
	\leq &\bar{T}^{-1} \sum_{s=1}^{\bar{T}}\frac{\left\|\boldsymbol{\bmv}_t-\boldsymbol{\bmv}_s\right\|^2}{\left\|\boldsymbol{Z}_t-\boldsymbol{Z}_s\right\|^2}\Big\|\frac{\left(\boldsymbol{Z}_{2s-1}^{\sigma}-\boldsymbol{Z}_{2s}^{\sigma}\right)\left(\boldsymbol{Z}_{2s-1}^{\sigma}-\boldsymbol{Z}_{2s}^{\sigma}\right)^{\top}-\left(\boldsymbol{\varepsilon}_{2s-1}^{\sigma}-\boldsymbol{\varepsilon}_{2s}^{\sigma}\right)\left(\boldsymbol{\varepsilon}_{2s-1}^{\sigma}-\boldsymbol{\varepsilon}_{2s}^{\sigma}\right)^{\top}}{\left\|\boldsymbol{\varepsilon}_{2s-1}^{\sigma}-\boldsymbol{\bmv}_{2s}^{\sigma}\right\|^2}\Big\|_{F}\n\\
	\leq &2\bar{T}^{-1} \sum_{s=1}^{\bar{T}}\Big\|\frac{\left(\boldsymbol{Z}_{2s-1}^{\sigma}-\boldsymbol{Z}_{2s}^{\sigma}\right)\left(\boldsymbol{Z}_{2s-1}^{\sigma}-\boldsymbol{Z}_{2s}^{\sigma}\right)^{\top}-\left(\boldsymbol{\varepsilon}_{2s-1}^{\sigma}-\boldsymbol{\varepsilon}_{2s}^{\sigma}\right)\left(\boldsymbol{\varepsilon}_{2s-1}^{\sigma}-\boldsymbol{\varepsilon}_{2s}^{\sigma}\right)^{\top}}{\left\|\boldsymbol{\varepsilon}_{2s-1}^{\sigma}-\boldsymbol{\bmv}_{2s}^{\sigma}\right\|^2}\Big\|_{F}\n\\
	\leq&2\bar{T}^{-1} \sum_{s=1}^{\bar{T}}\Big\|m_{s,\sigma}^{-2}(\vartheta_{2s-1}^{\sigma}-\vartheta_{2s}^{\sigma})^2\bm\alpha\bm\alpha^{\top}+m_{s,\sigma}^{-2}\bmv \f( \f\trans \f)^{-1}(\f_{2s-1}^{\sigma}-\f_{2s}^{\sigma})(\f_{2s-1}^{\sigma}-\f_{2s}^{\sigma})^{\top}( \f\trans \f)^{-1} \f^{\top}\bmv^{\top}\n\\
	&+m_{s,\sigma}^{-2}(\vartheta_{2s-1}^{\sigma}-\vartheta_{2s}^{\sigma})\bm\alpha(\bmv_{2s-1}^{\sigma}-\bmv_{2s}^{\sigma})^{\top}+m_{s,\sigma}^{-2}\bmv \f( \f\trans \f)^{-1}(\f_{2s-1}^{\sigma}-\f_{2s}^{\sigma})(\bmv_{2s-1}^{\sigma}-\bmv_{2s}^{\sigma})^{\top}\n\\
	&+m_{s,\sigma}^{-2}(\vartheta_{2s-1}^{\sigma}-\vartheta_{2s}^{\sigma})\bmv \f( \f\trans \f)^{-1}(\f_{2s-1}^{\sigma}-\f_{2s}^{\sigma})\bm\alpha^{\top}\n\\
	&+m_{s,\sigma}^{-2}(\vartheta_{2s-1}^{\sigma}-\vartheta_{2s}^{\sigma})(\bmv_{2s-1}^{\sigma}-\bmv_{2s}^{\sigma})\bm\alpha^{\top}+m_{s,\sigma}^{-2}(\bmv_{2s-1}^{\sigma}-\bmv_{2s}^{\sigma})(\f_{2s-1}^{\sigma}-\f_{2s}^{\sigma})^{\top}( \f\trans \f)^{-1} \f^{\top}\bmv^{\top}\n\\
	&+m_{s,\sigma}^{-2}(\vartheta_{2s-1}^{\sigma}-\vartheta_{2s}^{\sigma})\bm\alpha(\f_{2s-1}^{\sigma}-\f_{2s}^{\sigma})^{\top}( \f\trans \f)^{-1} \f^{\top}\bmv^{\top}	\Big\|_{F}=O_p(T^{-1/2}).
\end{align}
Hence, we can obtain that 
\begin{align*}
	&\frac{1}{N}\|\hat{\mathbf{\Gamma}}-\mathbf{\Gamma}\hat{\mathbf{H}}\|_F^2=O_p(\frac{1}{T}+\frac{1}{N^2})+o_p(1)\times	\frac{1}{N}\|\hat{\mathbf{\Gamma}}-\mathbf{\Gamma}\hat{\mathbf{H}}\|_F^2.
\end{align*}
  \hfill$\Box$
  
  \subsection{Proof of Lemma \ref{deltaWH}}
 Because $\hat{\mathbf{H}}^{\top}\mathbf{V}\hat{\mathbf{H}}\cp \mathbf{I}_r$, we have $\|\hat{\mathbf{H}}\|_F=O_p(1)$ and $\widehat{\mathbf{H}}$ is invertible with probability approaching to 1. By our robust estimation procedure,
  \begin{align*}
  	\widehat{\boldsymbol{W}}_t  =&\frac{1}{N} \widehat{\mathbf{\Gamma}}^{\top} \boldsymbol{Z}_t=\frac{1}{N} \widehat{\mathbf{\Gamma}}^{\top}\left(\mathbf{\Gamma} \boldsymbol{W}_t+\boldsymbol{\eta}_t+\vartheta_t\bm\alpha-\bmv \f( \f\trans \f)^{-1}\f_{t}\right) \\
  	=&\frac{1}{N} \widehat{\mathbf{\Gamma}}^{\top}\left(\widehat{\mathbf{\Gamma}} \widehat{\mathbf{H}}^{-1}-\left(\widehat{\mathbf{\Gamma}} \widehat{\mathbf{H}}^{-1}-\mathbf{\Gamma}\right)\right) \boldsymbol{W}_t+\frac{1}{N} \widehat{\mathbf{\Gamma}}^{\top} \boldsymbol{\eta}_t\\
  	&+\frac{1}{N} \widehat{\mathbf{\Gamma}}^{\top} \vartheta_t\bm\alpha-\frac{1}{N} \widehat{\mathbf{\Gamma}}^{\top} \bmv \f( \f\trans \f)^{-1}\f_{t}.
  \end{align*}
  Note that $N^{-1} \widehat{\mathbf{\Gamma}}^{\top} \widehat{\mathbf{\Gamma}}=\mathbf{I}_r$, then
  \begin{align*}
  	\widehat{\boldsymbol{W}}_t-\widehat{\mathbf{H}}^{-1} \boldsymbol{W}_t=&-\frac{1}{N} \widehat{\mathbf{\Gamma}}^{\top}(\widehat{\mathbf{\Gamma}}-\mathbf{\Gamma} \widehat{\mathbf{H}}) \widehat{\mathbf{H}}^{-1} \boldsymbol{W}_t+\frac{1}{N}(\widehat{\mathbf{\Gamma}}-\mathbf{\Gamma} \widehat{\mathbf{H}})^{\top} \boldsymbol{\eta}_t+\frac{1}{N} \widehat{\mathbf{H}}^{\top} \mathbf{\Gamma}^{\top} \boldsymbol{\eta}_t\\
  	&+\frac{1}{N} \widehat{\mathbf{\Gamma}}^{\top} \vartheta_t\bm\alpha-\frac{1}{N} \widehat{\mathbf{\Gamma}}^{\top} \bmv \f( \f\trans \f)^{-1}\f_{t}.
  \end{align*}
  Note that
  $$
  \frac{1}{N} \widehat{\mathbf{\Gamma}}^{\top}(\widehat{\mathbf{\Gamma}}-\mathbf{\Gamma} \widehat{\mathbf{H}})=\frac{1}{N}(\widehat{\mathbf{\Gamma}}-\mathbf{\Gamma} \widehat{\mathbf{H}})^{\top}(\widehat{\mathbf{\Gamma}}-\mathbf{\Gamma} \widehat{\mathbf{H}})+\frac{1}{N} \widehat{\mathbf{H}}^{\top} \mathbf{\Gamma}^{\top}(\widehat{\mathbf{\Gamma}}-\mathbf{\Gamma} \widehat{\mathbf{H}})
  $$
  while $N^{-1}\|\widehat{\mathbf{\Gamma}}-\mathbf{\Gamma} \widehat{\mathbf{H}}\|_F^2=O_p\left(T^{-1}+N^{-2}\right)$ and $\|\widehat{\mathbf{H}}\|_F^2=O_p(1)$. 
  By \eqref{A.1},
  \begin{align*}
  	\frac{1}{N} \mathbf{\Gamma}^{\top}(\widehat{\mathbf{\Gamma}}-\mathbf{\Gamma} \widehat{\mathbf{H}})=&\left(\frac{1}{N} \mathbf{\Gamma}^{\top} \mathbf{M}_2 \mathbf{\Gamma}^{\top} \widehat{\mathbf{\Gamma}}+\frac{1}{N} \mathbf{\Gamma}^{\top} \mathbf{\Gamma} \mathbf{M}_3 \widehat{\mathbf{\Gamma}}+\frac{1}{N} \mathbf{\Gamma}^{\top} \mathbf{M}_4 \widehat{\mathbf{\Gamma}}\right) \widehat{\boldsymbol{\Lambda}}^{-1}\n\\
  	&+\frac{1}{N}\mathbf{\Gamma}^{\top}\mathbf{J}_1\widehat{\mathbf{\Gamma}}\widehat{\boldsymbol{\Lambda}}^{-1}
  	+\frac{1}{N}\mathbf{\Gamma}^{\top}\mathbf{J}_2\widehat{\mathbf{\Gamma}}\widehat{\boldsymbol{\Lambda}}^{-1}
  \end{align*}
  where
  \begin{align*}
  	\mathbf{J}_1=\frac{2}{T(T-1)} \sum_{1 \leq t<s \leq T}\frac{\left(\boldsymbol{\varepsilon}_t-\boldsymbol{\varepsilon}_s\right)\left(\boldsymbol{\varepsilon}_t-\boldsymbol{\varepsilon}_s\right)^{\top}}{\left\|\boldsymbol{\bmv}_t-\boldsymbol{\bmv}_s\right\|^2}\Big( \frac{\left\|\boldsymbol{\bmv}_t-\boldsymbol{\bmv}_s\right\|^2}{\left\|\boldsymbol{Z}_t-\boldsymbol{Z}_s\right\|^2}-1\Big)
  \end{align*}
  and 
  \begin{align*}
  	\mathbf{J}_2=\frac{2}{T(T-1)} \sum_{1 \leq t<s \leq T} \frac{\left(\boldsymbol{Z}_t-\boldsymbol{Z}_s\right)\left(\boldsymbol{Z}_t-\boldsymbol{Z}_s\right)^{\top}-\left(\boldsymbol{\varepsilon}_t-\boldsymbol{\varepsilon}_s\right)\left(\boldsymbol{\varepsilon}_t-\boldsymbol{\varepsilon}_s\right)^{\top}}{\left\|\boldsymbol{\bmv}_t-\boldsymbol{\bmv}_s\right\|^2} \frac{\left\|\boldsymbol{\bmv}_t-\boldsymbol{\bmv}_s\right\|^2}{\left\|\boldsymbol{Z}_t-\boldsymbol{Z}_s\right\|^2}.
  \end{align*}
  According to Lemma S5 in \cite{he2022large}, we have $$\|\big(N^{-1} \mathbf{\Gamma}^{\top} \mathbf{M}_2 \mathbf{\Gamma}^{\top} \widehat{\mathbf{\Gamma}}+N^{-1} \mathbf{\Gamma}^{\top} \mathbf{\Gamma} \mathbf{M}_3 \widehat{\mathbf{\Gamma}}+N^{-1} \mathbf{\Gamma}^{\top} \mathbf{M}_4 \widehat{\mathbf{\Gamma}}\big) \widehat{\boldsymbol{\Lambda}}^{-1}\|_{F}^2=O_p(T^{-2}+N^{-2}).$$
  Then, 
  \begin{align*}
  	&\|N^{-1}\mathbf{\Gamma}^{\top}\mathbf{J}_1\widehat{\mathbf{\Gamma}}\widehat{\boldsymbol{\Lambda}}^{-1}\|_F^2\\
  	=&O_p(N^{-2})\|\mathbf{\Gamma}^{\top}\mathbf{J}_1\widehat{\mathbf{\Gamma}}\|_F^2\\
  	=&O_p(N^{-2})\|\mathbf{\Gamma}^{\top}\mathbf{J}_1\mathbf{\Gamma}\|_F^2+O_p(N^{-2})\|\mathbf{\Gamma}^{\top}\mathbf{J}_1\|_F^2\times\|\widehat{\mathbf{\Gamma}}-\mathbf{\Gamma}\widehat{\mathbf{H}}\|_F^2.
  \end{align*}
  Due the basic properties of norms, we have  
  \begin{align*}
  	\|\mathbf{\Gamma}^{\top}\mathbf{J}_1\mathbf{\Gamma}\|_F\leq \Big\|\frac{1}{\bar{T}} \sum_{s=1}^{\bar{T}}\mathbf{\Gamma}^{\top}\frac{\left(\boldsymbol{\varepsilon}_{2s-1}^{\sigma}-\boldsymbol{\varepsilon}_{2s}^{\sigma}\right)\left(\boldsymbol{\varepsilon}_{2s-1}^{\sigma}-\boldsymbol{\varepsilon}_{2s}^{\sigma}\right)^{\top}}{\left\|\boldsymbol{\bmv}_{2s-1}^{\sigma}-\boldsymbol{\bmv}_{2s}^{\sigma}\right\|^2}\Big( \frac{\left\|\boldsymbol{\bmv}_{2s-1}^{\sigma}-\boldsymbol{\bmv}_{2s}^{\sigma}\right\|^2}{\left\|\boldsymbol{Z}_{2s-1}^{\sigma}-\boldsymbol{Z}_{2s}^{\sigma}\right\|^2}-1\Big)\mathbf{\Gamma}\Big\|_F
  \end{align*}
  and 
  \begin{align*}
  	&\Big\|\frac{1}{\bar{T}} \sum_{s=1}^{\bar{T}}\mathbf{\Gamma}^{\top}\frac{\left(\boldsymbol{\varepsilon}_{2s-1}^{\sigma}-\boldsymbol{\varepsilon}_{2s}^{\sigma}\right)\left(\boldsymbol{\varepsilon}_{2s-1}^{\sigma}-\boldsymbol{\varepsilon}_{2s}^{\sigma}\right)^{\top}}{\left\|\boldsymbol{\bmv}_{2s-1}^{\sigma}-\boldsymbol{\bmv}_{2s}^{\sigma}\right\|^2}\Big( \frac{\left\|\boldsymbol{\bmv}_{2s-1}^{\sigma}-\boldsymbol{\bmv}_{2s}^{\sigma}\right\|^2}{\left\|\boldsymbol{Z}_{2s-1}^{\sigma}-\boldsymbol{Z}_{2s}^{\sigma}\right\|^2}-1\Big)\mathbf{\Gamma}\Big\|_F^2\\
  	= &\frac{1}{\bar{T}^2}\sum_{s=1}^{\bar{T}}\tr\Big(\mathbf{\Gamma}^{\top}\frac{\left(\boldsymbol{\varepsilon}_{2s-1}^{\sigma}-\boldsymbol{\varepsilon}_{2s}^{\sigma}\right)\left(\boldsymbol{\varepsilon}_{2s-1}^{\sigma}-\boldsymbol{\varepsilon}_{2s}^{\sigma}\right)^{\top}}{\left\|\boldsymbol{\bmv}_{2s-1}^{\sigma}-\boldsymbol{\bmv}_{2s}^{\sigma}\right\|^2}\Big( \frac{\left\|\boldsymbol{\bmv}_{2s-1}^{\sigma}-\boldsymbol{\bmv}_{2s}^{\sigma}\right\|^2}{\left\|\boldsymbol{Z}_{2s-1}^{\sigma}-\boldsymbol{Z}_{2s}^{\sigma}\right\|^2}-1\Big)\mathbf{\Gamma}\Big)^2\\
  	&+\frac{1}{\bar{T}^2}\underset{s\neq t}{\sum^{\bar{T}}\sum^{\bar{T}}}
  	\tr\Big(\mathbf{\Gamma}^{\top}\frac{\left(\boldsymbol{\varepsilon}_{2s-1}^{\sigma}-\boldsymbol{\varepsilon}_{2s}^{\sigma}\right)\left(\boldsymbol{\varepsilon}_{2s-1}^{\sigma}-\boldsymbol{\varepsilon}_{2s}^{\sigma}\right)^{\top}}{\left\|\boldsymbol{\bmv}_{2s-1}^{\sigma}-\boldsymbol{\bmv}_{2s}^{\sigma}\right\|^2}\Big( \frac{\left\|\boldsymbol{\bmv}_{2s-1}^{\sigma}-\boldsymbol{\bmv}_{2s}^{\sigma}\right\|^2}{\left\|\boldsymbol{Z}_{2s-1}^{\sigma}-\boldsymbol{Z}_{2s}^{\sigma}\right\|^2}-1\Big)\mathbf{\Gamma}\\
  	&\quad\times\mathbf{\Gamma}^{\top}\frac{\left(\boldsymbol{\varepsilon}_{2t-1}^{\sigma}-\boldsymbol{\varepsilon}_{2t}^{\sigma}\right)\left(\boldsymbol{\varepsilon}_{2t-1}^{\sigma}-\boldsymbol{\varepsilon}_{2t}^{\sigma}\right)^{\top}}{\left\|\boldsymbol{\bmv}_{2t-1}^{\sigma}-\boldsymbol{\bmv}_{2t}^{\sigma}\right\|^2}\Big( \frac{\left\|\boldsymbol{\bmv}_{2t-1}^{\sigma}-\boldsymbol{\bmv}_{2t}^{\sigma}\right\|^2}{\left\|\boldsymbol{Z}_{2t-1}^{\sigma}-\boldsymbol{Z}_{2t}^{\sigma}\right\|^2}-1\Big)\mathbf{\Gamma}\Big),
  \end{align*}
  where
  \begin{align*}
  	&\tr\Big(\mathbf{\Gamma}^{\top}\frac{\left(\boldsymbol{\varepsilon}_{2s-1}^{\sigma}-\boldsymbol{\varepsilon}_{2s}^{\sigma}\right)\left(\boldsymbol{\varepsilon}_{2s-1}^{\sigma}-\boldsymbol{\varepsilon}_{2s}^{\sigma}\right)^{\top}}{\left\|\boldsymbol{\bmv}_{2s-1}^{\sigma}-\boldsymbol{\bmv}_{2s}^{\sigma}\right\|^2}\Big( \frac{\left\|\boldsymbol{\bmv}_{2s-1}^{\sigma}-\boldsymbol{\bmv}_{2s}^{\sigma}\right\|^2}{\left\|\boldsymbol{Z}_{2s-1}^{\sigma}-\boldsymbol{Z}_{2s}^{\sigma}\right\|^2}-1\Big)\mathbf{\Gamma}\Big)^2\\
  	=&\Big( \frac{\left\|\boldsymbol{\bmv}_{2s-1}^{\sigma}-\boldsymbol{\bmv}_{2s}^{\sigma}\right\|^2}{\left\|\boldsymbol{Z}_{2s-1}^{\sigma}-\boldsymbol{Z}_{2s}^{\sigma}\right\|^2}-1\Big)^2\bigg(\frac{\left(\boldsymbol{\varepsilon}_{2s-1}^{\sigma}-\boldsymbol{\varepsilon}_{2s}^{\sigma}\right)^{\top}}{\left\|\boldsymbol{\bmv}_{2s-1}^{\sigma}-\boldsymbol{\bmv}_{2s}^{\sigma}\right\|}\mathbf{\Gamma}\mathbf{\Gamma}^{\top}\frac{\left(\boldsymbol{\varepsilon}_{2s-1}^{\sigma}-\boldsymbol{\varepsilon}_{2s}^{\sigma}\right)}{\left\|\boldsymbol{\bmv}_{2s-1}^{\sigma}-\boldsymbol{\bmv}_{2s}^{\sigma}\right\|}
  	\bigg)^2\\
  	=&O_p(N^2T^{-1})
  \end{align*}
  and 
  \begin{align*}
  	&\mathbb{E}\tr\Big(\mathbf{\Gamma}^{\top}\frac{\left(\boldsymbol{\varepsilon}_{2s-1}^{\sigma}-\boldsymbol{\varepsilon}_{2s}^{\sigma}\right)\left(\boldsymbol{\varepsilon}_{2s-1}^{\sigma}-\boldsymbol{\varepsilon}_{2s}^{\sigma}\right)^{\top}}{\left\|\boldsymbol{\bmv}_{2s-1}^{\sigma}-\boldsymbol{\bmv}_{2s}^{\sigma}\right\|^2}\mathbf{\Gamma}\mathbf{\Gamma}^{\top}\frac{\left(\boldsymbol{\varepsilon}_{2t-1}^{\sigma}-\boldsymbol{\varepsilon}_{2t}^{\sigma}\right)\left(\boldsymbol{\varepsilon}_{2t-1}^{\sigma}-\boldsymbol{\varepsilon}_{2t}^{\sigma}\right)^{\top}}{\left\|\boldsymbol{\bmv}_{2t-1}^{\sigma}-\boldsymbol{\bmv}_{2t}^{\sigma}\right\|^2}\mathbf{\Gamma}\Big)\\
  	=&\mathbb{E}\Big(\frac{\left(\boldsymbol{\varepsilon}_{2s-1}^{\sigma}-\boldsymbol{\varepsilon}_{2s}^{\sigma}\right)^{\top}}{\left\|\boldsymbol{\bmv}_{2s-1}^{\sigma}-\boldsymbol{\bmv}_{2s}^{\sigma}\right\|}\mathbf{\Gamma}\mathbf{\Gamma}^{\top}\frac{\left(\boldsymbol{\varepsilon}_{2t-1}^{\sigma}-\boldsymbol{\varepsilon}_{2t}^{\sigma}\right)}{\left\|\boldsymbol{\bmv}_{2t-1}^{\sigma}-\boldsymbol{\bmv}_{2t}^{\sigma}\right\|}\Big)^2.
  \end{align*}
  Because $\left(\boldsymbol{\varepsilon}_{2s-1}^{\sigma}-\boldsymbol{\varepsilon}_{2s}^{\sigma}\right)/{\left\|\boldsymbol{\bmv}_{2s-1}^{\sigma}-\boldsymbol{\bmv}_{2s}^{\sigma}\right\|}$ and $\bm{Q}_s/\|\bm Q_s\|$ are equally distributed, where for $1\leq s\leq \bar{T}$, $\bm Q_s\sim N(\bm 0,\bms_{\varepsilon})$ and $\bms_{\varepsilon}=\mathbf{\Gamma}\mathbf{\Gamma}^{\top}+\mathbf{A}\mathbf{A}^{\top}$, we have
  \begin{align*}
  	&\mathbb{E}\Big(\frac{\left(\boldsymbol{\varepsilon}_{2s-1}^{\sigma}-\boldsymbol{\varepsilon}_{2s}^{\sigma}\right)^{\top}}{\left\|\boldsymbol{\bmv}_{2s-1}^{\sigma}-\boldsymbol{\bmv}_{2s}^{\sigma}\right\|}\mathbf{\Gamma}\mathbf{\Gamma}^{\top}\frac{\left(\boldsymbol{\varepsilon}_{2t-1}^{\sigma}-\boldsymbol{\varepsilon}_{2t}^{\sigma}\right)}{\left\|\boldsymbol{\bmv}_{2t-1}^{\sigma}-\boldsymbol{\bmv}_{2t}^{\sigma}\right\|}\Big)^2\\
  	=&\mathbb{E}\Big(\frac{\bm Q_s^{\top}\mathbf{\Gamma}\mathbf{\Gamma}^{\top}\bm{Q}_t\bm{Q}_t^{\top}\mathbf{\Gamma}\mathbf{\Gamma}^{\top}\bm Q_s}{\|\bm Q_s\|^2\|\bm Q_t\|^2}\Big)\\
  	=&O\left(\frac{\tr(\mathbf{\Gamma}^{\top}\bms_{\varepsilon}\mathbf{\Gamma}\mathbf{\Gamma}^{\top}\bms_{\varepsilon}\mathbf{\Gamma})}{\tr^2(\bms_{\varepsilon})}\right)=O(N^2).
  \end{align*}
  Hence, we can conclude that $\|\mathbf{\Gamma}^{\top}\mathbf{J}_1\mathbf{\Gamma}\|_F^2=O(N^2T^{-1})$. Similarly, we can obtain that $\|\mathbf{\Gamma}^{\top}\mathbf{J}_1\|_F^2=O(NT^{-1})$, and  $\|N^{-1}\mathbf{\Gamma}^{\top}\mathbf{J}_2\widehat{\mathbf{\Gamma}}\widehat{\boldsymbol{\Lambda}}^{-1}\|_F^2=O_p(T^{-1})$ by \eqref{J2}.
  Hence, we can obtain that 
  \begin{align*}
  	\left\|\frac{1}{N} \widehat{\mathbf{\Gamma}}^{\top}(\widehat{\mathbf{\Gamma}}-\mathbf{\Gamma} \widehat{\mathbf{H}})\right\|_F^2 & =O_p\left(\frac{1}{T}+\frac{1}{N^2}\right) 
  \end{align*}
  Meanwhile, for any $t \leq T$, by Lemma A.2 in \cite{he2022large}, we have
  $
  \left\|\boldsymbol{W}_t\right\|^2=O_p(1).
  $
  We next consider $N^{-1}(\widehat{\mathbf{\Gamma}}-\mathbf{\Gamma} \widehat{\mathbf{H}})^{\top} \boldsymbol{\eta}_t$ and have 
  \begin{align*}
  	&\frac{1}{N}(\widehat{\mathbf{\Gamma}}-\mathbf{\Gamma} \widehat{\mathbf{H}})^{\top} \boldsymbol{\eta}_t\\
  	=&\frac{1}{N}(\hat{\boldsymbol{\Lambda}}^{-1})^{\top}\left(\widehat{\mathbf{\Gamma}}^{\top} \mathbf{\Gamma} \mathbf{M}_2^{\top} \boldsymbol{\eta}_t+\widehat{\mathbf{\Gamma}}^{\top} \mathbf{M}_3^{\top} \mathbf{\Gamma}^{\top} \boldsymbol{\eta}_t+\widehat{\mathbf{\Gamma}}^{\top} \mathbf{M}_4^{\top} \boldsymbol{\eta}_t+\widehat{\mathbf{\Gamma}}^{\top}\mathbf{J}_1\boldsymbol{\eta}_t
  	+\widehat{\mathbf{\Gamma}}^{\top}\mathbf{J}_2\boldsymbol{\eta}_t\right).
  \end{align*}
  By Lemma S6 in \cite{he2022large}, we can derive that 
  $$\Big\|\frac{1}{N}(\hat{\boldsymbol{\Lambda}}^{-1})^{\top}\left(\widehat{\mathbf{\Gamma}}^{\top} \mathbf{\Gamma} \mathbf{M}_2^{\top} \boldsymbol{\eta}_t+\widehat{\mathbf{\Gamma}}^{\top} \mathbf{M}_3^{\top} \mathbf{\Gamma}^{\top} \boldsymbol{\eta}_t+\widehat{\mathbf{\Gamma}}^{\top} \mathbf{M}_4^{\top} \boldsymbol{\eta}_t\right)\Big\|_F^2=O_p(T^{-2}+N^{-2}).$$
  By Lemma A.2 in \cite{he2022large}, for any $t\leq T$, $\|\bm\eta_t\|^2=O_p(N)$.
  Then, we have \begin{align*}
  	\|\widehat{\mathbf{\Gamma}}^{\top}\mathbf{J}_1\boldsymbol{\eta}_t\|_F^2=&\tr(\widehat{\mathbf{\Gamma}}^{\top}\mathbf{J}_1\boldsymbol{\eta}_t\boldsymbol{\eta}_t^{\top}\mathbf{J}_1\widehat{\mathbf{\Gamma}})\\
  	=&\boldsymbol{\eta}_t^{\top}\mathbf{J}_1\widehat{\mathbf{\Gamma}}\widehat{\mathbf{\Gamma}}^{\top}\mathbf{J}_1\boldsymbol{\eta}_t\\
  	\leq& \|\boldsymbol{\eta}_t\|^2\lambda_{\max}(\mathbf{J}_1\widehat{\mathbf{\Gamma}}\widehat{\mathbf{\Gamma}}^{\top}\mathbf{J}_1)\\
  	\leq& \|\boldsymbol{\eta}_t\|^2\lambda_{\max}(\mathbf{J}_1^2\widehat{\mathbf{\Gamma}}\widehat{\mathbf{\Gamma}}^{\top})\\
  	\leq& \|\boldsymbol{\eta}_t\|^2\lambda_{\max}(\mathbf{J}_1^2)\lambda_{\max}(\widehat{\mathbf{\Gamma}}\widehat{\mathbf{\Gamma}}^{\top})\\
  	\leq &\|\boldsymbol{\eta}_t\|^2\tr(\mathbf{J}_1^2)\lambda_{\max}(\widehat{\mathbf{\Gamma}}^{\top}\widehat{\mathbf{\Gamma}})=O(N^2)\|\mathbf{J}_1\|_F^2=O(N^2T^{-1}),
  \end{align*}
  where the last equality holds because \eqref{J1}. Similarly, due to \eqref{J2}, we also have $\|\widehat{\mathbf{\Gamma}}^{\top}\mathbf{J}_2\boldsymbol{\eta}_t\|_F^2=O_p(N^2 T^{-1})$.
  Hence, we can conclude that $$\Big\|\frac{1}{N}(\widehat{\mathbf{\Gamma}}-\mathbf{\Gamma} \widehat{\mathbf{H}})^{\top} \boldsymbol{\eta}_t\Big\|_F^2=O_p( T^{-1}).$$
  Due to the proof of Theorem 3.2 in \cite{he2022large}, we can derive that $\|N^{-1}\mathbf{\Gamma}^{\top}\bm \eta_t\|_{F}^2=O_p(N^{-1})$. Hence, we have 
  $\|\widehat{\boldsymbol{W}}_t-\widehat{\mathbf{H}}^{-1} \boldsymbol{W}_t\|^2\leq O_p(T^{-1}+N^{-1}) +\|N^{-1} \widehat{\mathbf{\Gamma}}^{\top} \vartheta_t\bm\alpha\|^2+\|N^{-1} \widehat{\mathbf{\Gamma}}^{\top} \bmv \f( \f\trans \f)^{-1}\f_{t}\|^2$. Obviously, $\|N^{-1} \widehat{\mathbf{\Gamma}}^{\top} \vartheta_t\bm\alpha\|^2=O_p(T^{-1})$ and $\|N^{-1} \widehat{\mathbf{\Gamma}}^{\top} \bmv \f( \f\trans \f)^{-1}\f_{t}\|^2=N^{-1}\|\bmv \f( \f\trans \f)^{-1}\f_{t}\|^2=O(T^{-1})$ by Theorem 1 in \cite{zhao2024}. Finally, by Cauchy-Schwartz inequality, we have $\|\hat{\mathbf{H}}\hat{\bm W}_t-\bm{W}_t\|^2=O_p(T^{-1}+N^{-1}).$ 
  \hfill$\Box$
  \subsection{Proof of Lemma \ref{maxdeltagammaH}}
  \begin{lemma}\label{lemtr}
  	Let $\mathbf{A}$ and $\mathbf{B}$ be the non-negative definite matrices, ${\rm tr}(\mathbf{A}\mathbf{B})\leq \lambda_{\max}(\mathbf{A}){\rm tr}(\mathbf{B})$.
  \end{lemma}
Let $\bm e_i\in \mathbb{R}^{N}$ be the unit vector with the $i$-th element equal to 1 and all other elements equal to 0, for $1\leq i\leq N$, and $\bm l_k\in\mathbb{R}^r$ be the unit vector with the $k$-th element equal to 1 and all other elements equal to 0, for $1\leq k\leq r$.
  Recall that \eqref{A.1}, we have 
  \begin{align*}
  	\hat{\bm \gamma}_i-\bm{\gamma}_i^{T}\hat{\H}=&\bm e_i^{\top}\mathbf{M}_2\mathbf{\Gamma}^{\top}\hat{\mathbf\Gamma}\hat{\mathbf\Lambda}^{-1}+\bm e_i^{\top}\mathbf{\Gamma}\mathbf{M}_3\hat{\mathbf\Gamma}\hat{\mathbf{\Lambda}}^{-1}+\bm e_i^{\top}\mathbf{M}_4\hat{\mathbf{\Gamma}}\hat{\mathbf{\Lambda}}^{-1}\\
  	&+\bm e_i^{\top}\frac{2}{T(T-1)} \sum_{1 \leq t<s \leq T}\frac{\left(\boldsymbol{\varepsilon}_t-\boldsymbol{\varepsilon}_s\right)\left(\boldsymbol{\varepsilon}_t-\boldsymbol{\varepsilon}_s\right)^{\top}}{\left\|\boldsymbol{\bmv}_t-\boldsymbol{\bmv}_s\right\|^2}\Big( \frac{\left\|\boldsymbol{\bmv}_t-\boldsymbol{\bmv}_s\right\|^2}{\left\|\boldsymbol{Z}_t-\boldsymbol{Z}_s\right\|^2}-1\Big)\hat{\mathbf{\Gamma}}\hat{\mathbf{\Lambda}}^{-1}\\
  	&+\bm e_i^{\top}\frac{2}{T(T-1)} \sum_{1 \leq t<s \leq T} \frac{\left(\boldsymbol{Z}_t-\boldsymbol{Z}_s\right)\left(\boldsymbol{Z}_t-\boldsymbol{Z}_s\right)^{\top}-\left(\boldsymbol{\varepsilon}_t-\boldsymbol{\varepsilon}_s\right)\left(\boldsymbol{\varepsilon}_t-\boldsymbol{\varepsilon}_s\right)^{\top}}{\left\|\boldsymbol{\bmv}_t-\boldsymbol{\bmv}_s\right\|^2} \frac{\left\|\boldsymbol{\bmv}_t-\boldsymbol{\bmv}_s\right\|^2}{\left\|\boldsymbol{Z}_t-\boldsymbol{Z}_s\right\|^2}\hat{\mathbf{\Gamma}}\hat{\mathbf{\Lambda}}^{-1}.
  \end{align*}
  
 We will consider $\max_{i}\|\bm e_i^{\top}\mathbf{M}_2\|^2.$
  Without loss of generality, we assume that $T$ is even and $\bar{T}=T / 2$. For any permutation of $\{1, \ldots, T\}$, denoted as $\sigma$, let $\boldsymbol{W}_t^\sigma, \boldsymbol{\eta}_t^\sigma$ and $\bmv_t^\sigma$ be the rearranged factors, random errors and error terms, further define
  $$
  \mathbf{M}_2^\sigma=\frac{1}{\bar{T}} \sum_{s=1}^{\bar{T}} \frac{\left(\boldsymbol{\eta}_{2 s-1}^\sigma-\boldsymbol{\eta}_{2 s}^\sigma\right)\left(\boldsymbol{W}_{2 s-1}^\sigma-\boldsymbol{W}_{2 s}^\sigma\right)^{\top}}{\left\|\bmv_{2 s-1}^\sigma-\bmv_{2 s}^\sigma\right\|^2},
  $$
  where $\mathcal{S}_{T}$ represents the set containing all the permutations of $\{1, \ldots, T\}$, then it's easy to prove that
  $$
  \sum_{\sigma \in \mathcal{S}_{T}} \frac{T}{2} \mathbf{M}_2^\sigma=T \times(T-2)!\times \frac{T(T-1)}{2} \mathbf{M}_2.
  $$
  Hence, for any $\sigma$, we have
  $$
  \mathbf{M}_2=\frac{1}{T!} \sum_{\sigma \in \mathcal{S}_{T}} \mathbf{M}_2^\sigma \quad \Rightarrow \quad \max_{i}\left\|\bm e_i^{\top}\mathbf{M}_2\right\| \leq \frac{1}{T!} \sum_{\sigma \in \mathcal{S}_{T}} \max_{i}\left\|\bm e_i^{\top}\mathbf{M}_2^\sigma\right\|\leq \max_{i}\left\|\bm e_i^{\top}\mathbf{M}_2^\sigma\right\| .
  $$
  Given $\sigma$, by the property of elliptical distribution, for any $s=1, \ldots, \bar{T}$,
  $$
  \binom{\boldsymbol{W}_{2 s-1}-\boldsymbol{W}_{2 s}}{\boldsymbol{\eta}_{2 s-1}-\boldsymbol{\eta}_{2 s}} \stackrel{d}{=} \nu^{\prime}\left(\begin{array}{cc}
  	\mathbf{I}_r & \mathbf{0} \\
  	\mathbf{0} & \mathbf{A}
  \end{array}\right) \frac{\boldsymbol{g}}{\|\boldsymbol{g}\|},
  $$
  where $\nu^\prime$ is determined by $\nu$, $\boldsymbol{g} \sim \mathcal{N}_{r+N}(\mathbf{0}, \mathbf{I})$ and $\boldsymbol{g}$ is independent of $\nu^\prime $, $\mathbf{A A}^{\top}=\boldsymbol{\Sigma}_\eta$. Hence,
  $$
  \mathbf{X}_s:=\frac{\left(\boldsymbol{\eta}_{2 s-1}^\sigma-\boldsymbol{\eta}_{2 s}^\sigma\right)\left(\boldsymbol{W}_{2 s-1}^\sigma-\boldsymbol{W}_{2 s}^\sigma\right)^{\top}}{\left\|\bmv_{2 s-1}^\sigma-\bmv_{2 s}^\sigma\right\|^2} \stackrel{d}{=} \frac{\mathbf{A} \boldsymbol{g}_2 \boldsymbol{g}_1^{\top}}{\left\|\mathbf{\Gamma} \boldsymbol{g}_1+\mathbf{A} \boldsymbol{g}_2\right\|^2},
  $$
  where $\boldsymbol{g}_1$ is composed of the first $r$ entries of $\boldsymbol g$ and $\boldsymbol{ g}_2$ is composed of the left ones. Define $\mathbf{\Sigma}_{\epsilon}=\mathbf{\Gamma}^{\top}\mathbf{\Gamma}+\mathbf{\Sigma}_{\eta}$, 
  $$
  \binom{\boldsymbol{u}_1}{\boldsymbol{u}_2}=\left(\begin{array}{cc}
  	\mathbf{I}_r & -\mathbf{\Gamma}^{\top} \boldsymbol{\Sigma}_\epsilon^{-1} \\
  	\mathbf{0} & \mathbf{I}_N
  \end{array}\right)\left(\begin{array}{cc}
  	\mathbf{I}_r & \mathbf{0} \\
  	\mathbf{\Gamma} & \mathbf{A}
  \end{array}\right)\binom{\boldsymbol{g}_1}{\boldsymbol{g}_2} \sim \mathcal{N}\left(\mathbf{0},\left(\begin{array}{cc}
  	\boldsymbol{\Sigma}_{\boldsymbol{u}_1} & \mathbf{0} \\
  	\mathbf{0} & \boldsymbol{\Sigma}_\epsilon
  \end{array}\right)\right),
  $$
  where $\boldsymbol{\Sigma}_{\boldsymbol{u}_1}=\mathbf{I}_r-\mathbf{\Gamma}^{\top} \boldsymbol{\Sigma}_\epsilon^{-1} \mathbf{\Gamma}$, then $\boldsymbol{u}_1$ and $\boldsymbol{u}_2$ are independent and
  $$
  \binom{\boldsymbol{g}_1}{\boldsymbol{g}_2}=\left(\begin{array}{cc}
  	\mathbf{I}_r & \mathbf{0} \\
  	-\mathbf{A}^{-1} \mathbf{\Gamma} & \mathbf{A}^{-1}
  \end{array}\right)\left(\begin{array}{cc}
  	\mathbf{I}_r & \mathbf{\Gamma}^{\top} \boldsymbol{\Sigma}_\epsilon^{-1} \\
  	\mathbf{0} & \mathbf{I}_N
  \end{array}\right)\binom{\boldsymbol{u}_1}{\boldsymbol{u}_2}=\left(\begin{array}{cc}
  	\mathbf{I}_r & \mathbf{\Gamma}^{\top} \boldsymbol{\Sigma}_\epsilon^{-1} \\
  	-\mathbf{A}^{-1} \mathbf{\Gamma} & \mathbf{A}^{\top} \boldsymbol{\Sigma}_\epsilon^{-1}
  \end{array}\right)\binom{\boldsymbol{u}_1}{\boldsymbol{u}_2} .
  $$
  as a result,
  $$
   \mathbf{X}_s\stackrel{d}{=}\frac{\mathbf{A} \boldsymbol{g}_2 \boldsymbol{g}_1^{\top}}{\left\|\mathbf{\Gamma} \boldsymbol{g}_1+\mathbf{A} \boldsymbol{g}_2\right\|^2}=\frac{\left(-\mathbf{\Gamma} \boldsymbol{u}_1+\mathbf{A} \mathbf{A}^{\top} \boldsymbol{\Sigma}_\epsilon^{-1} \boldsymbol{u}_2\right)\left(\boldsymbol{u}_1+\mathbf{\Gamma}^{\top} \boldsymbol{\Sigma}_\epsilon^{-1} \boldsymbol{u}_2\right)^{\top}}{\left\|\boldsymbol{u}_2\right\|^2},
  $$
  because $\boldsymbol{u}_1$ and $\boldsymbol{u}_2$ are zero-mean independent Gaussian vectors, we have
  $$
  \mathbb{E} \frac{\boldsymbol{u}_2 \boldsymbol{u}_1^{\top}}{\left\|\boldsymbol{u}_2\right\|^2}=\mathbf{0}, \quad \mathbb{E}\left\|\boldsymbol{u}_2\right\|^{-2} \leq \frac{1}{\lambda_N\left(\boldsymbol{\Sigma}_\epsilon\right)} \mathbb{E} \frac{1}{\chi_N^2} \asymp N^{-1},
  $$
  where $\chi_N^2$ is a Chi-square random variable with degree $N$. Moreover, we have 
  $\mathbb{E}(\bm e_i^{\top}\mathbf{X}_s\bm l_{m})=-\mathbb{E}(\|\bm u_2\|^{-2})\mathbb{E}(\bm e_i^{\top}\mathbf{\Gamma}\bm u_1\bm u_1^{\top}\bm l_m)+\mathbb{E}(\|\bm u_2\|^{-2}\bm e_i^{\top}\mathbf{A} \mathbf{A}^{\top} \boldsymbol{\Sigma}_\epsilon^{-1}\boldsymbol{u}_2 \boldsymbol{u}_2^{\top}
  \boldsymbol{\Sigma}_\epsilon^{-1}  \mathbf{\Gamma}\bm l_m   )$. According to Lemma S1 in \cite{he2022large}, we have $\|\boldsymbol{\Sigma}_{\boldsymbol{u}_1}\|_F^2=O(N^{-2})$ and $\|\mathbf{\Gamma}^{\top}\mathbf{\Sigma}_{\epsilon}\mathbf{\Gamma}\|_F^2=O(N^{-2})$. Hence, we can derive that $\{\mathbb{E}(\bm e_i^{\top}\mathbf{X}_s\bm l_{m})\}^2=\mathbb{E}^2(\|\bm u_2\|^{-2})(\bm e_i^{\top}\mathbf{\Gamma}\boldsymbol{\Sigma}_{\boldsymbol{u}_1}\bm l_m)^2+\mathbb{E}^2(\|\boldsymbol{u}_2\|^{-2}\bm e_i^{\top}\mathbf{A} \mathbf{A}^{\top} \boldsymbol{\Sigma}_\epsilon^{-1}\boldsymbol{u}_2\boldsymbol{u}_2^{\top}
  \boldsymbol{\Sigma}_\epsilon^{-1}  \mathbf{\Gamma}\bm l_m  )\leq O(N^{-2}\|\mathbf{\Gamma}\boldsymbol{\Sigma}_{\boldsymbol{u}_1}\|^2)+\{\bm e_i^{\top}\mathbf{A} \mathbf{A}^{\top} \boldsymbol{\Sigma}_\epsilon^{-1}\mathbb{E}(\boldsymbol{u}_2\boldsymbol{u}_2^{\top}/\|\boldsymbol{u}_2\|^{2})
  	\boldsymbol{\Sigma}_\epsilon^{-1}  \mathbf{\Gamma}\bm l_m \}^2=O(N^{-3}),$
 where the last inequality holds because 
 $$
 \mathbb{E}(\boldsymbol{u}_2\boldsymbol{u}_2^{\top}/\|\boldsymbol{u}_2\|^{2})\leq \frac{1}{\lambda_N(\mathbf{\Sigma}_{\epsilon})}\frac{\boldsymbol{\Sigma}_\epsilon^{1/2}\bm g \bm g^{\top}\boldsymbol{\Sigma}_\epsilon^{1/2}}{\|\bm g\|^2}=\frac{\boldsymbol{\Sigma}_\epsilon}{\lambda_N(\mathbf{\Sigma}_{\epsilon})N},\,\,\bm g \sim{N}(\mathbf{0}, \mathbf{I}).
 $$
   Note that $\mathbf{X}_s$ and $\mathbf{X}_t$ are independently and identically distributed when $s\neq t$, and
  $	\|\bm e_i^{\top}\mathbf{M}_2^\sigma\|^2=\sum_{m=1}^{r}\big({\bar T}^{-1}\sum_{s=1}^{\bar{T}}\bm e_i^{\top}\mathbf{X}_s\bm l_{m}\big)^2.
$
  So, we have
  \begin{align*}
  	&\max_{i}\left\|\bm e_i^{\top}\mathbf{M}_2^\sigma\right\|^2\\
  	=&\max_{i}\sum_{m=1}^{r}\left(\frac{1}{\bar T}\sum_{s=1}^{\bar{T}}\bm e_i^{\top}\mathbf{X}_s\bm l_{m}\right)^2\\
  	\leq &O(r)\max_i\left(\frac{1}{\bar T}\sum_{s=1}^{\bar{T}}\bm e_i^{\top}\mathbf{X}_s\bm l_{m}\right)^2\\
  	\leq &O(r)\left(\max_i\left|\frac{1}{\bar T}\sum_{s=1}^{\bar{T}}\bm e_i^{\top}\mathbf{X}_s\bm l_{m}\right|\right)^2\\
  	\leq &O(r)\left(\max_i\left|\frac{1}{\bar T}\sum_{s=1}^{\bar{T}}\bm e_i^{\top}\mathbf{X}_s\bm l_{m}-\mathbb{E}(\bm e_i^{\top}\mathbf{X}_s\bm l_{m})\right|\right)^2+O(r)\left(\max_i\left|\mathbb{E}(\bm e_i^{\top}\mathbf{X}_s\bm l_{m})\right|\right)^2\\
  	\leq &O(r)\left(\max_i\left|\frac{1}{\bar T}\sum_{s=1}^{\bar{T}}\bm e_i^{\top}\mathbf{X}_s\bm l_{m}-\mathbb{E}(\bm e_i^{\top}\mathbf{X}_s\bm l_{m})\right|\right)^2+o_p(N^{-2}/\log N).
  \end{align*}
Similarly, we have \begin{align*}
  	&\var(\bm e_i^{\top}\mathbf{X}_s\bm l_{m})\\
  	\leq& \mathbb{E}\{(\bm e_i^{\top}\mathbf{X}_s\bm l_{m})^2\}\\
  	\leq &O(1)\{\mathbb{E}\|\bm u_2\|^{-4}(\bm e_i^{\top}\mathbf{\Gamma}\bm u_1\bm u_1^{\top}\bm l_m)^2+\|\bm u_2\|^{-4}(\bm e_i^{\top}\mathbf{A} \mathbf{A}^{\top} \boldsymbol{\Sigma}_\epsilon^{-1}\boldsymbol{u}_2 \boldsymbol{u}_2^{\top}
  	\boldsymbol{\Sigma}_\epsilon^{-1}  \mathbf{\Gamma}\bm l_m   )^2\\
  	&+\|\bm u_2\|^{-4}(\bm e_i^{\top}\mathbf{A} \mathbf{A}^{\top}\boldsymbol{\Sigma}_\epsilon^{-1}\boldsymbol{u}_2 \boldsymbol{u}_1^{\top})^2+\|\bm u_2\|^{-4}(\bm e_i^{\top}\mathbf{\Gamma}\bm u_1\boldsymbol{u}_2^{\top}\boldsymbol{\Sigma}_\epsilon^{-1}\mathbf{\Gamma}\bm l_m)^2\}\\
  	=&O(N^{-2}).
  	  \end{align*}
  By classical Cram\'er type large deviation results for independent random variables in \cite{Statu1966}, we have for any $\varepsilon>0$ and some constant $C>0$,
  \begin{align*}
  	&P\left(\left|N\sum_{s=1}^{\bar{T}}\big[\bm e_i^{\top}\mathbf{X}_s\bm l_{m}-\mathbb{E}(\bm e_i^{\top}\mathbf{X}_s\bm l_{m})\big]\right|>Cx{\bar T}^{1/2}\right)	\\
  	&P\left(\left|N\sum_{s=1}^{\bar{T}}\big[\bm e_i^{\top}\mathbf{X}_s\bm l_{m}-\mathbb{E}(\bm e_i^{\top}\mathbf{X}_s\bm l_{m})\big]\right|>x\sqrt{N^2\bar T\var(\bm e_i^{\top}\mathbf{X}_s\bm l_{m})}\right)\leq C\exp(-\frac{x^2}{2}(1-\varepsilon)),
  \end{align*}
  uniformly in $x \in [0,o({\bar T}^{1/2})]$. Taking $x=C\log N$, we have 
  \begin{align*}
  	\max_i\left|N\sum_{s=1}^{\bar{T}}\big[\bm e_i^{\top}\mathbf{X}_s\bm l_{m}-\mathbb{E}(\bm e_i^{\top}\mathbf{X}_s\bm l_{m})\big]\right|=O_p(\sqrt{\bar T}\log N).
  \end{align*}
  Hence, we can conclude that $\max_i\left|{\bar T}^{-1}\sum_{s=1}^{\bar{T}}\big[\bm e_i^{\top}\mathbf{X}_s\bm l_{m}-\mathbb{E}(\bm e_i^{\top}\mathbf{X}_s\bm l_{m})\big]\right|^2=O_p\{N^{-2}T^{-1}(\log N)^2\}$ 
and $\max_i\|\bm e_i^{\top}\mathbf{M}_2\mathbf{\Gamma}^{\top}\hat{\mathbf\Gamma}\hat{\mathbf\Lambda}^{-1}\|=o_p(1/\sqrt{{\rm{log}} N}).$
  Because $\mathbf{M}_3$ is $\mathbf{M}_2$'s transpose, we also can prove that $\max_i\|\bm e_i^{\top}\mathbf{\Gamma}\mathbf{M}_3\hat{\mathbf\Gamma}\hat{\mathbf{\Lambda}}^{-1}\|=o_p(1/\sqrt{\log N})$. Similarly, we can derive that 
  $\max_i\|\bm e_i^{\top}\mathbf{M}_4\mathbf{\Gamma}\|^2=O_p(N^{-1}T^{-1}\log N+N^{-1}\log N)$
  and $\max_i\|\bm e_i^{\top}\mathbf{M}_4\|^2=O_p(N^{-1}T^{-1}\log N+N^{-1}\log N).$
  Hence, we can derive that
  $$\max_i\left\|\bm e_i^{\top}\mathbf{M}_4 \hat{\mathbf{\Gamma}}\right\|^2 \lesssim \left\|\bm e_i^{\top}\mathbf{M}_4 \mathbf{\Gamma}\right\|^2\|\widehat{\mathbf{H}}\|_F^2+\left\|\bm e_i^{\top}\mathbf{M}_4\right\|^2 \times \|\widehat{\mathbf{L}}-\mathbf{L} \widehat{\mathbf{H}}\|_F^2=o_p(1/\log N).$$
 Then, we will focus on $$\bm e_i^{\top}\frac{2}{T(T-1)} \sum_{1 \leq t<s \leq T}\frac{\left(\boldsymbol{\varepsilon}_t-\boldsymbol{\varepsilon}_s\right)\left(\boldsymbol{\varepsilon}_t-\boldsymbol{\varepsilon}_s\right)^{\top}}{\left\|\boldsymbol{\bmv}_t-\boldsymbol{\bmv}_s\right\|^2}\Big( \frac{\left\|\boldsymbol{\bmv}_t-\boldsymbol{\bmv}_s\right\|^2}{\left\|\boldsymbol{Z}_t-\boldsymbol{Z}_s\right\|^2}-1\Big)\hat{\mathbf{\Gamma}}\hat{\mathbf{\Lambda}}^{-1}.$$
  We have 
  \begin{align*}
  	&\max_i\left\|\bm e_i^{\top}\frac{2}{T(T-1)} \sum_{1 \leq t<s \leq T}\frac{\left(\boldsymbol{\varepsilon}_t-\boldsymbol{\varepsilon}_s\right)\left(\boldsymbol{\varepsilon}_t-\boldsymbol{\varepsilon}_s\right)^{\top}}{\left\|\boldsymbol{\bmv}_t-\boldsymbol{\bmv}_s\right\|^2}\Big( \frac{\left\|\boldsymbol{\bmv}_t-\boldsymbol{\bmv}_s\right\|^2}{\left\|\boldsymbol{Z}_t-\boldsymbol{Z}_s\right\|^2}-1\Big)\right\|^2\\
  	\leq &\max_i\left\|\bm e_i^{\top}\frac{1}{\bar{T}} \sum_{s=1}^{\bar{T}}\frac{\left(\boldsymbol{\varepsilon}_{2s-1}^{\sigma}-\boldsymbol{\varepsilon}_{2s}^{\sigma}\right)\left(\boldsymbol{\varepsilon}_{2s-1}^{\sigma}-\boldsymbol{\varepsilon}_{2s}^{\sigma}\right)^{\top}}{\left\|\boldsymbol{\bmv}_{2s-1}^{\sigma}-\boldsymbol{\bmv}_{2s}^{\sigma}\right\|^2}\Big( \frac{\left\|\boldsymbol{\bmv}_{2s-1}^{\sigma}-\boldsymbol{\bmv}_{2s}^{\sigma}\right\|^2}{\left\|\boldsymbol{Z}_{2s-1}^{\sigma}-\boldsymbol{Z}_{2s}^{\sigma}\right\|^2}-1\Big)\right\|^2\\
  	\leq & \max_i\left\|\frac{\bm e_i^{\top}\left(\boldsymbol{\varepsilon}_{2s-1}^{\sigma}-\boldsymbol{\varepsilon}_{2s}^{\sigma}\right)\left(\boldsymbol{\varepsilon}_{2s-1}^{\sigma}-\boldsymbol{\varepsilon}_{2s}^{\sigma}\right)^{\top}}{\left\|\boldsymbol{\bmv}_{2s-1}^{\sigma}-\boldsymbol{\bmv}_{2s}^{\sigma}\right\|^2}\right\|^2\left| \frac{\left\|\boldsymbol{\bmv}_{2s-1}^{\sigma}-\boldsymbol{\bmv}_{2s}^{\sigma}\right\|^2}{\left\|\boldsymbol{Z}_{2s-1}^{\sigma}-\boldsymbol{Z}_{2s}^{\sigma}\right\|^2}-1\right|^2\\
  	=&O_p(T^{-1/2})\max_i\left\|\frac{\bm e_i^{\top}\left(\boldsymbol{\varepsilon}_{2s-1}^{\sigma}-\boldsymbol{\varepsilon}_{2s}^{\sigma}\right)\left(\boldsymbol{\varepsilon}_{2s-1}^{\sigma}-\boldsymbol{\varepsilon}_{2s}^{\sigma}\right)^{\top}}{\left\|\boldsymbol{\bmv}_{2s-1}^{\sigma}-\boldsymbol{\bmv}_{2s}^{\sigma}\right\|^2}\right\|^2\\
  	\stackrel{d}{=}&O_p(T^{-1/2})\max_i\frac{\bm e_i^{\top}\bm u_2\bm u_2^{\top }\bm e_i}{\left\|\bm u_2\right\|^2}=O_p\{N^{-1}T^{-1/2}(\log N)^2\},
  \end{align*}
  where the last equality holds because $\max_i\bm e_i^{\top}\mathbf{\Sigma}_{\epsilon}\bm e_i=O(1)$.
  Hence, we can obtain that $$\max_i\|\bm e_i^{\top}\frac{2}{T(T-1)} \sum_{1 \leq t<s \leq T}\frac{\left(\boldsymbol{\varepsilon}_t-\boldsymbol{\varepsilon}_s\right)\left(\boldsymbol{\varepsilon}_t-\boldsymbol{\varepsilon}_s\right)^{\top}}{\left\|\boldsymbol{\bmv}_t-\boldsymbol{\bmv}_s\right\|^2}\Big( \frac{\left\|\boldsymbol{\bmv}_t-\boldsymbol{\bmv}_s\right\|^2}{\left\|\boldsymbol{Z}_t-\boldsymbol{Z}_s\right\|^2}-1\Big)\hat{\mathbf{\Gamma}}\hat{\mathbf{\Lambda}}^{-1}\|^2=O_p\{T^{-1/2}(\log N)^2\}.$$
 Finally, we have
  \begin{align*}
  	&\max_i\left\|\bm e_i^{\top}\frac{2}{T(T-1)} \sum_{1 \leq t<s \leq T} \frac{\left(\boldsymbol{Z}_t-\boldsymbol{Z}_s\right)\left(\boldsymbol{Z}_t-\boldsymbol{Z}_s\right)^{\top}-\left(\boldsymbol{\varepsilon}_t-\boldsymbol{\varepsilon}_s\right)\left(\boldsymbol{\varepsilon}_t-\boldsymbol{\varepsilon}_s\right)^{\top}}{\left\|\boldsymbol{\bmv}_t-\boldsymbol{\bmv}_s\right\|^2} \frac{\left\|\boldsymbol{\bmv}_t-\boldsymbol{\bmv}_s\right\|^2}{\left\|\boldsymbol{Z}_t-\boldsymbol{Z}_s\right\|^2}\right\|^2\\
  	\leq &\max_{t<s}\frac{\left\|\boldsymbol{\bmv}_t-\boldsymbol{\bmv}_s\right\|^2}{\left\|\boldsymbol{Z}_t-\boldsymbol{Z}_s\right\|^2}^2\max_i\left\|\bm e_i^{\top} \frac{\left(\boldsymbol{Z}_t-\boldsymbol{Z}_s\right)\left(\boldsymbol{Z}_t-\boldsymbol{Z}_s\right)^{\top}-\left(\boldsymbol{\varepsilon}_t-\boldsymbol{\varepsilon}_s\right)\left(\boldsymbol{\varepsilon}_t-\boldsymbol{\varepsilon}_s\right)^{\top}}{\left\|\boldsymbol{\bmv}_t-\boldsymbol{\bmv}_s\right\|^2}\right. \\
  	&\left.-\frac{(\vartheta_{t}-\vartheta_{s})\bm\alpha(\bmv_{t}-\bmv_{s})^{\top}}{\left\|\boldsymbol{\bmv}_t-\boldsymbol{\bmv}_s\right\|^2}\right\|^2\\
  	&+\max_{i}\left\|\bm e_i^{\top}\frac{2}{T(T-1)} \sum_{1 \leq t<s \leq T} \frac{(\vartheta_{t}-\vartheta_{s})\bm\alpha(\bmv_{t}-\bmv_{s})^{\top}
  	}{\left\|\boldsymbol{\bmv}_t-\boldsymbol{\bmv}_s\right\|^2} \frac{\left\|\boldsymbol{\bmv}_t-\boldsymbol{\bmv}_s\right\|^2}{\left\|\boldsymbol{Z}_t-\boldsymbol{Z}_s\right\|^2}\right\|^2\\
  	\leq & O(1)\max_i\left\|\bm e_i^{\top} \frac{\left(\boldsymbol{Z}_t-\boldsymbol{Z}_s\right)\left(\boldsymbol{Z}_t-\boldsymbol{Z}_s\right)^{\top}-\left(\boldsymbol{\varepsilon}_t-\boldsymbol{\varepsilon}_s\right)\left(\boldsymbol{\varepsilon}_t-\boldsymbol{\varepsilon}_s\right)^{\top}-(\vartheta_{t}-\vartheta_{s})\bm\alpha(\bmv_{t}-\bmv_{s})^{\top}}{\left\|\boldsymbol{\bmv}_t-\boldsymbol{\bmv}_s\right\|^2} \right\|^2\\
  	&+\max_i (\bm e_i^{\top}\bm \alpha)^2\left\|\frac{1}{\bar{T}}\sum_{s=1}^{\bar{T}}\frac{(\vartheta_{2s-1}^{\sigma}-\vartheta_{2s}^{\sigma})(\bmv_{2s-1}^{\sigma}-\bmv_{2s}^{\sigma})^{\top}
  	}{\left\|\boldsymbol{\bmv}_{2s-1}^{\sigma}-\boldsymbol{\bmv}_{2s}^{\sigma}\right\|^2}\frac{\left\|\boldsymbol{\bmv}_{2s-1}^{\sigma}-\boldsymbol{\bmv}_{2s}^{\sigma}\right\|^2}{\left\|\boldsymbol{Z}_{2s-1}^{\sigma}-\boldsymbol{Z}_{2s}^{\sigma}\right\|^2}\right\|^2\\
  	\leq &O(1)\max_i\big\|m_{t,s}^{-2}(\vartheta_{t}-\vartheta_{s})^2\bm e_i^{\top}\bm\alpha\bm\alpha^{\top}+m_{t,s}^{-2}\bm e_i^{\top}\bmv \f( \f\trans \f)^{-1}(\f_{t}-\f_{s})(\f_{t}-\f_{s})^{\top}( \f\trans \f)^{-1} \f^{\top}\bmv^{\top}\n\\
  	&+m_{t,s}^{-2}(\vartheta_{t}-\vartheta_{s})\bm e_i^{\top}\bm\alpha(\bmv_{t}-\bmv_{s})^{\top}+m_{t,s}^{-2}\bm e_i^{\top}\bmv \f( \f\trans \f)^{-1}(\f_{t}-\f_{s})(\bmv_{t}-\bmv_{s})^{\top}\n\\
  	&+m_{t,s}^{-2}(\vartheta_{t}-\vartheta_{s})\bm e_i^{\top}\bmv \f( \f\trans \f)^{-1}(\f_{t}-\f_{s})\bm\alpha^{\top}\n\\
  	&+m_{t,s}^{-2}(\vartheta_{t}-\vartheta_{s})\bm e_i^{\top}(\bmv_{t}-\bmv_{s})\bm\alpha^{\top}+m_{t,s}^{-2}\bm e_i^{\top}(\bmv_{t}-\bmv_{s})(\f_{t}-\f_{s})^{\top}( \f\trans \f)^{-1} \f^{\top}\bmv^{\top}\n\\
  	&+m_{t,s}^{-2}(\vartheta_{t}-\vartheta_{s})\bm e_i^{\top}\bm\alpha(\f_{t}-\f_{s})^{\top}( \f\trans \f)^{-1} \f^{\top}\bmv^{\top}\big\|^2+O(1)\left\|\frac{1}{\bar{T}}\sum_{s=1}^{\bar{T}}\frac{(\vartheta_{2s-1}^{\sigma}-\vartheta_{2s}^{\sigma})(\bmv_{2s-1}^{\sigma}-\bmv_{2s}^{\sigma})^{\top}
  	}{\left\|\boldsymbol{\bmv}_{2s-1}^{\sigma}-\boldsymbol{\bmv}_{2s}^{\sigma}\right\|^2}\right\|^2\\
  	=&o_p(N^{-1}/\log N)
  \end{align*}
  
  Hence, we can conclude that $\max_i	\|\hat{\bm \gamma}_i-\bm{\gamma}_i^{T}\hat{\H}\|=o_p(1/\sqrt{\log N})$.
  
  \hfill$\Box$

\end{document}